# A shear cell study on oral and inhalation grade lactose powders

G. Cavalli[1], R. Bosi[2], A. Ghiretti[2], C. Cottini[2], A. Benassi[2,3*] and R. Gaspari[2*]

1 - Early product development, Chiesi Limited, Chippenham, Wiltshire (UK)

2 - DP Manufacturing & Innovation, Chiesi Farmaceutici SpA, Parma (Italy)

3 - International School for Advanced Studies (SISSA), Trieste (Italy)

[*]Corresponding author address: Chiesi Farmaceutici S.p.A. Largo Belloli 11A– *43122 Parma (Italy)*

**Declarations of interest:** The authors declare the absence of any conflict of interest, including any financial, personal or other relationships with other people or organizations that could inappropriately influence, or be perceived to influence, their work.

**Keywords:** powder rheology; shear cell; lactose; stick-slip; particle morphology

## Abstract

Shear cell tests have been conducted on twenty different lactose powders, most of which commercially available for oral or inhalation purposes, spanning a wide range of particle sizes, particle morphologies, production processes. The aims of the investigation were: i) to verify the reliability of the technique in evaluating and classifying the flowability of powders; ii) to understand the connection between the flowability of a powder and the morphological properties of its particles; iii) to find a general mathematical relationship able to predict the yield locus shape given the particle size, shape and consolidation state of a lactose powder. These aspects and their limitations are detailed in the manuscript together with other interesting findings on the stick-slip behavior observed in most of the lactose powders examined.

## 1. Introduction

Flowability is a key property of powders and its understanding and characterization are crucial for several manufacturing processes in industry sectors such as food, pharmaceuticals and cosmetics. The flow properties of powders can be quantified by a number of rheological methods, the shear cell test standing out for its capability to provide a very rich physical information as well as a direct connection to equipment design [1–5]. Therefore the shear cell test has been applied on several materials including minerals [6,7], metal oxides [8] as well as food [9,10] and pharmaceutical powders [11–13]. The response of the powders to the test depends on the size and shape [11,14] of their particles and can be history dependent being influenced by aspects such as powder storage condition and consolidation [15,16] or by the preparation of the powder sample itself [17]. In this regard, lactose powders offer a very intriguing case study as they are produced for food and pharma industries using a wide range of manufacturing processes. Lactose powders can be produced by direct precipitation [18,19] their particles being tomahawk-like shaped single crystals or may be obtained by spray drying [20], their particles taking a rather spherical shape. Larger composite structures are created by granulation processes [21,22] or by fast crystallization, yielding for instance granules with multiple fused tomahawk single crystals. Particle size reduction can be obtained by several milling processes down to submicron scale [23,24], while a size range selection can be pursued by sieving [25].

Examples of shear test methods applied to lactose powders can be found in previous literature [14,15,26–29]. Shear testing was performed to address the properties of lactose at low consolidation stresses [28] and the influence of granule shape and size effects on flowability [14].

Both aspects are relevant since the low consolidation regime is typical of many industrial applications, such as manufacturing and storage of small batches and handling of small powder dosages, while the granule size and shape can be engineered to improve flowability and tune other physical properties functional to process and product quality, safety and efficacy [30,31].

In this work we focused on the shear cell testing of 17 different commercial lactose powders, they are listed in Table and grouped according to their morphology. Single tomahawk, multiple tomahawk and spherical-like lactose powders, ranging from micronized to coarse particle size were considered. Shear cell experiments were first conducted in the low consolidation regime, using the FT4 powder rheometer [32].

The aim of this analysis is to provide a comprehensive picture of the response of lactose powders to the shear test method, focusing on the relationship between the yield locus parameters and the morphological features of the powder particles, such as shape and size. While a similar study is available [14], the number of lactose grades analyzed in this paper allows an extensive study of trends in the flowability parameters within the distinct geometrical families.

A special focus is devoted to the development of a shear cell test method for low consolidation stresses, aiming to reduce the variability of the yield locus parameters for a given powder. Such low variability is pivotal to detect significant differences in the rheological behavior of powders with similar size but different morphology or, conversely, similar morphology but slightly different size. One class of powders that are often considered not adequate for shear cell testing are the coarser, free-flowing grades. For these powders, very large variability of yield locus parameters have been found [33], yielding also non-physical results such as negative cohesion values [8,34]. It has been suggested [33] that the use of a larger number of points in the yield locus determination could be helpful to increase the precision of the parameter estimation. Such solution has been evaluated in the present work, showing how the number of shear-to-failure experiments at a given critical consolidation state impacts on the result variability. The possibility to obtain precise estimation of powder flowability for the coarser powders is discussed.

We also provide an insight on general rules to predict the yield locus of powders based on the physical properties of the granules and of the critical consolidation conditions. Universal relationships between density, pre-shear stress, particle diameter and cohesion were suggested in previous literature [13,15,28,35]. Here we apply these relationships and provide a rationale for modifications that could be further generalize the approach. Furthermore, we provide power laws for the dependency of the yield locus parameters on the geometrical properties of the powder granules and discuss the non-monotonous relationship between internal friction and granule size already reported for limestone powders in previous works [13].

The response of the powders is not only characterized in terms of the yield locus parameters, we also analyzed the shear stress profiles during a shear-to-failure experiments. Previous works have shown the presence of stick-slip behavior in food powders, such as diary [36], wheat flour [37], maltodextrin [38] and starch [10] powders. Stick-slip phenomena are highly sensitive to external environmental conditions [36,38–40] and shear cell parameters [37] and offer additional tools for powder characterization. For instance the slope of the slip-stick amplitude vs normal load may be used to identify powders [40] besides the more commonly measured quantities such as the bulk internal friction or the extrapolated tensile strength. A study of the stick-slip phenomenon in lactose powders and its dependence on the particle morphology is included in the last part of the present paper.

## 2. Materials and Methods

### 2.1. Materials

17 different lactose powders for inhalation and oral use, from different suppliers, produced with different techniques (milling, wet sieving, granulation, direct precipitation, spray-drying) have been analyzed in this work, each of them has been identified by a number code as illustrated in Table 1. The powders have been grouped into four different morphology categories: micronized powders, single tomahawk particles, fused tomahawk particles and spherical particles. Except for three samples (Pharmatose DCL11, Capsulac 60 and Pharmatose 50M), which have been sieved to reach the desired particle size distribution and poly-dispersion degree using a sieve shaker AS200 (Retsch, Germany), all other powders were characterized in terms of rheological properties, particle size distribution and morphology as received by the supplier.

### 2.2. Methods

#### 2.2.1. Particle size characterization

The particle size distribution was measured by laser diffraction using a Sympatec Helos BR equipped with Rodos dry powder module combined with Aspiros dispersion unit. A sampling rate of 30 mm/s and a dispersion pressure of 0.5 bar were used. Each powder was tested in duplicate using lens R1 for InhaLac 500, Lactohale 300, Lacto-Sphere MM3 and lens R5 for all other samples. The full particle size distributions are collected in the Supplementary material, in the manuscript every powder is labelled according to the median volume equivalent diameter $dv_{50}$.

#### 2.2.2. Morphology characterization

The particle morphology was analysed by scanning electron microscopy (Phenom XL, Thermo Fisher Scientific) and optical microscopy (AZ100M Nikon, Japan). Some examples are shown in Figure 1. Before imaging, each SEM sample was sputter coated with gold for 180 seconds at 30 mA, SEM images have then been taken at 10kV with 10 Pa pressure.

#### 2.2.3. The shear test methodology

The shear cell equipment of the FT4 Powder rheometer (Freeman Technology Inc., UK) was employed [32,41], the powder samples being collected in a cylindrical glass vessel of 25 mm diameter. The testing procedure consists of four steps: conditioning, consolidation, pre-shear and shear-to-failure [42]. The powder was first poured into the glass vessel, during the conditioning step any inhomogeneities present were removed through a gentle downward and upward motion of a helix. In the consolidation step a piston applies a normal stress $\sigma_{pre}$, the powder bed is then released and split into a controlled volume so that, weighing it, the powder bed density can be measured. In the pre-shear phase a bladed piston is pressed on the specimen again with normal stress $\sigma_{pre}$ and kept in motion around its axis with angular velocity $\omega$ (6 degrees/min unless otherwise specified). The pre-shear phase ends when a steady state is reached and the measured shear stress $\tau$ becomes constant $\tau = \tau_{pre}$, in this condition (critical consolidation state), the powder flowability should not depend anymore on its previous mechanics history. The so-called pre-shear point is given by the couple of values ($\sigma_{pre}, \tau_{pre}$) and constitutes the upper bound of the measured yield locus curve. During the subsequent shear-to-failure measurements normal stresses $\sigma < \sigma_{pre}$ are applied to the sample, with the bladed head still rotating at the same angular velocity $\omega$, while measuring the shear stress $\tau$ exerted by the powder on the piston. After each shear-to-failure experiment the rotation is first stopped and then the system is brought again to critical consolidation at the pre-shearing normal load. At stresses smaller than $\sigma_{pre}$ the powder behavior is different from the pre-shear phase: the shear stress $\tau$ will initially increase as the powder bed deforms rigidly, up to the nucleation of a shear band, to drop down again while the powder bed start shearing. Each yield locus point corresponds to the maximum shear stress recorded and the corresponding applied normal stress. After the failure the pre-shear procedure must be repeated before the next point can be acquired. In the present work we employed two different testing procedures featuring 5 or 7 yield locus points, stressing the powder specimens at $\sigma_{pre}$ = 1,3,6,9 and 15 kPa and $\omega$ = 6, 12 and 18 degrees/min.

## 3. Results

### 3.1. Warren-Spring equation for the yield locus fitting

Few examples of the measured yield loci are shown in Figure 2, different data collected for the same powder at increasing pre-shear stresses are presented in panel (a) while the response of different powders prepared at the same pre-shear stress of 1 KPa are collected in panel (b).

The general Warren-Spring form for the yield locus was used to fit experimental data for all powder investigated [43,44],

$$\left(\frac{\tau}{C}\right)^n = \frac{(\sigma + T)}{T} \quad (1)$$

where $\tau$ is the shear stress, $\sigma$ is the normal stress, $C$ is the cohesion, $T$ is the tensile strength and $n$ is the shear index of the powder. In the linear approximation, i.e. when $n = 1$, the angle of internal friction ($\varphi$) can be defined as

$$\varphi = arctan(C/T) \quad (2).$$

The variability of the shear index *n* is larger for cohesive powders and a clearer picture can be obtained by plotting the average values $< n >$, over 3 distinct realizations (i.e performed in triplicate), as a function of the powder $dv_{50}$. As it can be evinced in Figure 3, the shear index $n$ sizably decreases when moving from micronized to coarse powders. This indicates that the fit based on the Warren-Spring functional form detects the larger curvature of the yield locus of cohesive powders. Such result is not evident by visual inspection as a linear fit often appears to well model the yield locus, see Figure 2. To quantify this, the $R^2$ values for the fit obtained with an optimized shear index *n* were compared to those relative to linear fit ($n = 1$). The data are shown in Table 3,

Table 4 and demonstrate that using a general Warren-Spring function allows a better fit to the experimental data in comparison to the linear fit, especially for cohesive powders. However, the linear fit provides also low $R^2$ values and the difference between the general and the linear approach vanishes for coarse powders. As a result, while the linear approach appears optimal for coarser powders, it does model well also cohesive powders, at least in the region of normal consolidation stresses considered in this work. Extrapolation to zero consolidation stress could be problematic using the linear approximation and leads to the well-known overestimation of the powder tensile strength. Such problem is however not necessarily sorted out using a general Warren-Spring model, as the larger number of unknowns ($C$, $T$ and $n$) may lead to data overfitting and unphysical values of the tensile strength [45,46]. Regarding the powders considered in this work, for instance, a general Warren-Spring model tends to equate the tensile strength of cohesive and free-flowing powders. A better fit with the Warren-Spring model could be obtained by measuring the tensile strength with a separate experiment [45]. Considering these limitations and the after all satisfactory quality of the linear model, the latter will be used throughout this study.

The raw data of normal and shear stresses for all the tests performed on all the powders are collected in the Supplementary material as well as the yield loci and their fittings.

### 3.2. Importance of the shear cell testing protocol

The variability of the yield locus parameters $C$ and $\varphi$ was tested as a function of the number of shear-to-failure measurement $n_p$ (either 3,5 or 7) acquired for each test on each single sample. Since we used a linear form to fit the experimental yield loci, we focused on free-flowing powders ($dv_{50} >$ 200 µm) whose yield locus was shown to be linear, within the range of normal stresses considered in this work, even in the general Warren-Spring approximation. An inspection of Table 5 shows that an increase of $n_p$ mostly corresponds to a decrease in the variability of the yield locus parameters. The root mean square (RSD) decreases mostly when increasing $n_p$ from 3 to 5 (on average by 30.7

% for *C* and 3.5 % for $\varphi$ values). On the other hand, increasing $n_p$ from 5 to 7 mostly improves the results on *C* by 7.3 % while the decrease of the variability of $\varphi$ is small (0.3 %). It can be argued that the residual variability on $\varphi$ cannot be further decreased by improving the construction of the yield locus and only depends on other factors, such as the powder sample size used for the measurements or instrumentational effects. Overall our analysis indicates that, for the setup given, performing a number of shear-to-failure experiments between 5 and 7 is recommended to obtain converged results on yield loci parameters, a value of 7 being advised for critical applications where the result variability must be minimized. It is worth noting that even for free-flowing powders all the cohesion values are well-defined and above 0, meaning that the shear cell method is able to provide precise results also on coarser grades.

### 3.3. Yield loci of the powders at 1 kPa pre-shear

Low consolidation stress is typically approached in small bins. As a reference value one could take the maximum normal stress given by the classical Janssen equation [2], using a bulk density typical of lactose powders (0.6 g/cm³), and common values for the Janssen constant and wall friction (0.4). With these values we obtain that the maximum normal load in a bin with diameter 0.1 m is roughly 1 kPa. Data reported in this chapter refer to analysis performed at 1 kPa pre-shear. Linear yield loci can be analyzed in terms of parameters *C, T* and $\varphi$, their behavior as a function of the $dv_{50}$ is presented in Figure 4 and Table 4. For both *C* and *T* a power law behavior in the form $\frac{\beta}{dv_{50}^{\alpha}} + \gamma$ was used to fit the experimental values with $\alpha, \beta$ and $\gamma$ fitting parameters. The fit was first performed on all grades, and then only on the family of single tomahawk powders, to analyze shape effects. In all cases the exponent $\alpha$ was significantly smaller than one, while the $\gamma$ parameter was vanishing. In the ideal case of mono-disperse perfectly spherical particles Rumpf predicted a value of $\alpha = 1$ [47], we thus tested also the simpler power law $\frac{\beta}{dv_{50}} + \gamma$. With this choice the $\gamma$ parameter is non-vanishing, however the $R^2$ decreases.

The difference between the theoretical value $\alpha = 1$ and the fitted value can be partly traced back to the difficulties in measuring the tensile strength, as this is obtained by an extrapolation of the yield loci to the negative region of normal stress. On the other hand, the family of single tomahawk lactose grades appear to have a similar behavior to the complete set of powders investigated, suggesting size as the property that mainly rules the cohesion and tensile strength parameters.

By inspection of the different classes of powders it appears that the clearest distinction occurs between micronized ($dv_{50}$ < 20 µm) and non-micronized powders. For micronized powders, $\varphi$ is always larger than 42° and C is above 0.3 kPa. Conversely, non-micronized powders correspond to an $\varphi$ below 42° and C smaller than 0.3 kPa.

One of the most notable effects on flow properties is given by sieving. Removing the fine particles has two clear consequences: a) the angle of internal friction increases, b) the cohesion parameter decreases. This can be observed in Table 7. This result compares with the finding of Shi et al. [13] that reports an increase of the angle of internal friction and typically an increase of bulk density for coarser Eskal powder grades. This result was suggested to be a consequence of the increased interlocking between the larger granules. However, at variance with their findings, we report the sieving does indeed affect the bulk friction of powders, at least for the lactose grades investigated in the present study. This suggests that the powder chemical composition may play a role in determining the onset of this effect. Additional data on flowability parameters related to stress circles are reported in Supplementary Information.

### 3.4. Higher pre-shear stresses and general fitting formula for the yield locus

As anticipated in the previous sections, tests at higher values of the pre-shear stress have also been performed up to a maximum of 15 kPa. The effect of increasing the powder compaction, and thus the average coordination number of the powder particles, is to make the specimen more resistant to the shear motion. At the same applied normal stress, a larger shear stress is required to provoke the shear failure and initiate the powder motion, i.e. the yield loci curves will shift upward in the $\sigma - \tau$ plane. This is clearly visible in Figure 2(a) as well as the increase of both the intercepts $C$ and $T$ with increasing $\sigma_{pre}$.

Exploring the possibility to define a general fitting formula, capable of predicting the yield locus behavior given the $dv_{50}$ and $\sigma_{pre}$ of any lactose powder sample, several forms for the $\tau(\sigma)$ function have been tested. Two of them proved to be particularly effective:

$$\tau(\sigma) = C(dv_{50})\left[1 + \frac{\sigma}{T(dv_{50})}\right] + \vartheta\, \sigma_{pre} \qquad (3)$$

$$\tau(\sigma) = C(dv_{50})\left[1 + \frac{\sigma}{T(dv_{50})}\right] + \left(\frac{\theta}{dv_{50}}\right)^\chi \sigma_{pre} \qquad (4)$$

where the dependence on $dv_{50}$ and $\sigma_{pre}$ appears in two separate terms, $\vartheta, \theta$ and $\chi$ are fitting parameter, $C$ and $T$ have the same shape previously introduced with $\alpha, \beta$ and $\gamma$ fitting parameters. Equation (3) is slightly better in fitting the low $\sigma_{pre}$ yield loci while, Figure 5 (a) shows the fits for the micronized and single tomahawk powders at 1 kPa pre-shear stress. Equation (4) improves the fitting of 9 and 15 kPa data, as shown for micronized and single tomahawk powders at 9 kPa in Figure 5 (b).

The fits have been performed with the software *R* version 3.6.1 with the package *minpack.lm* [48] implementing a non-linear regression fit adopting the Levenberg-Marquardt algorithm [49]. Fitting parameters and numerical details are given in the Supplementary Material.

*3.5. Stick-slip and its dependence on particle size and shape*

As originally demonstrated by Prandtl and Tomlinson [50] the stick-slip motion emerges when an elastic body is driven at constant velocity over another one and their interaction energy landscape is periodic, quasi-periodic or disordered, featuring many maxima and minima. When the driven system is in a local energy minimum the two sliding bodies are macroscopically at rest sticking to each other, although the driver is moving at constant velocity. Being sticking on one side and driven at the same time on the other side results in an elastic deformation of the bodies (or their sliding interface), the accumulation of elastic energy changes the energy landscape of the sliding system pushing it out of the local minimum and thus out of equilibrium, here is where the slip between the two bodies starts. The elastic energy is first converted into kinetic energy of the sliding bodies and must then be dissipated in some way so that they return at rest laying in a new local minimum of the energy landscape. This alternation of stick and slip events can be observed at the micro and nano-scales, where single micro or nano-contacts can be slide in a controlled way on clean substrates (using for instance an atomic force microscope or a surface force apparatus) [51], and at the macro-scale if a single rigid body is pulled by a spring (dynamometer) or if a confined bulk solid is sheared [2,52]. The physics generating the sticking can be very different ranging from the mechanical interlocking of single asperities in rough bodies sliding on each other to the adhesion forces between particles in a confined, sheared bulk solid. If a lubricant is strongly confined between two sliding bodies it can solidify sticking the system, slip occurs when its local elastic deformation promotes its temporary melting. The nature of the kinetic energy dissipation during the slip events is eventually always the same no matter if one deals with sliding micro-asperities, sliding particles and grains or sliding lubricant molecular layers: phonons and vibrational modes are excited distributing irreversibly the energy over an infinite number of atoms and molecules, i.e. promoting Joule heating.

In our powder rheometer the shear head rotates at constant radial velocity $\omega$ on the vertical axis of a cylindrical powder bed. As illustrated in Figure 6 (a) during the stick phase the upper part of the powder cylinder moves together with the rotating head while the bottom part remains at rest, in such condition the powder bed can be regarded as an elastic solid cylinder deforming under a torque. Before the slip event takes place, a shear band nucleates and expands in the powder bed locally decreasing the powder density. While the upper part of the powder bed still moves at constant radial velocity the bottom part slips back releasing the elastic stress in the form of particle kinetic energy, this kinetic energy is then dissipated by the particle-particle friction. The blades mounted on the rotating head ensure that the shear bend nucleates and expands inside the powder bed and not at the powder-head interface.

The typical fingerprint of stick-slip dynamics in our sheared powder beds is the appearance of a saw-tooth profile in both shear and normal stress curves measured as a function of time. An example is given by the red curve of Figure 6 (b) for the shear-to-failure stress curves plotted as a function of time for a coarse lactose at pre-shear of 1 KPa and normal stress 0.4 kPa. The slowly growing ramps in the shear stress curve are the stick phases, here the powder resists to internal motion, upon reaching a maximum shear stress the powder column breaks, i.e. the shear band is formed, and slip motion starts, this results in the quick drops in the stress curve. Few more considerations must be done:

- Both the shear and normal stresses fluctuate perfectly in phase during motion, they increase during stick phases and drop during slips, compare red and blue lines in Figure 6 (b). This is easily explained considering the powder dynamics along the direction normal to the shear plane. To promote the slip event and nucleate the shear band the powder must decrease locally its density (Reynolds dilatancy principle) to do this it must exert a larger counter-force normal to the shear head surface. This force is maximized at the switch point between stick and slip phases. During the slip events the particles energy is dissipated, and the shear band disappears recompacting the powder to a higher density, during this rearrangement the powder bed exerts a weaker force on the shear head plane.

- While in the shear-to-failure tests stick-slip has always been observed for all the powders in all shearing conditions, in the preliminary pre-shear part this is not always the case, it is not present for instance in the data collected on the fine Excipure DPI 211.

- The height variation of the powder bed during stick and slip events is smaller than the instrument sensitivity, thus it appears to be constant.

- Finally, as Figure 6 (b) shows, the duration of the stick phase grows in time during each sequence and this seems even more pronounced with micronized powder than free flowing ones. The cause of that might be the occurrence of changes in the microstructure of the powder (compaction, agglomeration) induced by the shearing motion. Anyway, the time scale of such phenomenon is much longer than the recording capability of the instrument and for this reason it is not possible to capture it until the actual steady state is reached and the stick-slip regularizes.

Dealing with a rotational shear motion, rather than a one-dimensional one, it is difficult to define a characteristic length scale for the stick-slip events. An estimation is still possible plotting the stress profiles as a function of the length of the arc swept by a point on the bladed piston boundary, such length can be calculated from the sliding time $t$ knowing the piston radius $r_0$

$$\ell = \omega t r_0 \qquad (6)$$

Panel (c) of Figure 6 shows two typical stick-slip profiles for a micronized and a coarse powder: the characteristic length decreases from 150 µm for the coarse case (comparable to the particle size) to 40 - 50 µm for the micronized one (much larger than the single particle size).

Two characteristic features of stick-slip dynamics, no matter its nature and scale, are: i) the increase in stress fluctuations, i.e. in the depth of the saw-tooth profile, as the normal stress increases; ii) the disappearance of the saw-tooth profile as the driving velocity increases significantly [36,37,40]. Both these features are present in our measurements, in Figure 6 (d) the stick-slip depth is clearly increasing with increasing normal load while panel (e) shows how the asymmetric (triangular) saw-tooth gradually transforms into a symmetric sinusoid increasing the driving velocity. Sliding friction is known to have a logarithmic dependence on the driving velocity [53] thus to observe a compete disappearance of the stick-slip the latter should increase by one order of magnitude rather than just a factor 3, however this is not possible due to the instrument limitations. A detailed analysis of the stick-slip depth as a function of the normal load is presented in the Supplementary material.

To conclude the analysis on the stick-slip behavior the stick and slip times have been plotted as a function of their $dv_{50}$. Figure 7 (a) shows, for every powder pre-consolidated at 1 kPa, the average stick time measured at σ = 0.7 kPa. Indeed, there is a clear non-trivial dependence of the stick time on the particle size and the curves seem to be quite sensitive to the particle shape as well. Interestingly, the slip time has a completely different behavior as illustrated in panel (a), it is completely insensitive to the $dv_{50}$. The slip time is a measure of how efficiently the shear band region dissipates the particle kinetic energy, for bulk solids this dissipation occurs via particle-particle and particle-wall friction. Two are the fundamental ingredients: i) the friction coefficient, which is a material parameter and is thus constant for us considering only lactose powders; ii) the average number of sliding asperities per particles, i.e. the coordination number. The latter is extremely sensitive to the powder $dv_{50}$, large particles have a better packing and thus a larger coordination number, fine cohesive powders have a small coordination number. If the slip time is the same regardless of the $dv_{50}$ it means that fine cohesive powders must compensate for the small coordination number by increasing the shear band thickness. Free flowing powders made of large particles can dissipate the same energy with thinner shear bands. This direct consequence of a constant slip time is in agreement with the experimental observations reporting shear bands 5 to 20 times larger than the $dv_{50}$ for coarse powders having $dv_{50}$ >100 µm and shear bands with thickness 200 times the $dv_{50}$ for micronized powders [2].

A detailed study of the stick and slip time as a function of the normal load is presented in the Supplementary material. The stick time is found to increase almost linearly with the normal load, a larger normal load compacts the powder bed allowing it to resist larger shear stresses before nucleating the shear band and slide. The slip time remains constant or decreases slightly with increasing normal load, the latter in fact tends to squeeze the particles limiting the life-time and the extension of the shear band.

As a matter of fact, shear cell testing requires that normal stresses are constant throughout the pre-shear or shear-to failure process [54]. In our case, normal stresses oscillate in phase with shear stresses as the result of a combination of powder response to the tester, powder-piston interactions, sample size and mechanical response of the instrumentation. For this reason, the testing instrumentation used in this work should be referred to as "powder rheometer" rather than "shear cell tester", this second name being usually employed for testing systems where constant normal stresses on the sample are enforced by appropriate weights. Finally, slip-stick evidence was found in the specific conditions tested in this work and we cannot exclude effects due to sample size, instrumentation, test parameters and environmental condition. These aspects may be investigated by further work on the rheological properties of lactose powders.

## 4. Conclusions

The work focused on the shear cell analysis of 21 lactose powders across a wide range of shape and size features.

We first developed a method for the shear cell test, by evaluating the variability of the flow parameters with respect to the number $n_p$ of shear-to-failure points in the yield locus. Construction of the yield locus with $n_p = 3,5,7$ showed a marked difference of the $\varphi$ and C variability, that is minimized by increasing $n_p$. Overall our analysis indicates that, for the setup given, performing a number of shear-to-failure experiments between 5 and 7 is recommended to obtain converged results on yield loci parameters, a value of 7 being advised for critical application where the result variability must be minimized.

Using the optimized shear cell method, we measured the flow parameters (T, C and $\varphi$) of all powders and their dependence on size and morphology. It turns out that size is by far the most relevant feature determining the flow behavior, while little influence of morphology is observed. The $T(dv_{50})$ and $C(dv_{50})$ functions could be fitted using a simple power law $\propto = \frac{\beta}{dv_{50}^{\alpha}}$. The exponent $\alpha$ was smaller than the theoretical value $\alpha = 1$ expected for T, possibly indicating that the linear approximation of the yield locus may underestimate the decrease of T over increasing $dv_{50}$.

Besides $dv_{50}$, the parameter with largest impact on powder flowability was the particle size polydispersion. It was observed that removal of the fine particle fraction by powder sieving dramatically increases the $\varphi$, possibly as a consequence of the enhanced interlocking between coarser particles. This effect is evident for powders with $dv_{50} > 100$ µm while for finer powders adhesion is expected to rule the flow properties.

Shear cell measurements were performed at different pre-shearing conditions. We thereby proposed an empirical model that could accurately describe the yield locus of all powders analyzed, with the only starting knowledge of $\sigma_{pre}$ and $dv_{50}$. The model is a first effort towards a strategy of flow parameters estimation for cases where a full experimental powder characterization is not possible or desired. The model is able to adequately represent lactose, un-sieved powders but it could be extended to different materials with varying size spans.

Finally, we analyzed the shear-to-failure patterns of all powders and found a stick-slip behavior for all lactose grades investigated. Our study generally reports that for increasing pre-shear stresses the stick-slip amplitude and stick times increase, while on the contrary the slip time tends to decrease. The dependence of the stick-slip parameters to the powder size, morphology and loading is not trivial and a true understanding of the powder behavior could be obtained by bridging the knowledge at the micro and macro scales. While our work mainly deals with a description of the powders at a macro scale level, insights on a particle-by-particle description of the system can be obtained by discrete element modeling simulations. This has already been undertaken in previous studies [55,56] and could as well represent a next step to unravel the non-trivial flow properties of lactose powders.

List of Figures

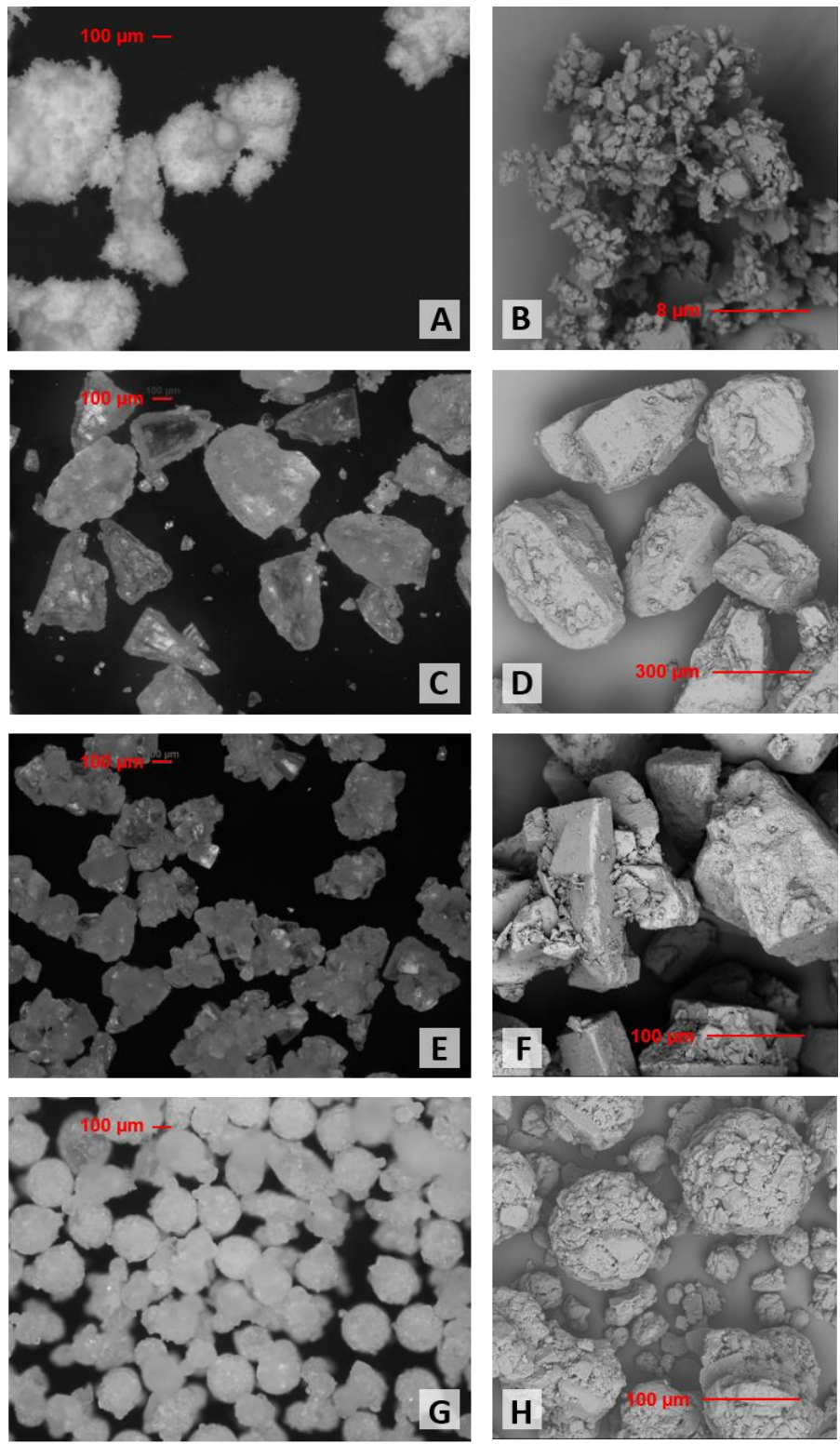

*Figure 1. Optical and SEM micrographies of different lactose samples. (a) and (b) Micronized Lacto-Sphere MM3, (c) and (d) single tomahawk particles of Pharmatose 50M, (e) and (f) fused tomahawk particles of Capsulac 60, (g) and (h) spherical spray-dried particles of Pharmatose DCL 11.*

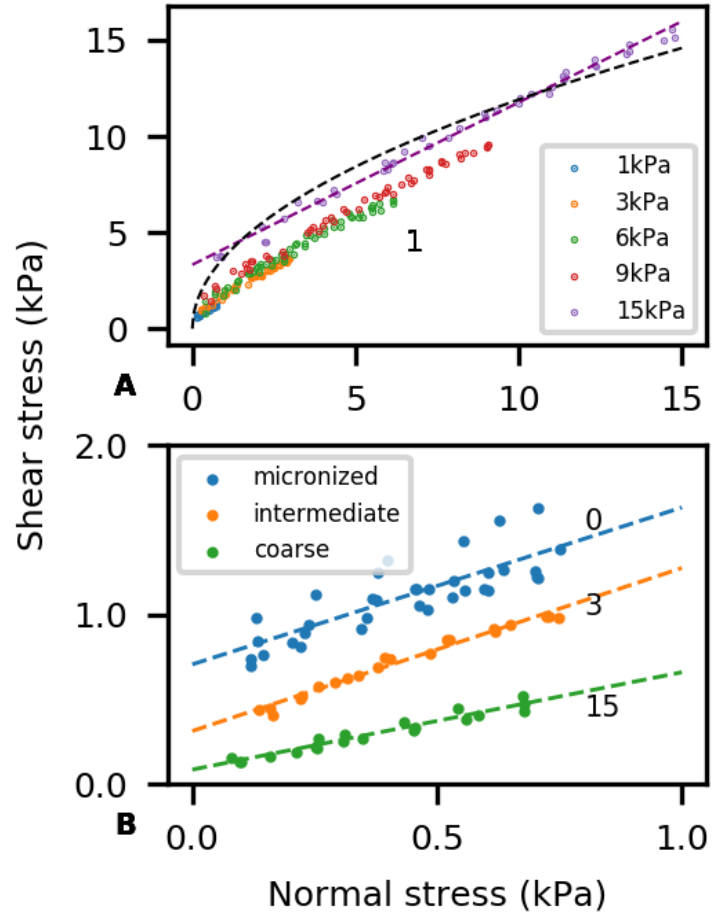

*Figure 2. Yield locus plots, i.e. normal versus shear stresses during shear to failure tests. (a) Lacto-Hale 300 yield loci at different pre-shear stresses (different colors), dots represent the measured stresses in three different tests; dashed black and purple lines show respectively the quadratic and linear fits described below. (b) Yield loci measured at 1 kPa pre-shear stress for powder having different particle size, Inhalac 500 (blue), Excipure 221 (orange) and Pharmatose 50M (green). Numbers close to lines label the different powders according to Table 1 and 2.*

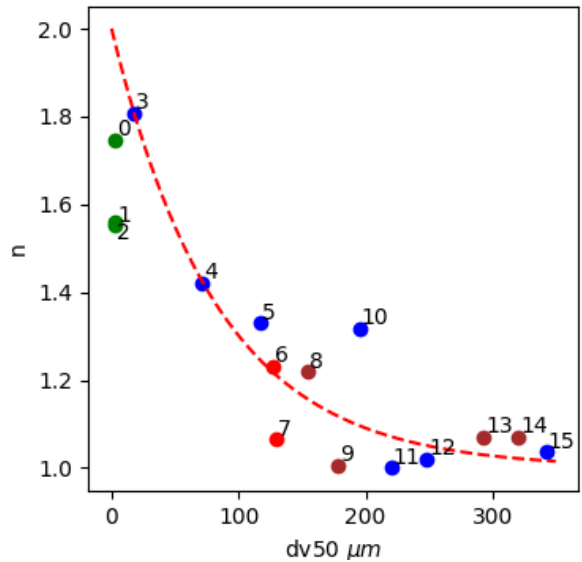

Figure 3. Shear index $n$ as a function of the particle size, numbers refer to powder grades as described in Table 1 and 2. The red curve represents a fitted exponential function of the form $n(dv_{50}) = 0.7\,exp(-\beta dv_{50}) + 1$, $\beta = 0.01$. The chosen exponential form ensures that $n$ is always bound between 1 and 2 and that it converges to 1 for a large $dv_{50}$ (linear limit for free-flowing powders). Data refer to 1 kPa pre-shear stress levels.

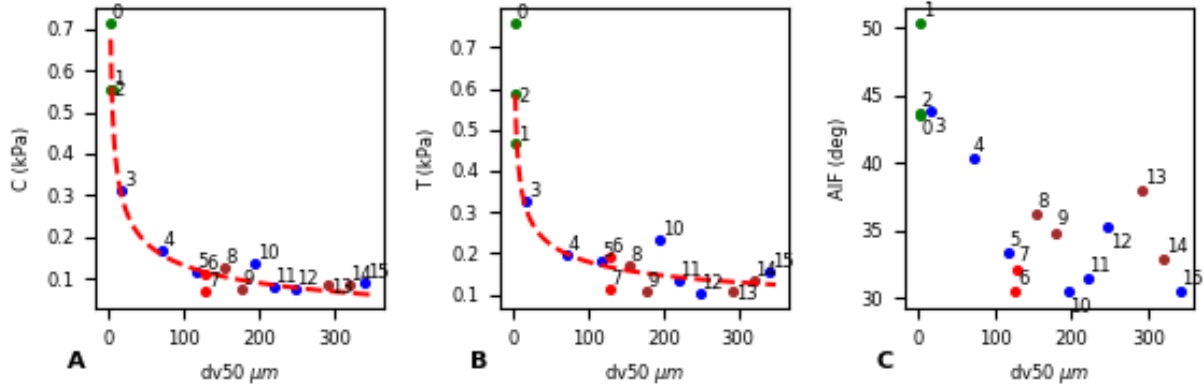

Figure 4. Cohesion parameter C (a) and Tensile strength T (b) has a function of $dv_{50}$, the red curves represent a fitting function of the form $\frac{\beta}{dv_{50}^{\alpha}} + \gamma$, the fitting parameters are listed in Table. (c) Angle of internal friction as a function of $dv_{50}$. Numbers close to points label the different powders according to Table and Table.

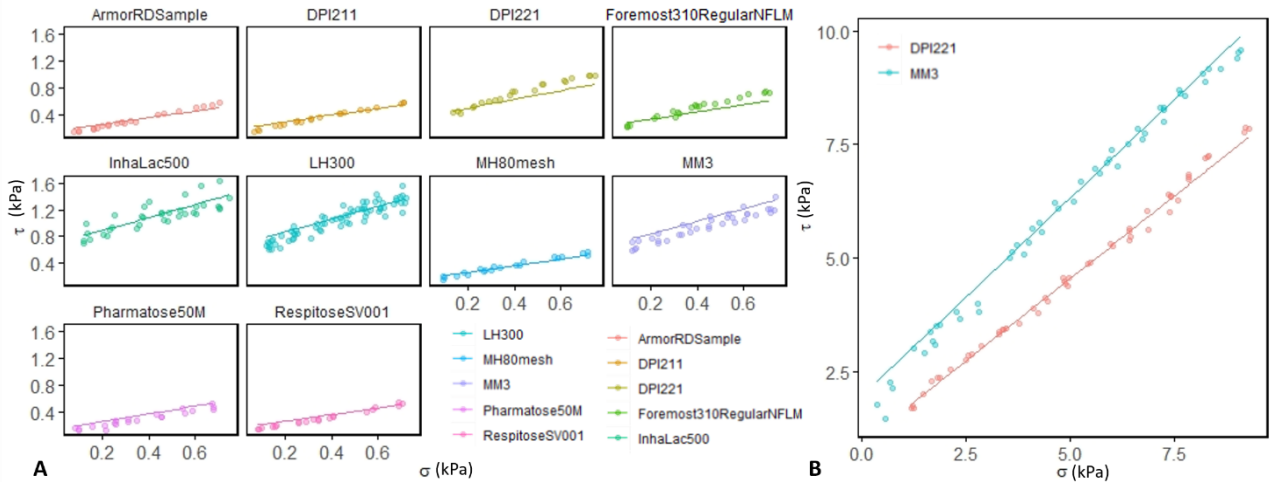

Figure 5. (a) fits of the yield loci for micronized and single tomahawk powders pre-consolidated at 1 kPa using equation (3), dots represent the experimental data of 3-4 different tests. (b) fits of the yield loci for micronized and single tomahawk powders pre-consolidated at 9 kPa using equation (4), dots represent the experimental data of 3-4 different tests.

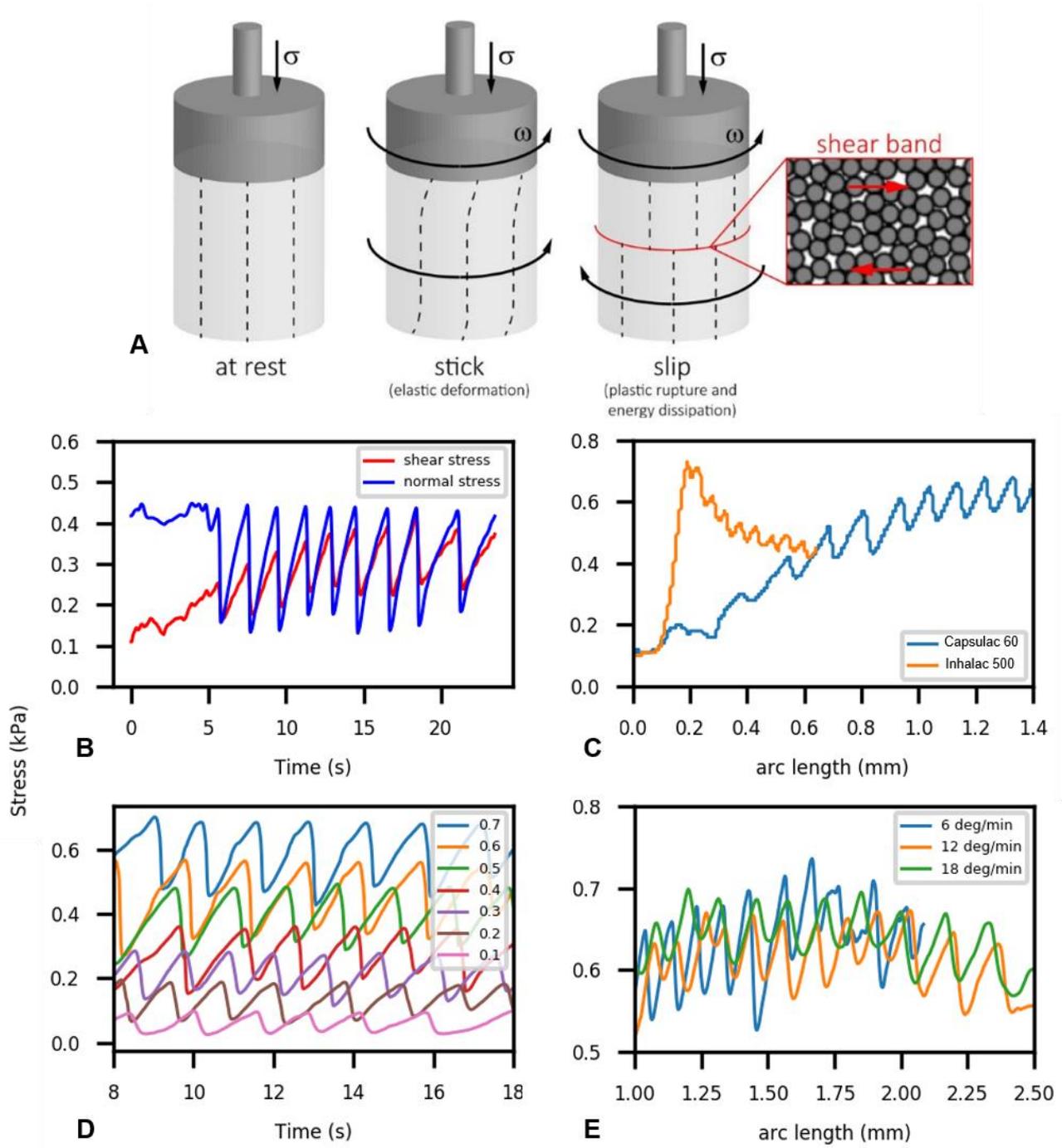

*Figure 6. (a) Sketch of the occurrence of the stick and slip events during the shearing of a cylindrical powder bed. Dashed vertical lines illustrate the deformation of the powder bed, the red line represents the shear band. (b) Normal and shear stresses as a function of time for Capsulac 60 during a shear-to-failure measure showing the typical stick-slip saw-tooth profile. (c) Shear stress as a function of the arc length traveled by the piston during sliding for a micronized and a coarse powder, the normal load is 0.5 kPa. (d) Stick-slip dependence on the normal load for Capsulac 60, different colors represent different normal load in kPa as indicated by the legend. Normal stress is plotted as a function of time. (e) Driving velocity dependence of the stick-slip profile for Capsulac 60. Normal stress is plotted as a function of arc length.*

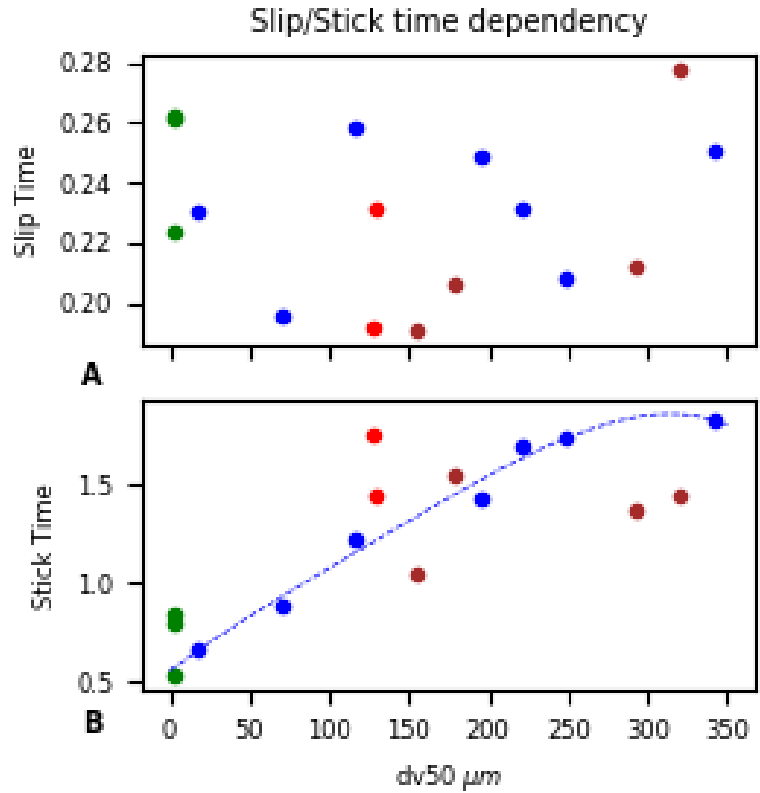

*Figure 7. Average slip (a) and stick (b) times (seconds) as a function of the particle $dv_{50}$ for all the powders analyzed, colors represent the different morphology groups. Data have been collected for the tests at $\sigma_{pre}$ =1 kPa and $\sigma$ =0.7 kPa.*

List of Tables

| Product Name | $dv_{50}$ (µm) | # | Particle shape |
|---|---|---|---|
| Inhalac500 | 2.5 | 0 | Micronized |
| Lacto-Hale 300 | 2.7 | 1 | Micronized |
| Lacto-Sphere MM3 | 3.0 | 2 | Micronized |
| Excipure DPI 221 | 17.0 | 3 | Single Tomahawk |
| Foremost 310 regular NFLM | 71.0 | 4 | Single Tomahawk |
| Excipure DPI 211 | 117.0 | 5 | Single Tomahawk |
| Excipress SD2 150 | 127.0 | 6 | Spherical |
| FlowLac 90 | 129.0 | 7 | Spherical |
| Excipress GR 150 | 154.0 | 8 | Fused Tomahawk |
| Tablettose 70 | 178.0 | 9 | Fused Tomahawk |
| Lactose Monohydrate 80 Mesh | 195.0 | 10 | Single Tomahawk |
| Respitose SV 001 | 220.0 | 11 | Single Tomahawk |
| Excipure DPI 111 | 248.0 | 12 | Single Tomahawk |
| Capsulac 60 | 293.0 | 13 | Fused Tomahawk |
| SacheLac 80 | 320.0 | 14 | Fused Tomahawk |
| Pharmatose 50M | 342.0 | 15 | Single Tomahawk |

Table 1. Particle size ($dv_{50}$), lactose grade and number code for each powder analyzed in the present paper.

| Product Name | $dv_{50}$ (µm) unsieved | Retained sieved fraction | # |
|---|---|---|---|
| Pharmatose DCL 11 | 105 | > 63 µm | 16 |
| | | > 212 µm | 17 |
| Capsulac 60 | 293 | > 250 µm | 18 |
| | | > 300 µm | 19 |
| Pharmatose 50M | 342 | > 250 µm | 20 |

Table 2. Particle size for the sieved grades.

| Name | $dv_{50}$ (µm) | n | σ | $R^2$ (n=$n_{opt}$) | $R^2$ (n=1) | $R^2$ (n=2) |
|---|---|---|---|---|---|---|
| Inhalac 500 | 2.5 | 1.74 | 0.38 | 0.9882 | 0.9788 | 0.9866 |
| Lacto-Hale 300 | 2.7 | 1.55 | 0.32 | 0.9911 | 0.9841 | 0.9876 |
| Lacto-Sphere MM3 | 3.0 | 1.56 | 0.40 | 0.9964 | 0.9916 | 0.9924 |
| Excipure DPI 221 | 17.0 | 1.80 | 0.03 | 0.9954 | 0.9820 | 0.9876 |
| Foremost 310 regular NFLM | 71.0 | 1.42 | 0.14 | 0.9858 | 0.9819 | 0.9500 |
| Excipure DPI 211 | 117.0 | 1.32 | 0.07 | 0.9996 | 0.9953 | 0.9649 |

| Name | | | | | | |
|---|---|---|---|---|---|---|
| Excipress SD2 150 | 127.0 | 1.23 | 0.08 | 0.9948 | 0.9935 | 0.9598 |
| FlowLac 90 | 129.0 | 1.06 | 0.07 | 0.9959 | 0.9956 | 0.9065 |
| Excipress GR 150 | 154.0 | 1.21 | 0.14 | 0.9996 | 0.9971 | 0.9473 |
| Tablettose 70 | 178.0 | 1.00 | 0.00 | 0.9943 | 0.9943 | 0.8897 |
| Lactose Monohydrate 80 Mesh | 195.0 | 1.31 | 0.13 | 0.9985 | 0.9942 | 0.9781 |
| Respitose SV 001 | 220.0 | 1.0 | 0.0 | 0.9982 | 0.9982 | 0.9199 |
| Excipure DPI 111 | 248.0 | 1.02 | 0.01 | 0.9977 | 0.9977 | 0.8967 |
| Capsulac 60 | 293.0 | 1.08 | 0.13 | 0.9963 | 0.9958 | 0.9172 |
| SacheLac 80 | 320.0 | 1.06 | 0.04 | 0.9974 | 0.9971 | 0.9353 |
| Pharmatose 50M | 342.0 | 1.03 | 0.05 | 0.9967 | 0.9967 | 0.9356 |

*Table 3. The shear index n is shown per each powder grade, together with standard deviation σ. $R^2$ values for the yield loci fit are reported for the fitted value of n ($n_{opt}$) for the linear case n=1 and for n=2.*

| Name | $d_{v50}$ (μm) | $C_{opt}$ (kPa) | $T_{opt}$ (kPa) | $C_{lin}$ (kPa) | $T_{lin}$ (kPa) | φ (deg) |
|---|---|---|---|---|---|---|
| Inhalac 500 | 2.5 | 0.6559+-0.0912 | 0.3102+-0.0848 | 0.7069+-0.0858 | 0.7786+-0.1374 | 42.485+-4.2124 |
| Lacto-Hale 300 | 2.7 | 0.4615+-0.0827 | 0.1350+-0.0446 | 0.5711+-0.0552 | 0.4800+-0.0789 | 50.206+-2.2659 |
| Lacto-Sphere MM3 | 3.0 | 0.4886+-0.0583 | 0.1967+-0.0437 | 0.5569+-0.0515 | 0.5868+-0.0776 | 43.615+-2.2767 |
| Excipure DPI 221 | 17.0 | 0.1938+-0.0365 | 0.0593+-0.0178 | 0.3126+-0.0189 | 0.3247+-0.0238 | 43.931+-0.4070 |
| Foremost 310 NFLM | 71.0 | 0.1105+-0.0070 | 0.0596+-0.0053 | 0.1645+-0.0063 | 0.1946+-0.0042 | 40.194+-0.6623 |
| Excipure DPI 211 | 117.0 | 0.1022+-0.0081 | 0.0999+-0.0102 | 0.1227+-0.0077 | 0.1883+-0.0134 | 33.101+-0.4615 |

| Name | | | | | |
|---|---|---|---|---|---|
| Excipress SD 2150 | 127.0 | 0.0944+-0.0056 | 0.1079+-0.0053 | 0.1104+-0.0041 | 0.1891+-0.0030 | 30.285+-0.5532 |
| FlowLac 90 | 129.0 | 0.0590+-0.0179 | 0.0609+-0.0224 | 0.0794+-0.0161 | 0.1298+-0.0284 | 31.537+-1.0457 |
| Excipress GR 150 | 154.0 | 0.0998+-0.0230 | 0.1004+-0.0249 | 0.1180+-0.0215 | 0.1642+-0.0289 | 35.683+-0.7123 |
| Tablettose 70 | 178.0 | 0.0653+-0.0090 | 0.0692+-0.0092 | 0.0796+-0.0084 | 0.1120+-0.0098 | 35.334+-0.7138 |
| Monohydrate 80 Mesh | 195.0 | 0.1215+-0.0213 | 0.1684+-0.0339 | 0.1293+-0.0205 | 0.2162+-0.0375 | 30.944+-0.7139 |
| Respitose SV 001 | 220.0 | 0.0768+-0.0031 | 0.1070+-0.0053 | 0.0832+-0.0033 | 0.1368+-0.0060 | 31.348+-1.2275 |
| Excipure DPI 111 | 248.0 | 0.0689+-0.0125 | 0.0854+-0.0178 | 0.0753+-0.0119 | 0.1067+-0.0188 | 35.320+-0.5887 |
| Capsulac 60 | 293.0 | 0.0820+-0.0106 | 0.0965+-0.0126 | 0.0859+-0.0105 | 0.1095+-0.0132 | 38.086+-0.8121 |
| SacheLac 80 | 320.0 | 0.0807+-0.0066 | 0.1196+-0.0144 | 0.0830+-0.0065 | 0.1305+-0.0150 | 32.585+-1.1915 |
| Pharmatose 50M | 342.0 | 0.0850+-0.0144 | 0.1410+-0.0168 | 0.0866+-0.0144 | 0.1502+-0.0171 | 29.815+-1.3890 |

*Table 4. Linear ($C_{lin}$, $T_{lin}$ and $\varphi$) and general Warren-Spring ($C_{opt}$, $T_{opt}$) yield locus parameters.*

| Name | C (3) | C (5) | C (7) | φ (3) | φ (5) | φ (7) |
|---|---|---|---|---|---|---|
| Pharmatose 50M | 60.9 | 15.8 | 7.03 | 12.3 | 5.09 | 8.54 |
| Pharmatose DCL11 >212µm | 71.7 | 8.86 | 4.17 | 11.0 | 7.05 | 5.97 |
| Respitose SV 001 | 1.52 | 2.01 | 3.96 | 4.02 | 3.91 | 3.91 |
| SacheLac 80 | 36.6 | 16.3 | 7.83 | 9.51 | 5.29 | 3.65 |
| Excipure DPI 221 | 26.3 | 19.6 | 15.8 | 3.32 | 2.04 | 1.66 |
| Tablettose 70 | 102. | 12.0 | 10.6 | 8.07 | 1.64 | 2.02 |
| FlowLac 90 | 24.0 | 24.4 | 20.2 | 6.78 | 3.23 | 3.31 |
| Capsulac 60 | 60.1 | 39.4 | 10.7 | 6.74 | 5.19 | 2.06 |

*Table 5. Dimensionless relative standard deviation of flowability parameters (C and $\varphi$) as a function of the number of points $n_p$ in the yield locus. The number of points is reported in brackets.*

| $C(dv_{50})$ | $\alpha$ | $\beta(\mu m^{\alpha} \cdot kPa)$ | $\gamma(kPa)$ | $R^2$ |
|---|---|---|---|---|
| All grades | $0.32 \pm 0.09$ | $0.95 \pm 0.08$ | $-0.08 \pm 0.06$ | 0.91 |
| Only tomahawk shape | $0.36 \pm 0.19$ | $1.11 \pm 0.40$ | $-0.07 \pm 0.11$ | 0.92 |
| All grades | 1 | $1.67 \pm 0.37$ | $0.09 \pm 0.01$ | 0.80 |
| Only tomahawk shape | 1 | $5.90 \pm 1.03$ | $0.06 \pm 0.01$ | 0.85 |
| $T(dv_{50})$ | $\alpha$ | $\beta(\mu m^{\alpha} \cdot kPa)$ | $\gamma(kPa)$ | $R^2$ |
| All grades | $0.31 \pm 0.17$ | $0.71 \pm 0.18$ | $0.01 \pm 0.10$ | 0.82 |
| Only tomahawk shape | $0.32 \pm 0.31$ | $0.80 \pm 0.41$ | $-0.01 \pm 0.17$ | 0.74 |
| All grades | 1 | $2.00 \pm 0.59$ | $0.17 \pm 0.01$ | 0.72 |
| Only tomahawk shape | 1 | $5.62 \pm 0.83$ | $0.12 \pm 0.01$ | 0.72 |

*Table 6. Parameters of the power law used to fit the C and T behavior as a function of $dv_{50}$.*

| Name | $d_{v50}$ (µm) pre-sieving | C (kPa) | T (kPa) | $\varphi$ (deg) |
|---|---|---|---|---|
| Pharmatose DCL11 > 63 µm | 105 | 0.1025+-0.0052 | 0.1648+-0.0117 | 31.926+-1.1597 |
| Pharmatose DCL11 > 212 µm |  | 0.0811+-0.0024 | 0.1102+-0.0091 | 36.462+-2.0422 |
| Capsulac 60 | 293 | 0.0859+-0.0105 | 0.1095+-0.0132 | 38.086+-0.8121 |
| Capsulac 60 > 250 µm |  | 0.0786+-0.0042 | 0.0893+-0.0080 | 41.423+-1.5699 |
| Capsulac 60 > 300 µm |  | 0.0298+-0.0368 | 0.0335+-0.0415 | 42.025+-0.6942 |
| Pharmatose 50M |  | 0.0866+-0.0144 | 0.1502+-0.0171 | 29.815+-1.3890 |
| Pharmatose 50M > 250 µm | 342 | 0.0543+-0.0184 | 0.0589+-0.0214 | 43.035+-1.0438 |

*Table 7. C, T and $\varphi$ values of sieved lactose grades.*

# List of symbols

| Symbol | Meaning | Units |
|---|---|---|
| $\sigma_{pre}$ | normal stress applied during the pre-shear | kPa |
| $\tau_{pre}$ | Maximum shear stress reached during the pre-shear | kPa |
| $\sigma$ | normal stress | kPa |
| $\tau$ | Shear stress | kPa |
| $C$ | Cohesion (yield locus intercept on the $\tau$ axis) | kPa |
| $T$ | Tensile strength (yield locus intercept on the $\sigma$ axis) | kPa |
| $\varphi$ | internal friction angle | deg |
| $n$ | shear index | / |
| $dv_{50}$ | median of the volume equivalent diameter distribution | µm |

# Supporting Information

## A shear cell study on oral and inhalation grade lactose powders

G. Cavalli[1], R. Bosi[2], A. Ghiretti[2], C. Cottini[2], A. Benassi[2,3*] and R. Gaspari[2*]

1 - Early product development, Chiesi Limited, Chippenham, Wiltshire (UK)

2 - DP Manufacturing & Innovation, Chiesi Farmaceutici SpA, Parma Italy

3 - International School for Advanced Studies (SISSA), Trieste (Italy)

[*]Corresponding author address: Chiesi Farmaceutici S.p.A. Largo Belloli 11A– *43122 Parma (Italy)*

## Contents

# Particle size distribution of the lactose samples

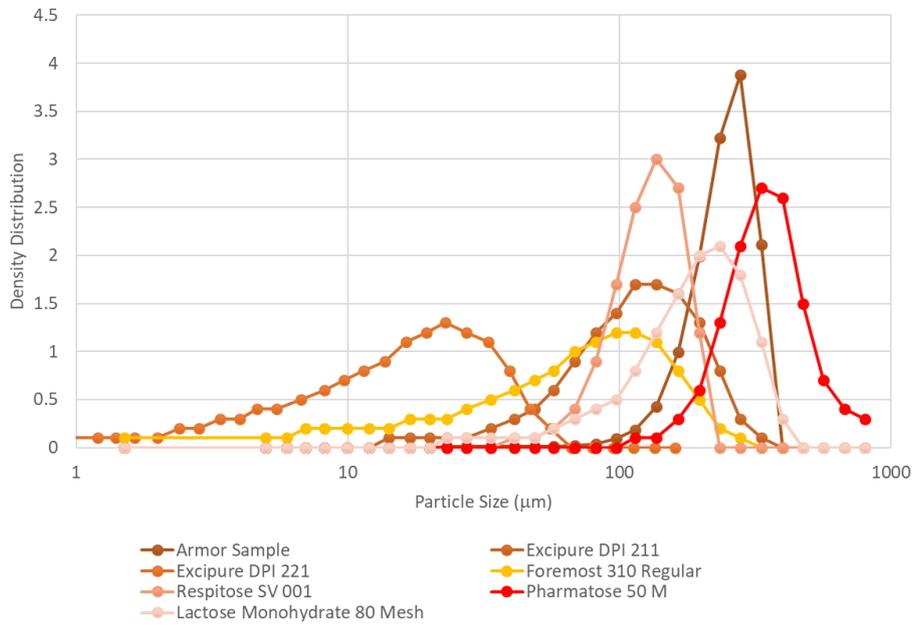

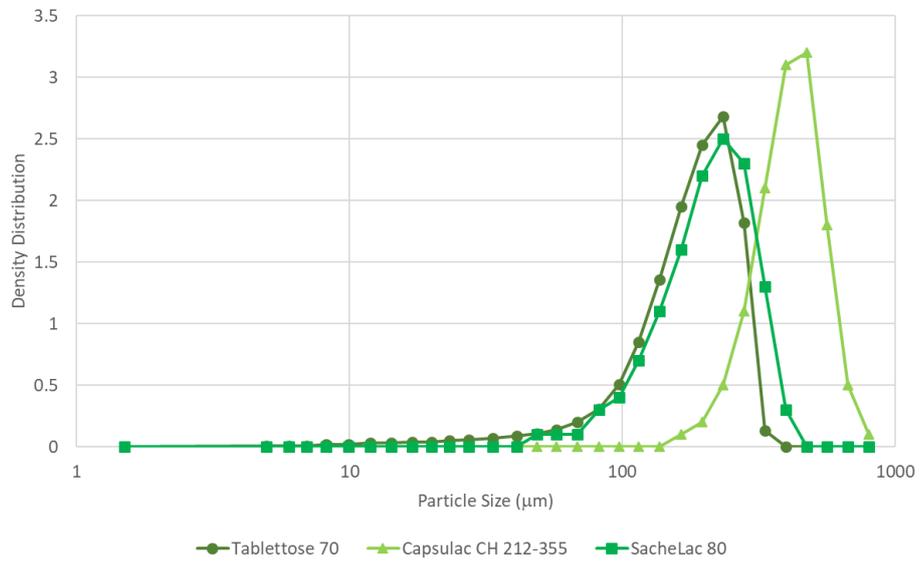

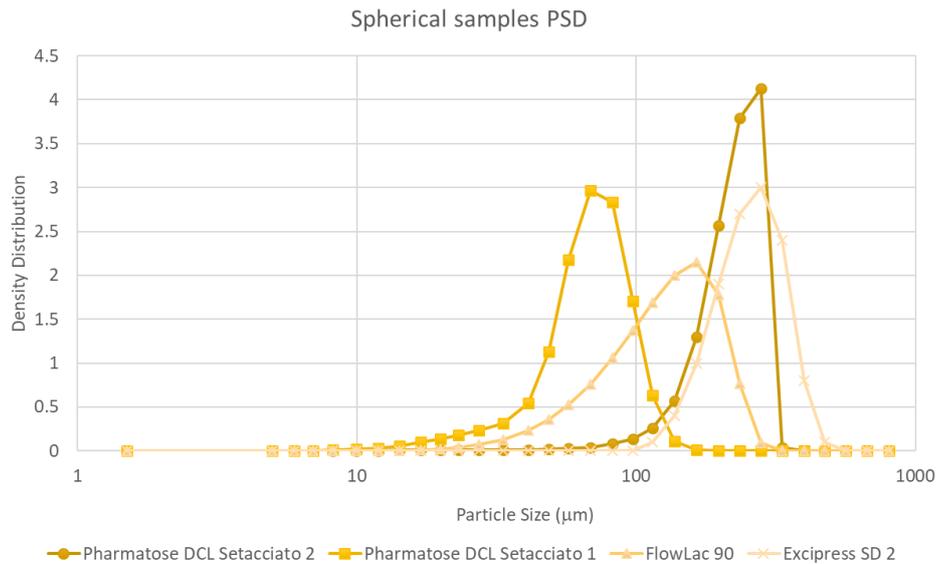

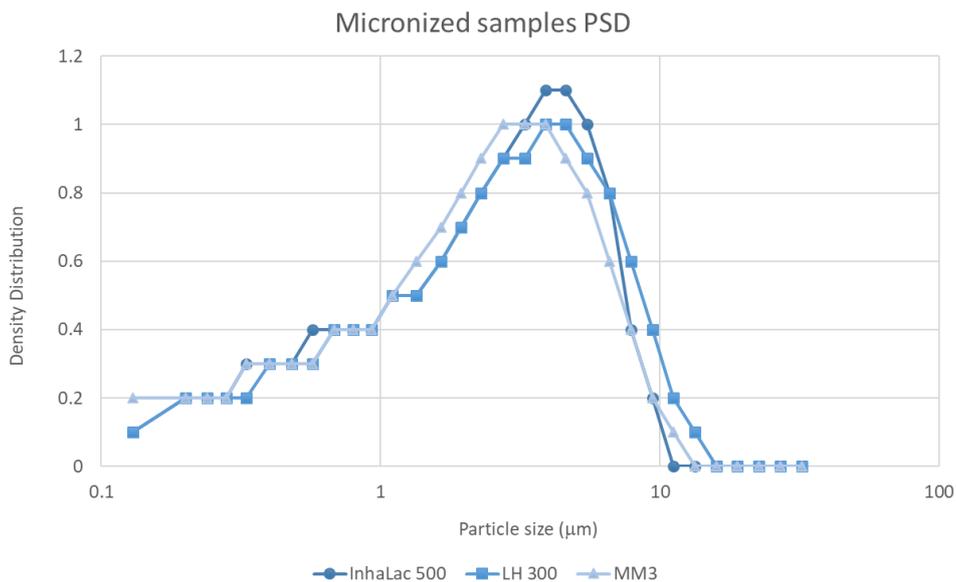

The term *density distribution* on the y-axis of the plots above refers to the so called $q_3lg$, i.e. a volume weighted distribution (see Sympatec Manual https://www.sympatec.com/en/particle-measurement/sensors/laser-diffraction/helos/).

## Shear to Failure profiles

### Normal Stress vs Time

Normal stress profiles are plotted for all powders investigated and are shown grouped by shape categories (single tomahawks, fused tomahawks, micronized, spherical). Three normal stress profiles are obtained from different samples of the same batch and are numbered progressively.

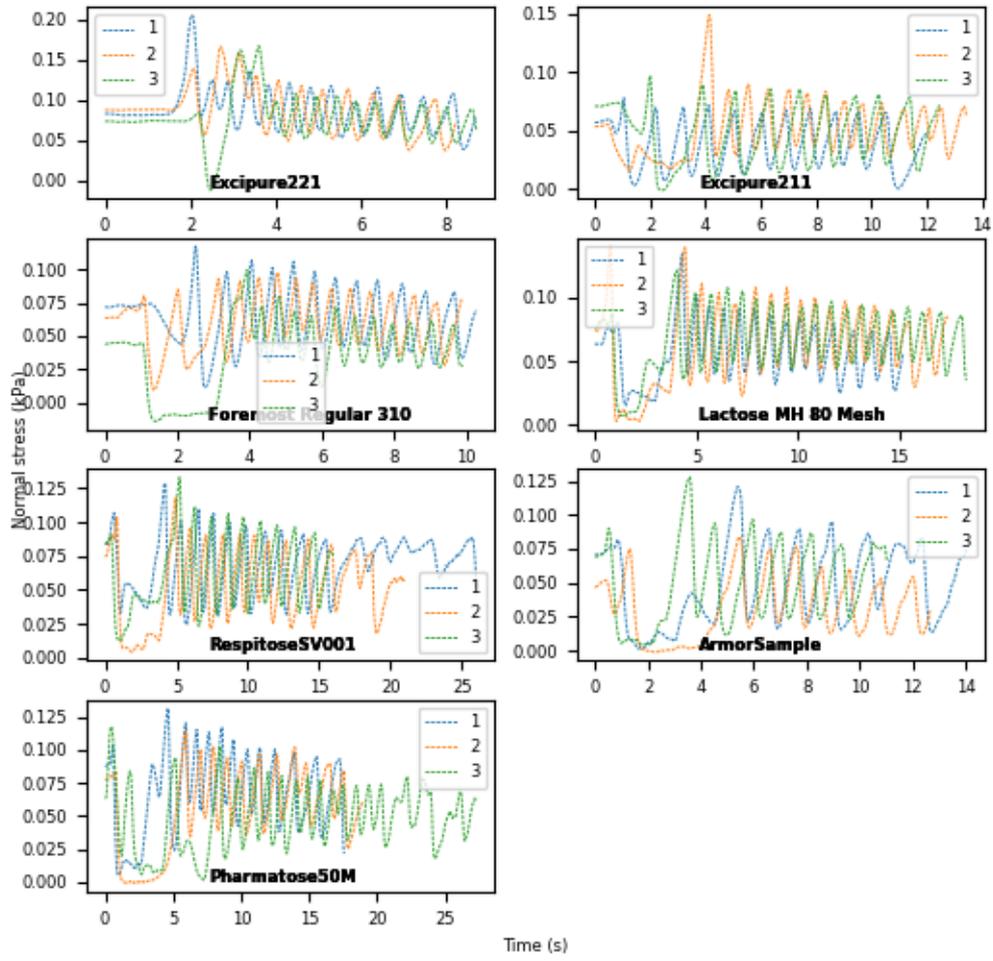

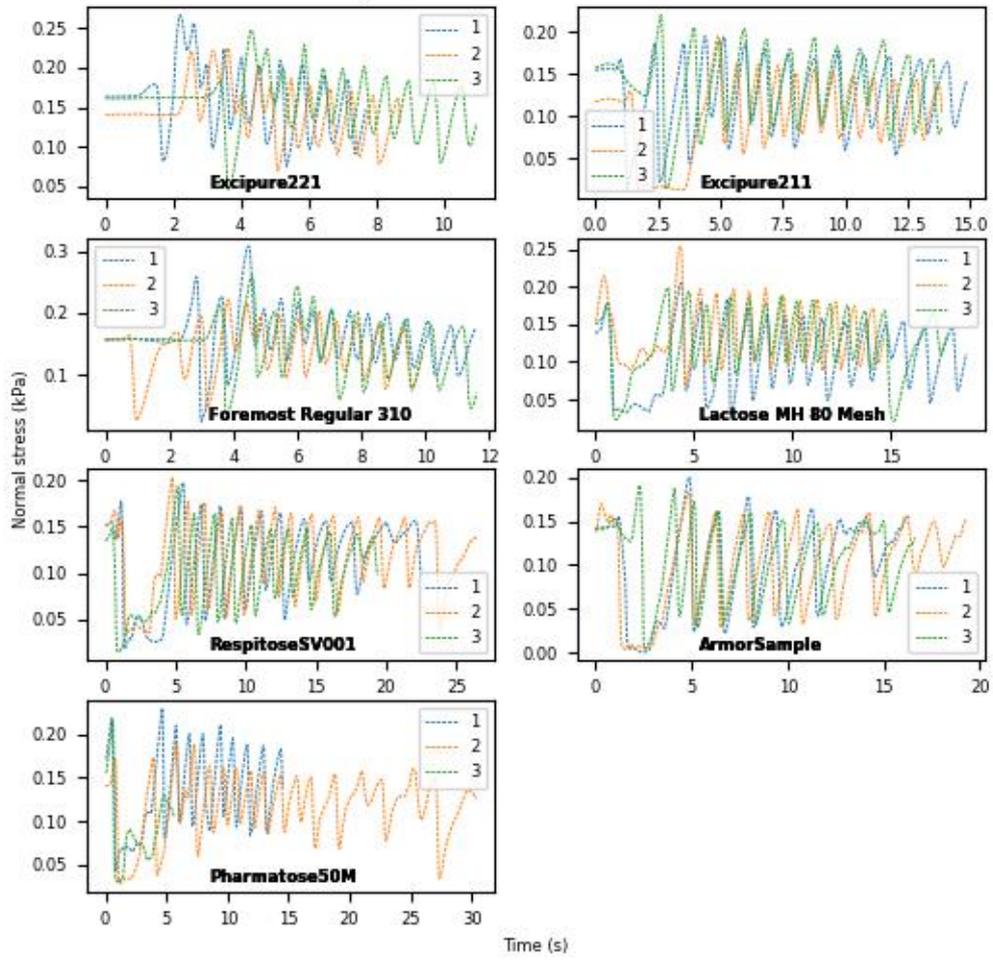

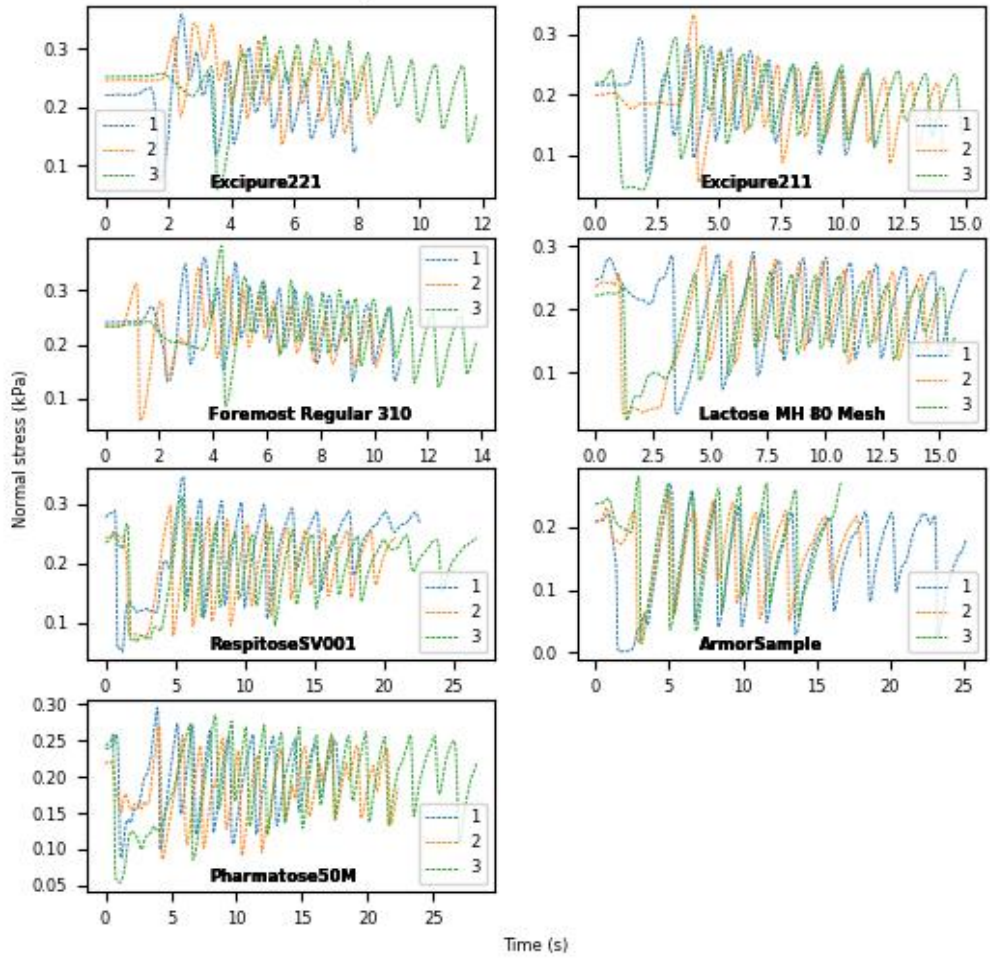

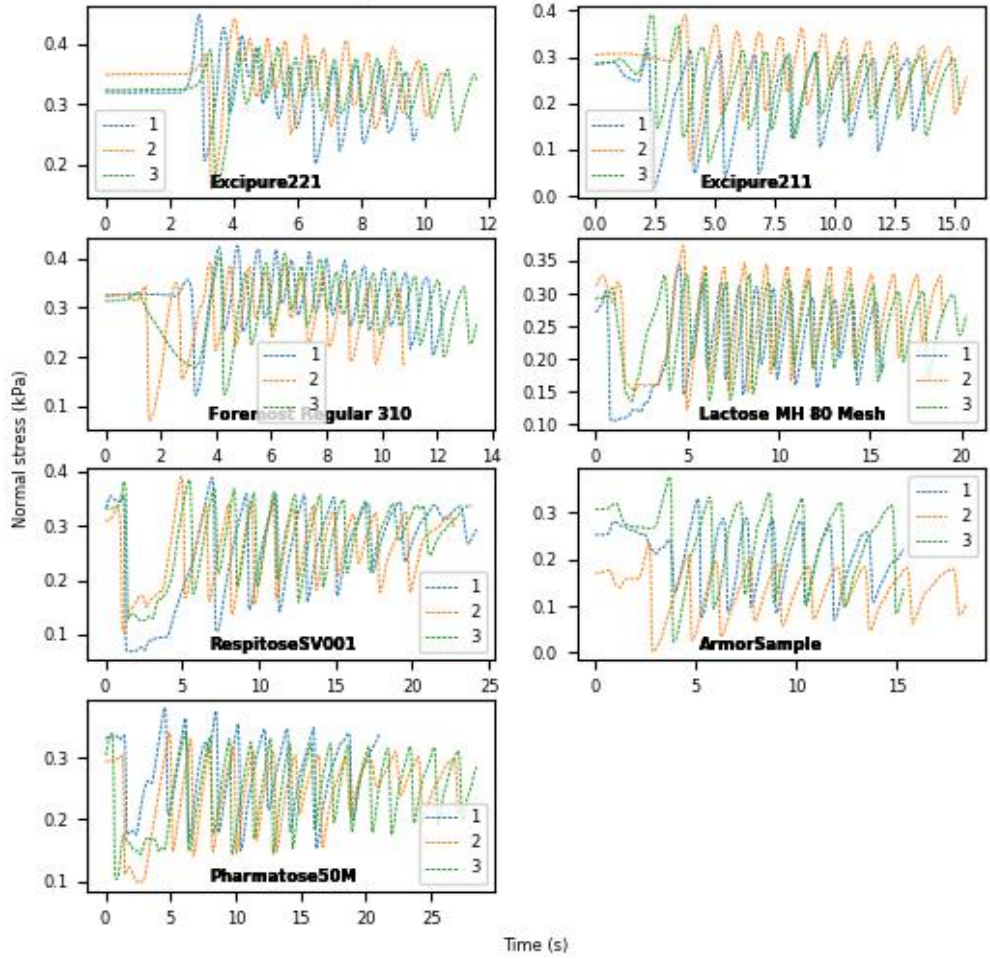

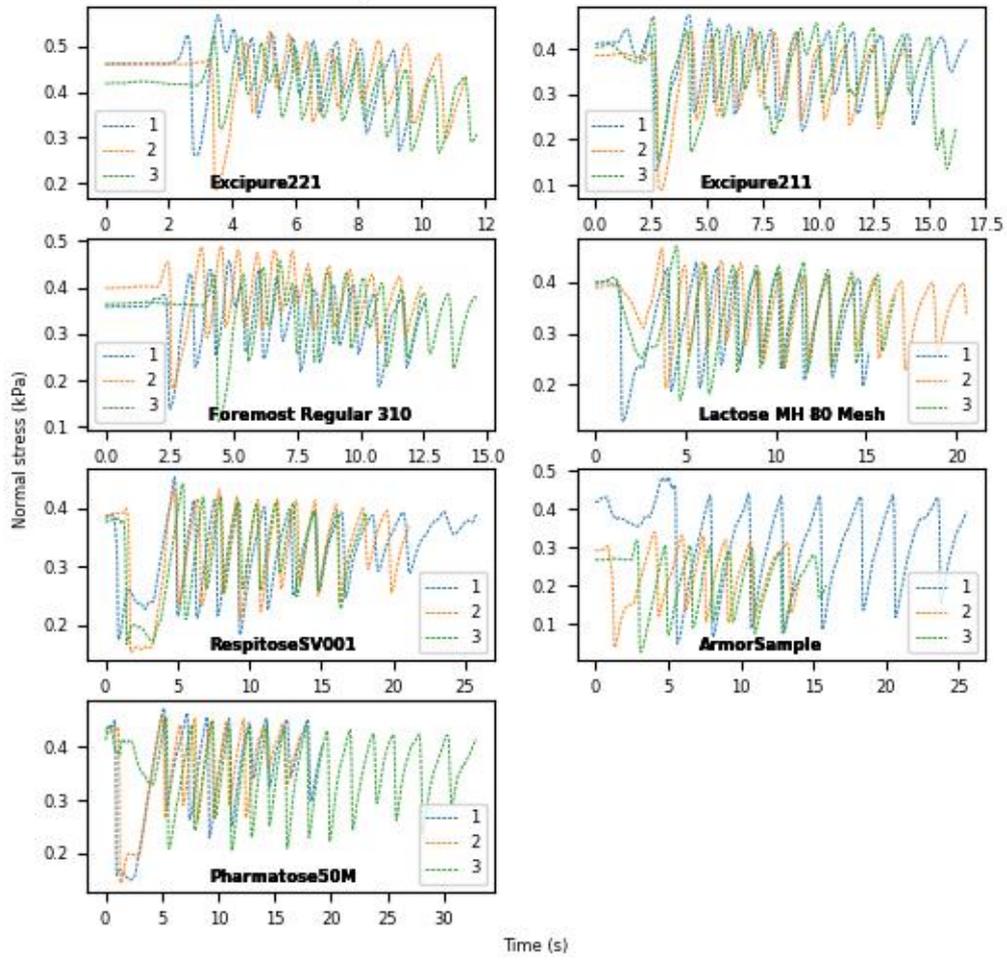

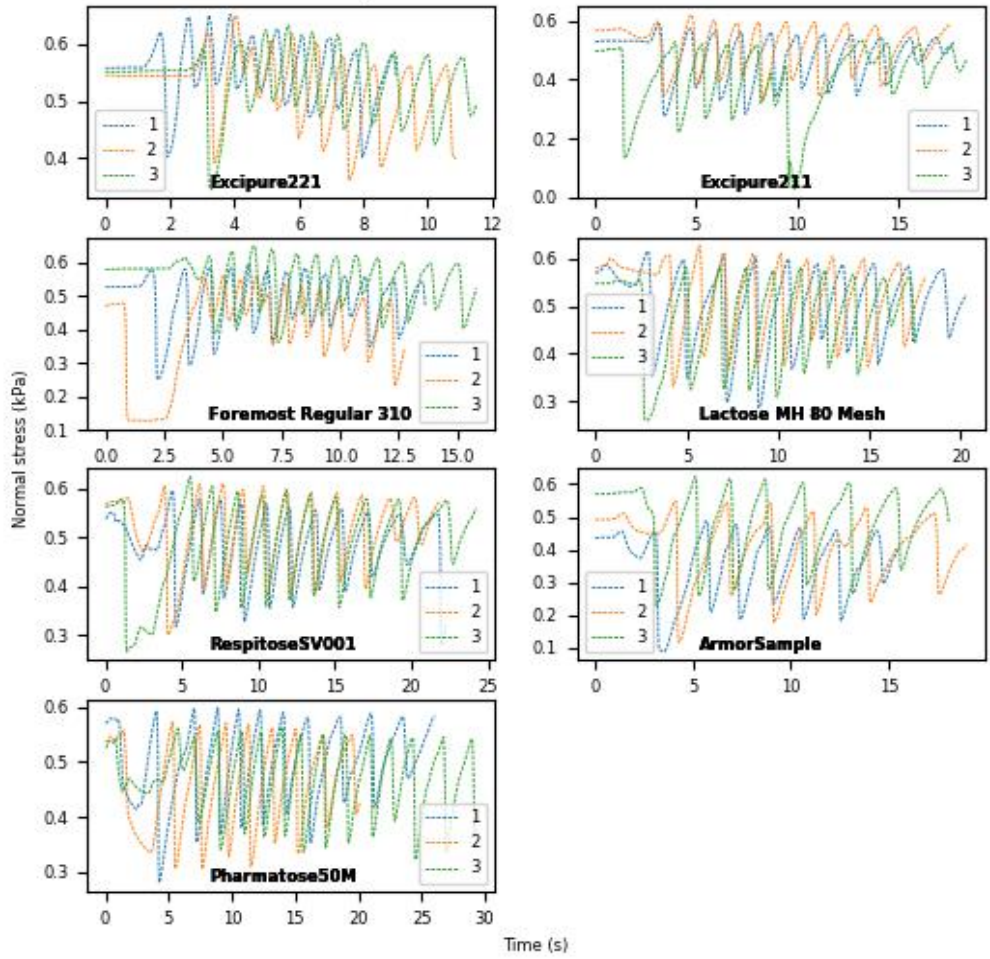

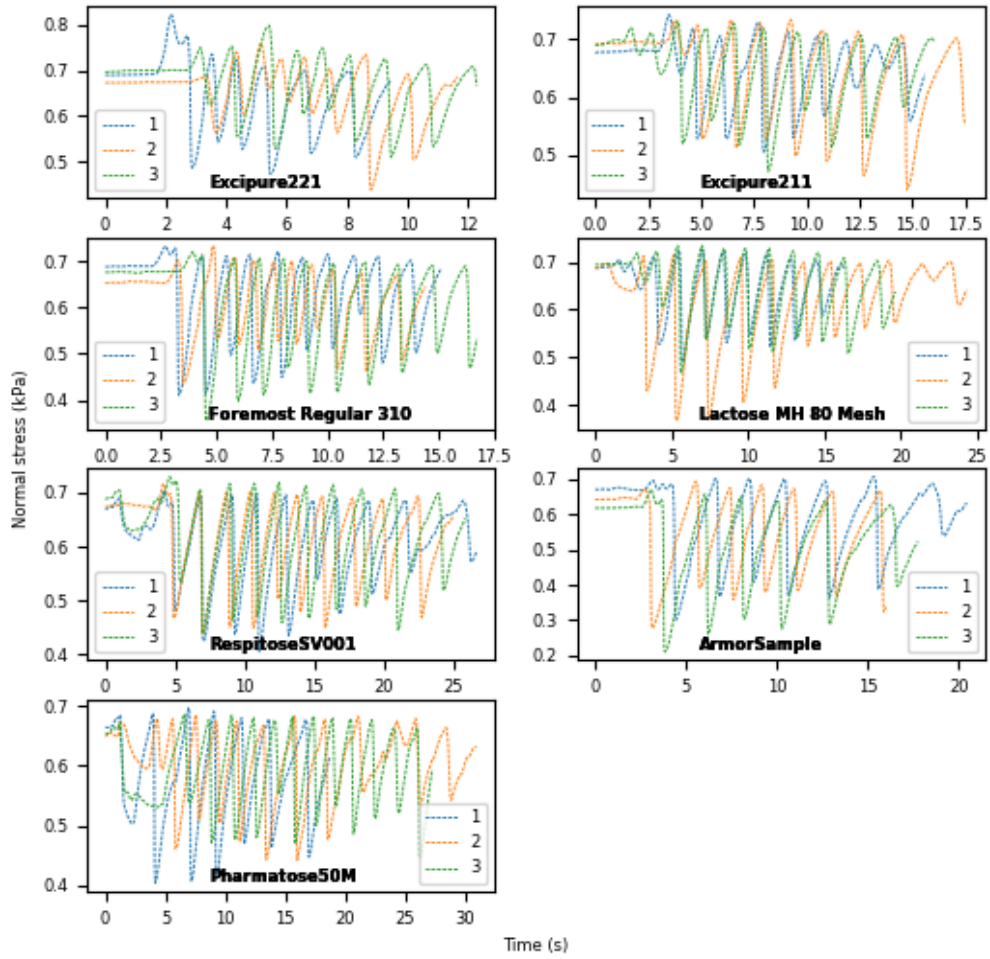

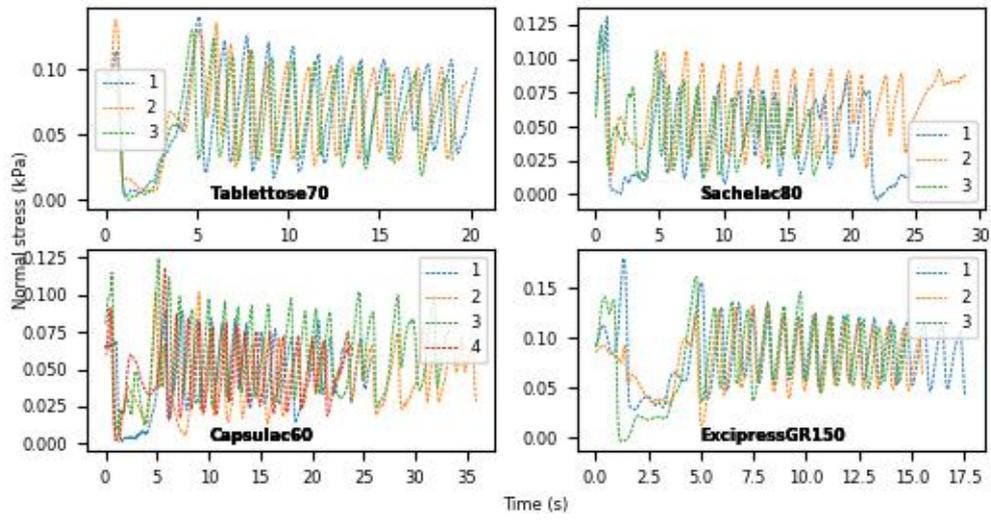

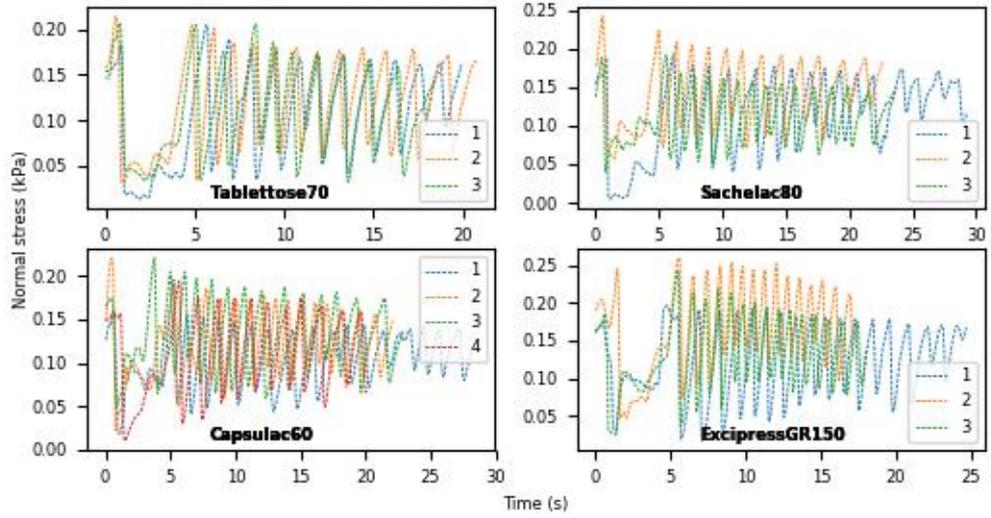

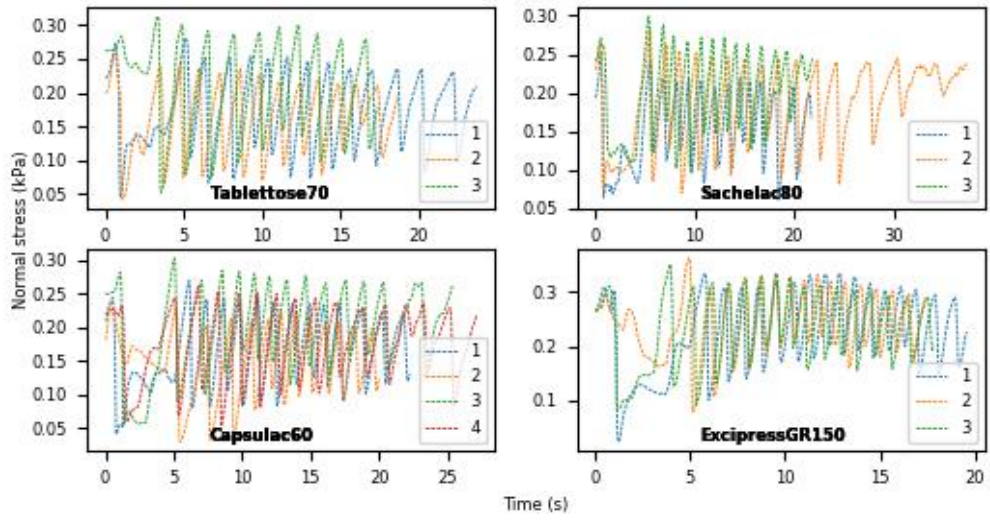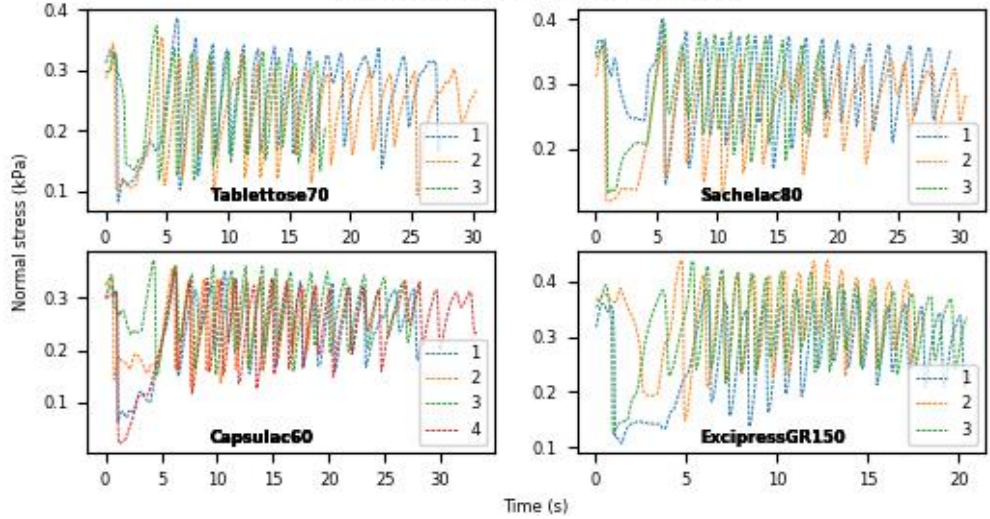

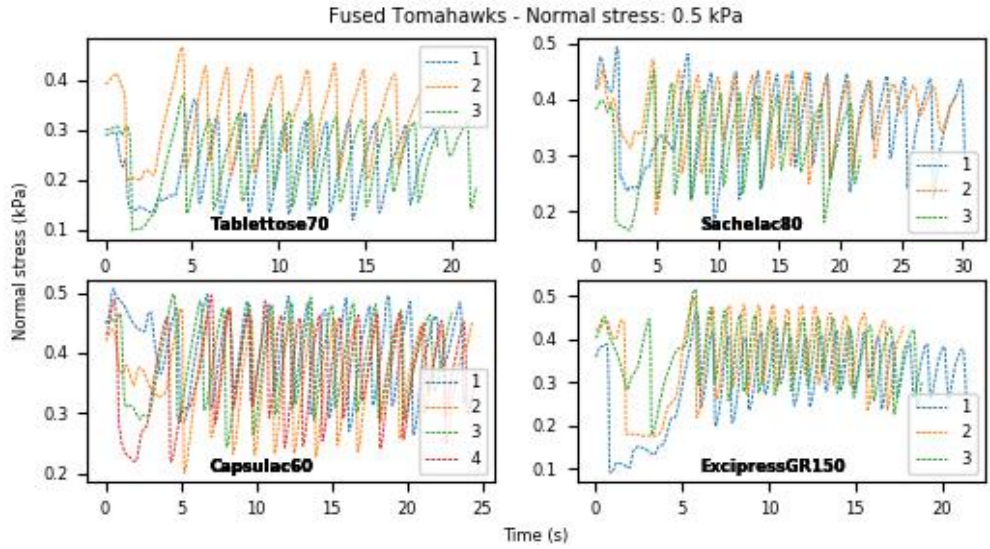

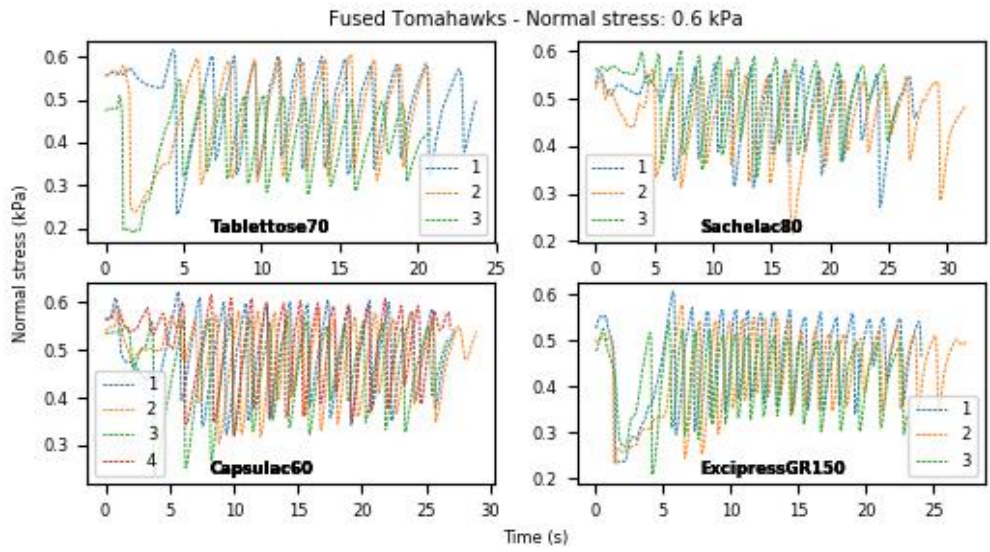

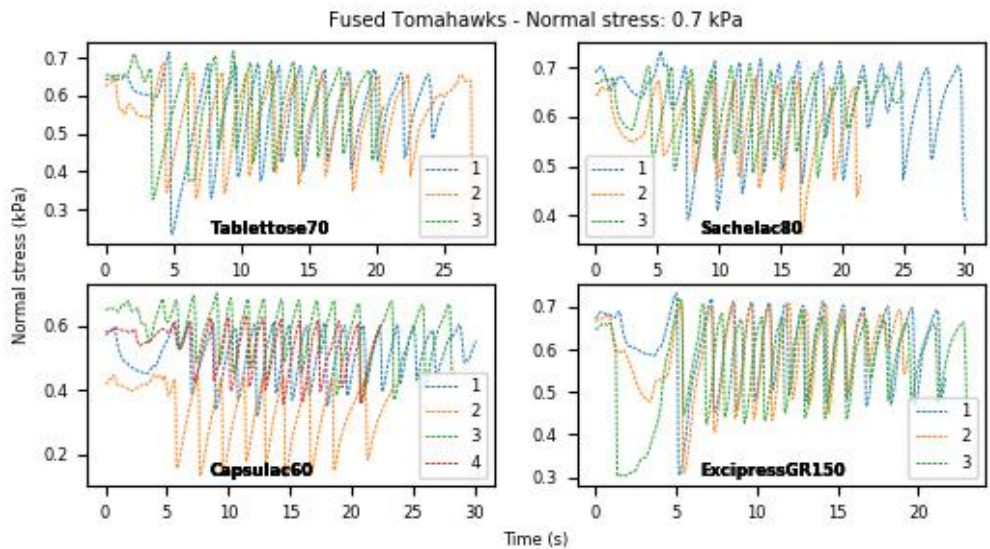

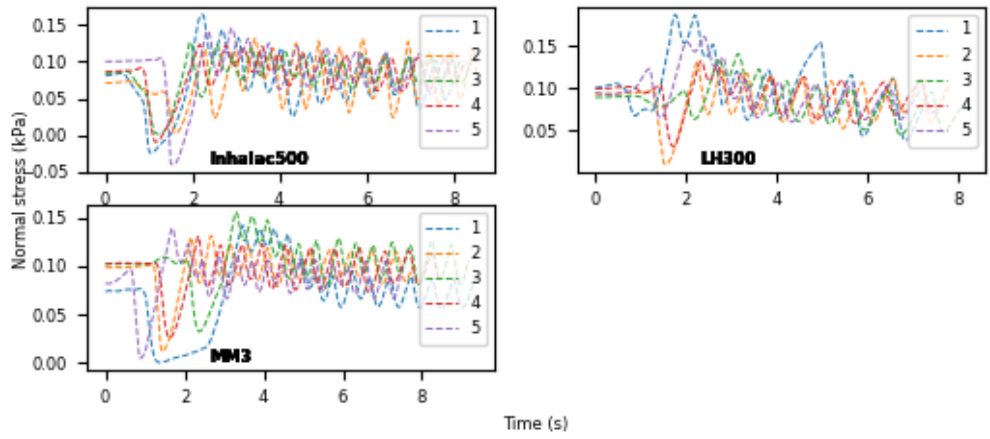

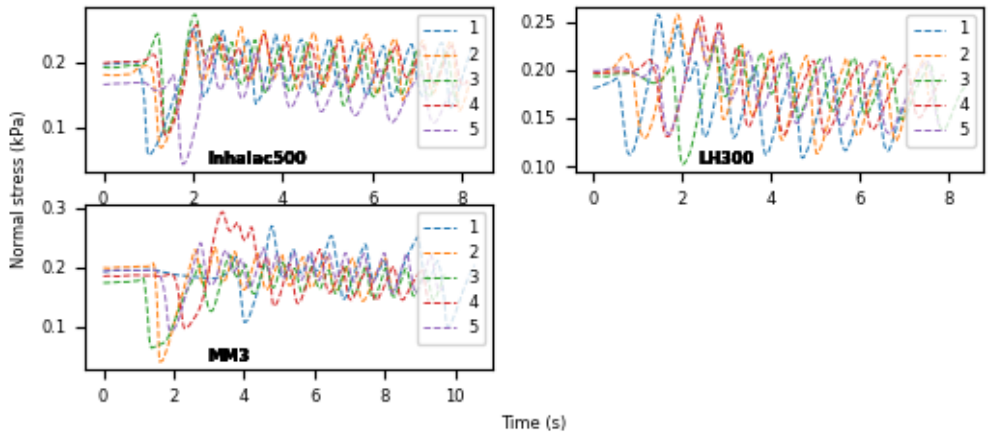

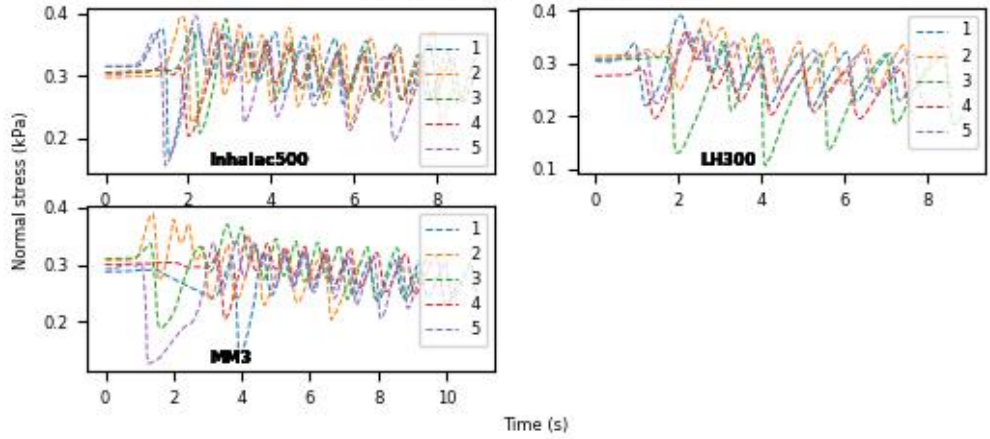

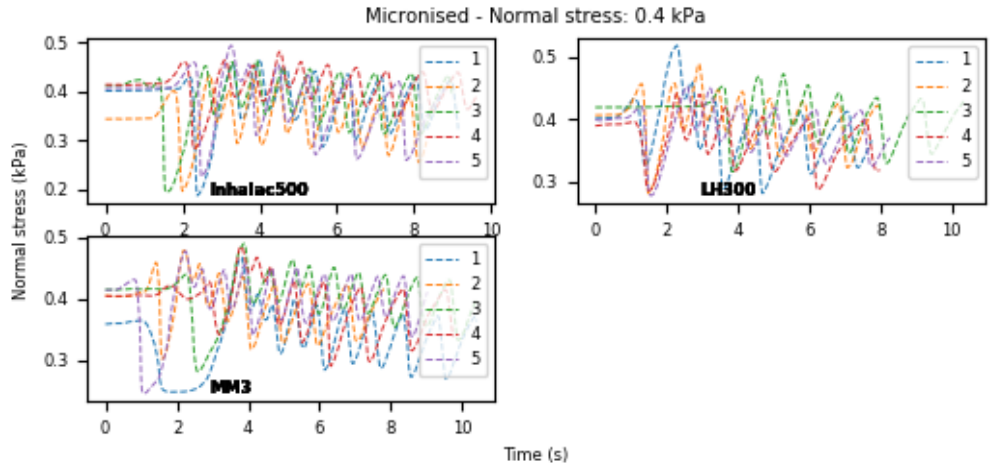
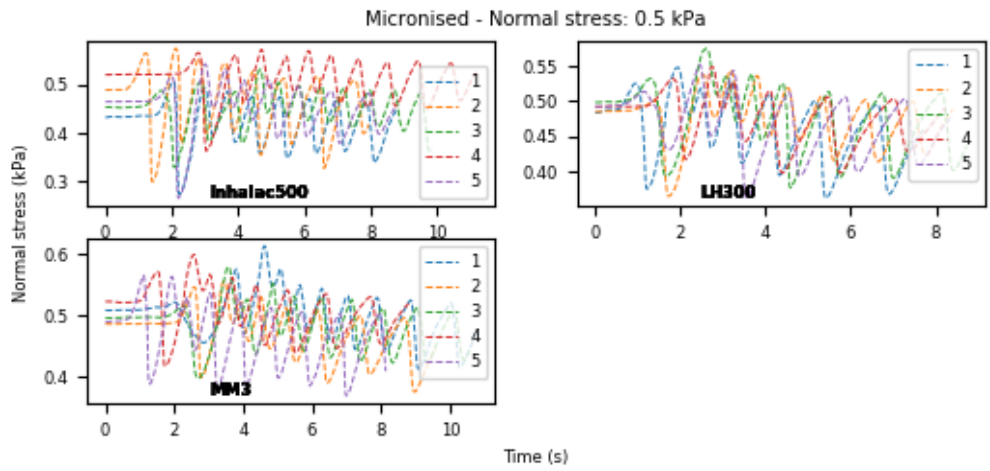
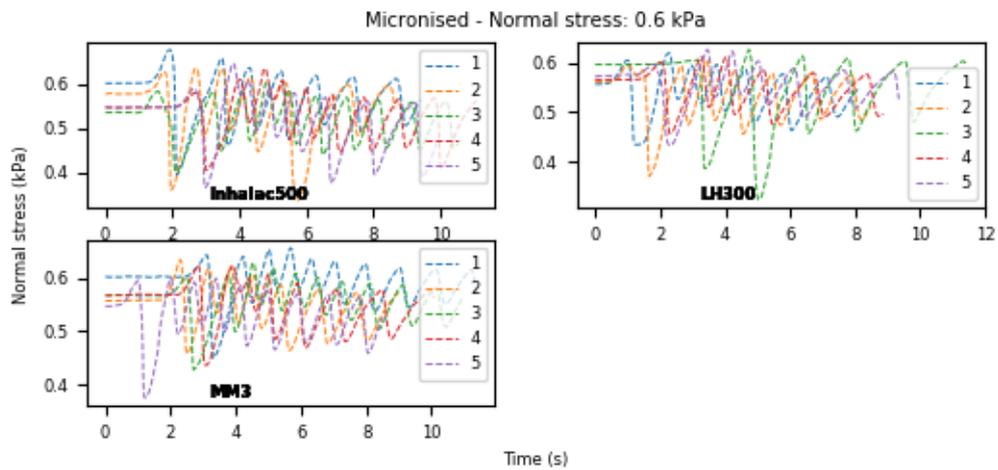

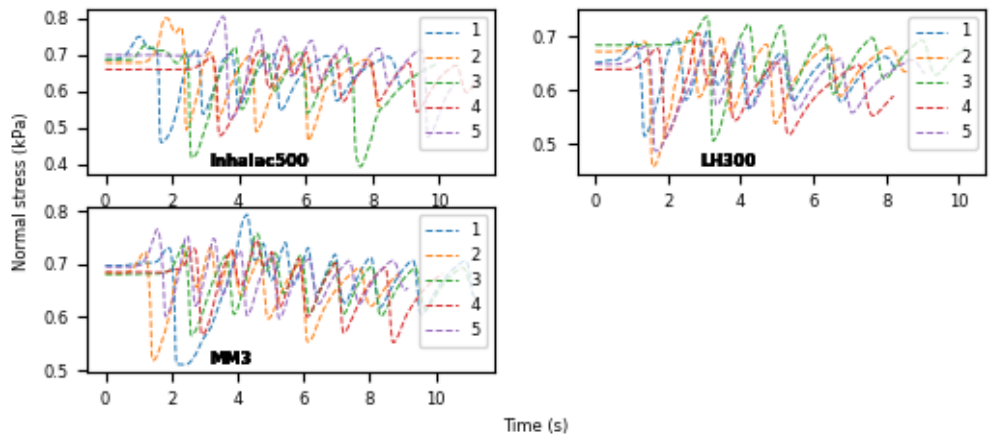
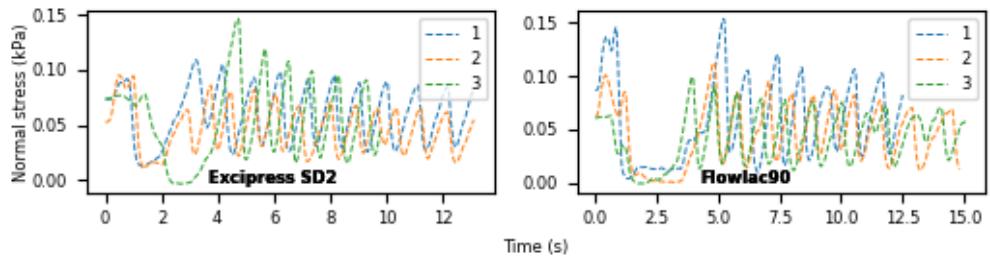
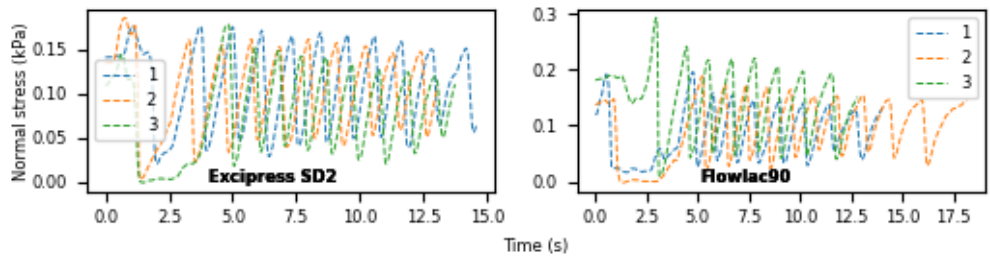
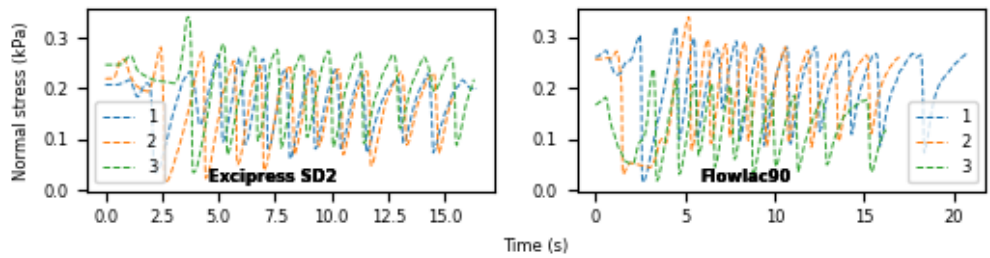

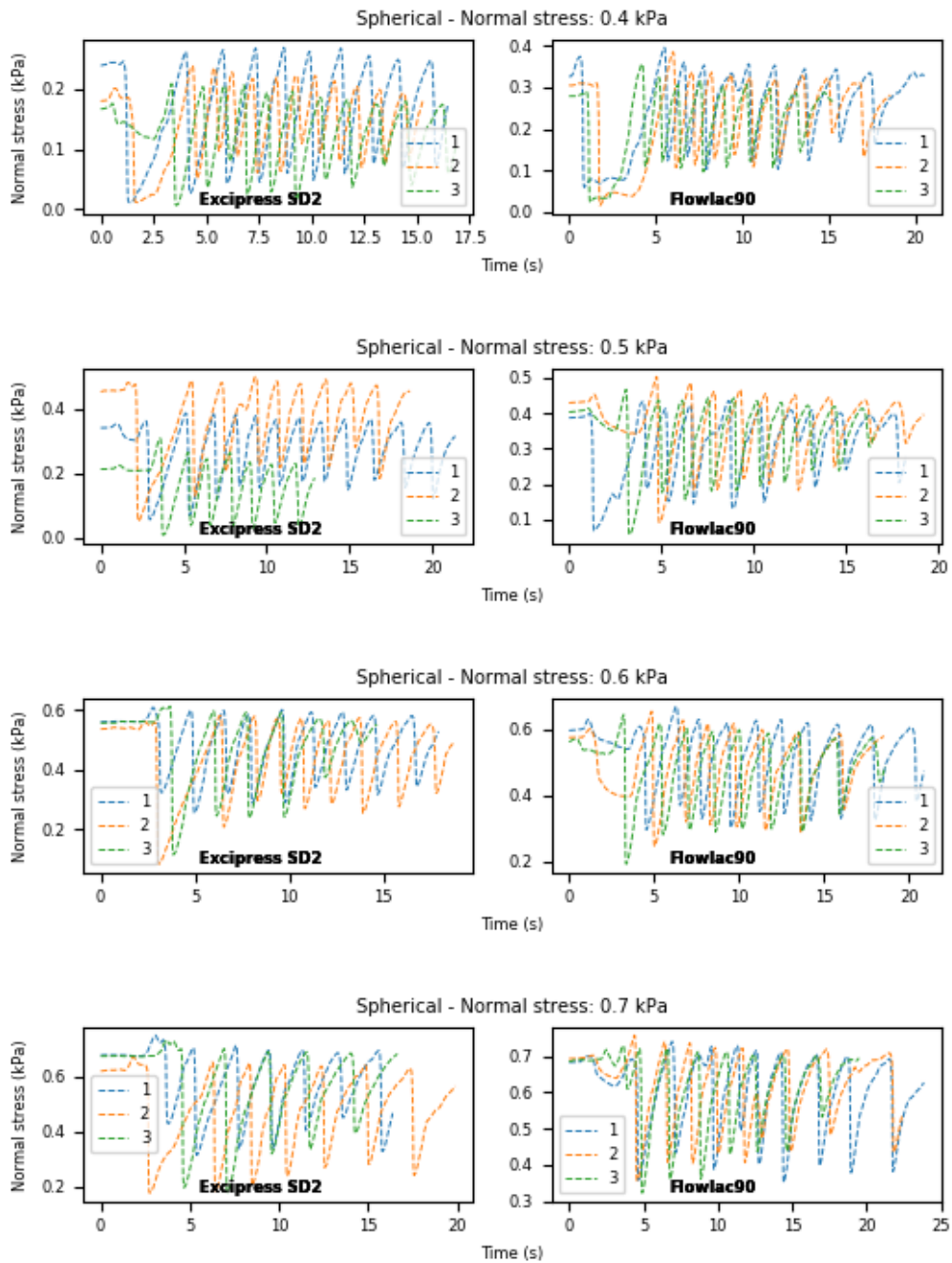

## Shear Stress vs Time

Shear stress profiles are plotted for all powders investigated and are shown grouped by shape categories (single tomahawks, fused tomahawks, spherical). Shear stress profiles are obtained from different samples of the same batch and are numbered accordingly.

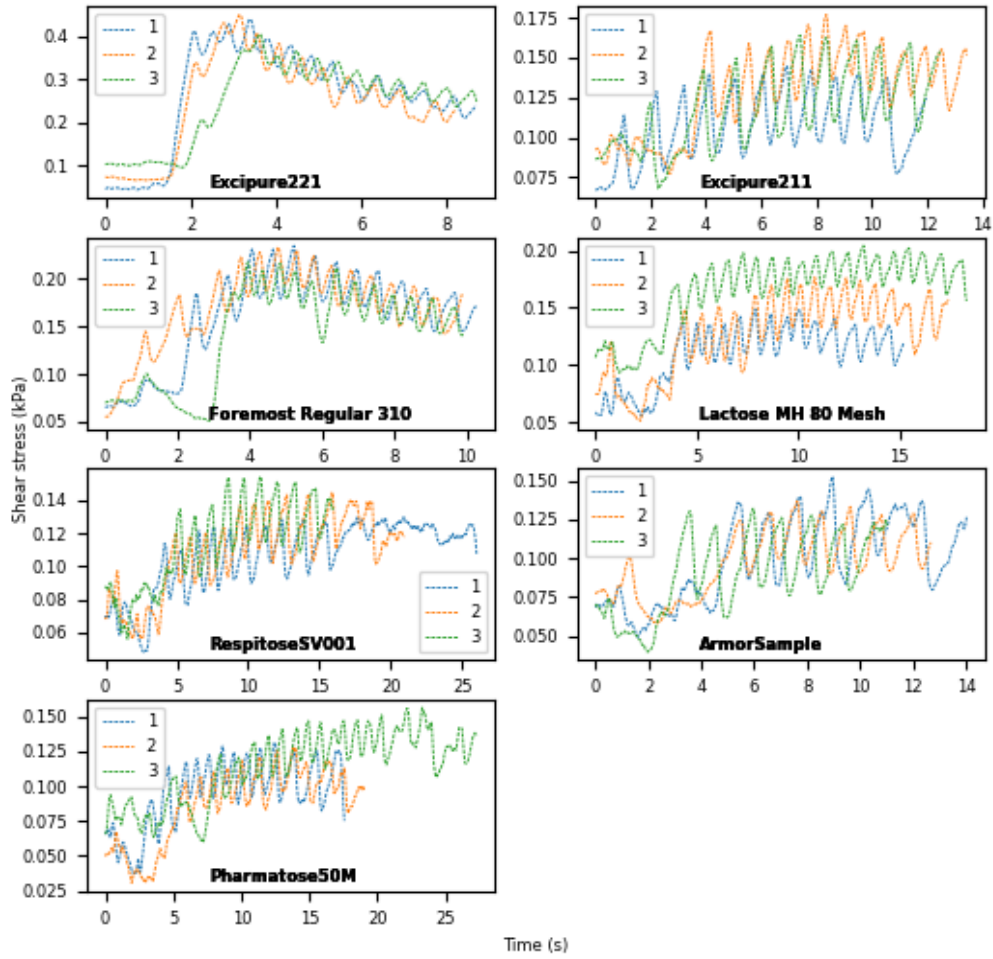

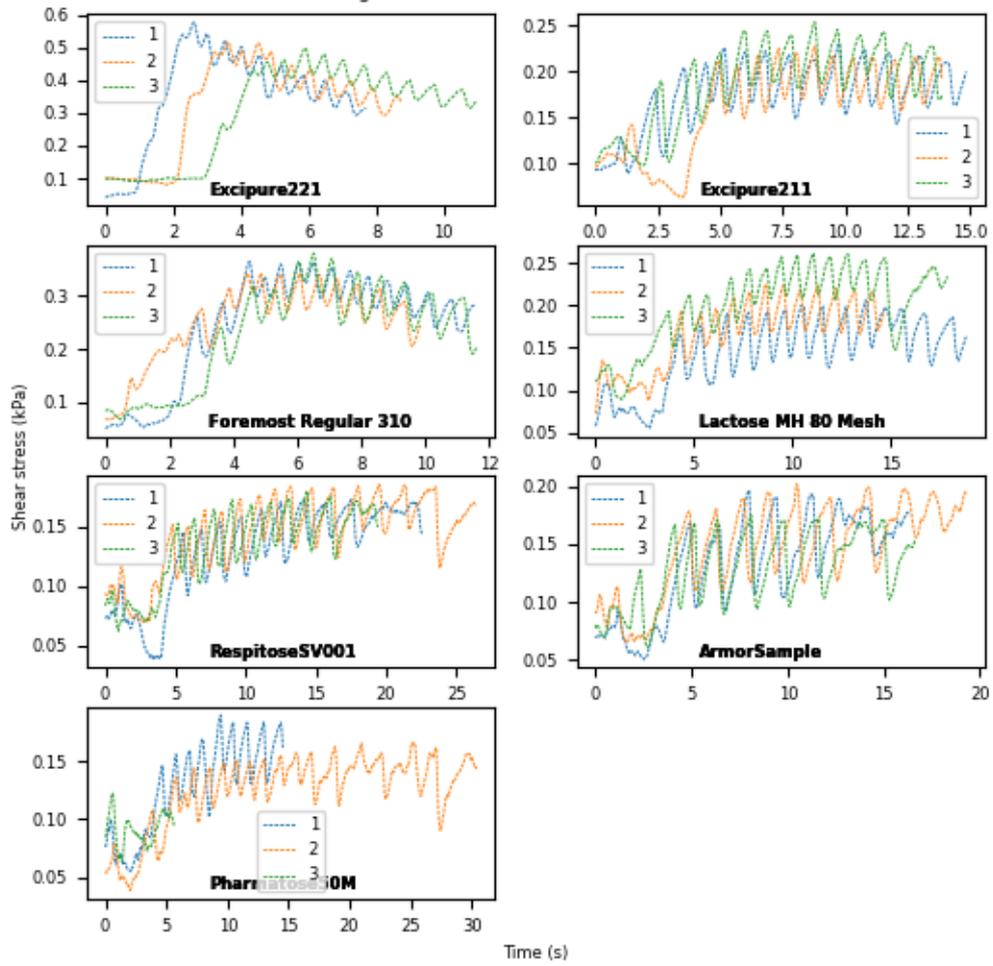

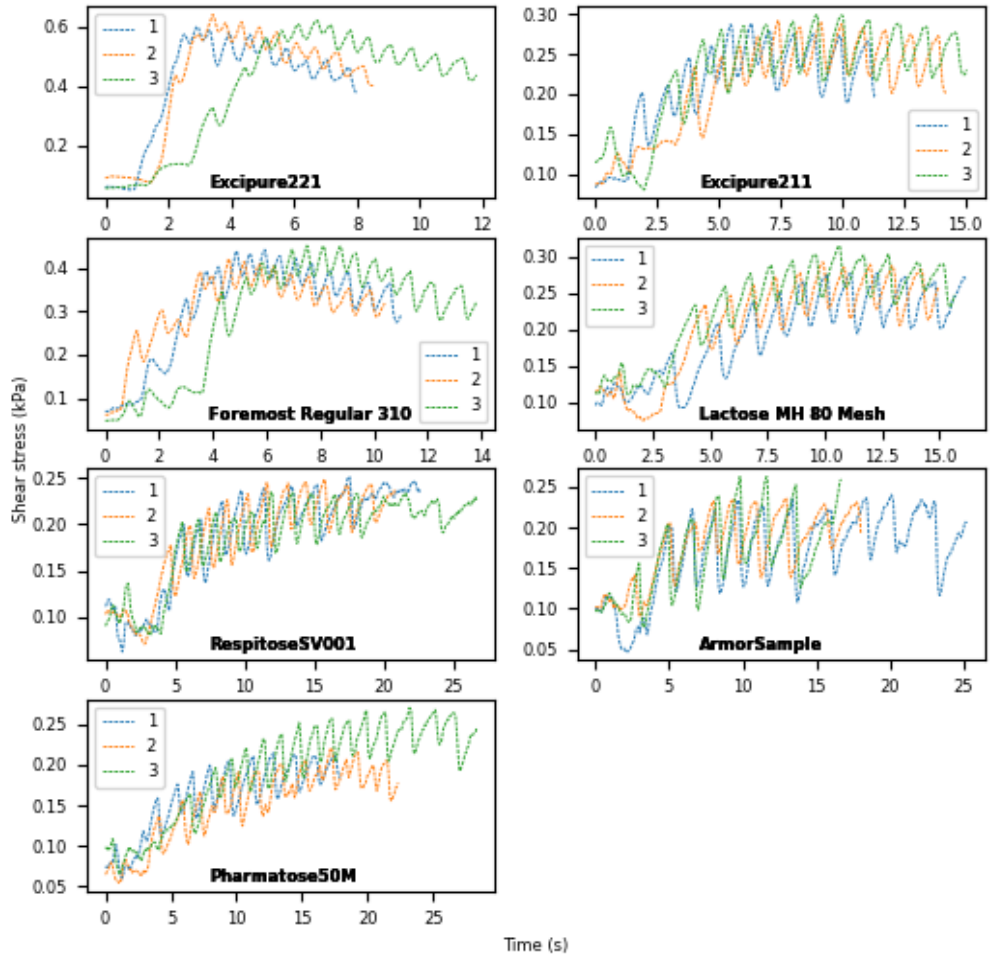

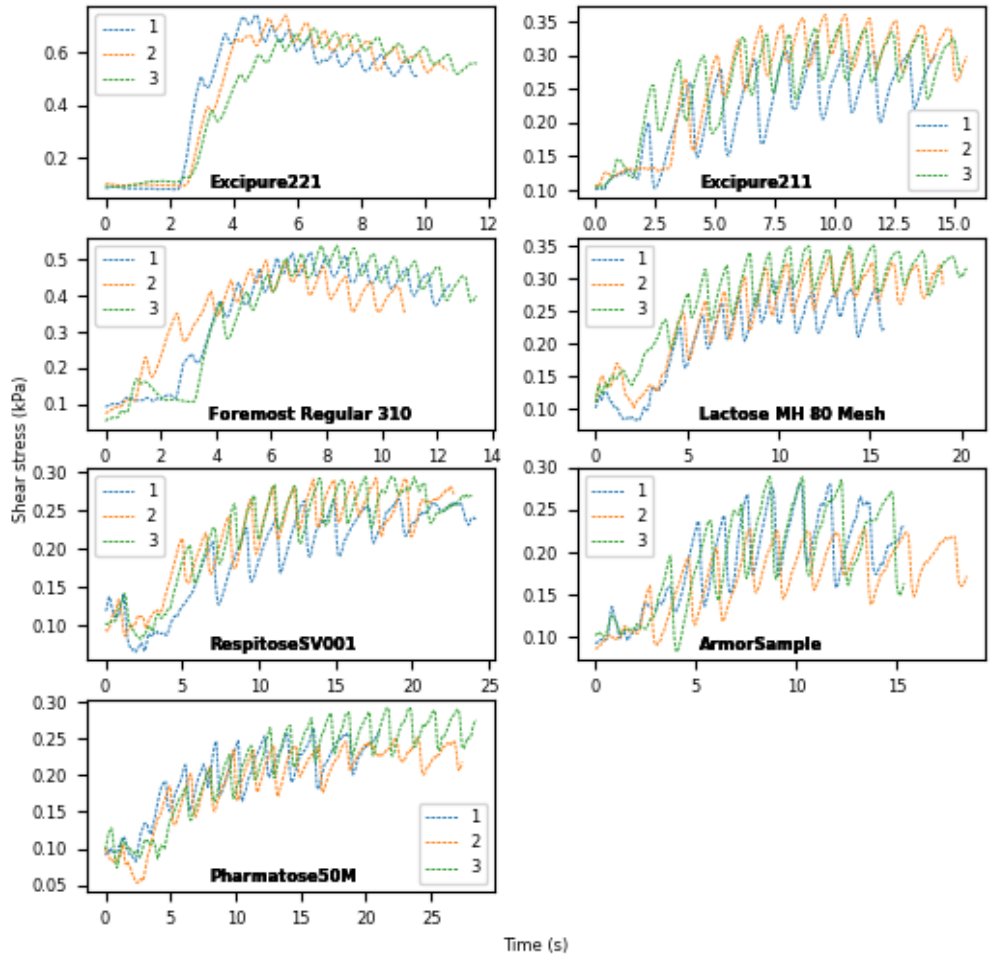

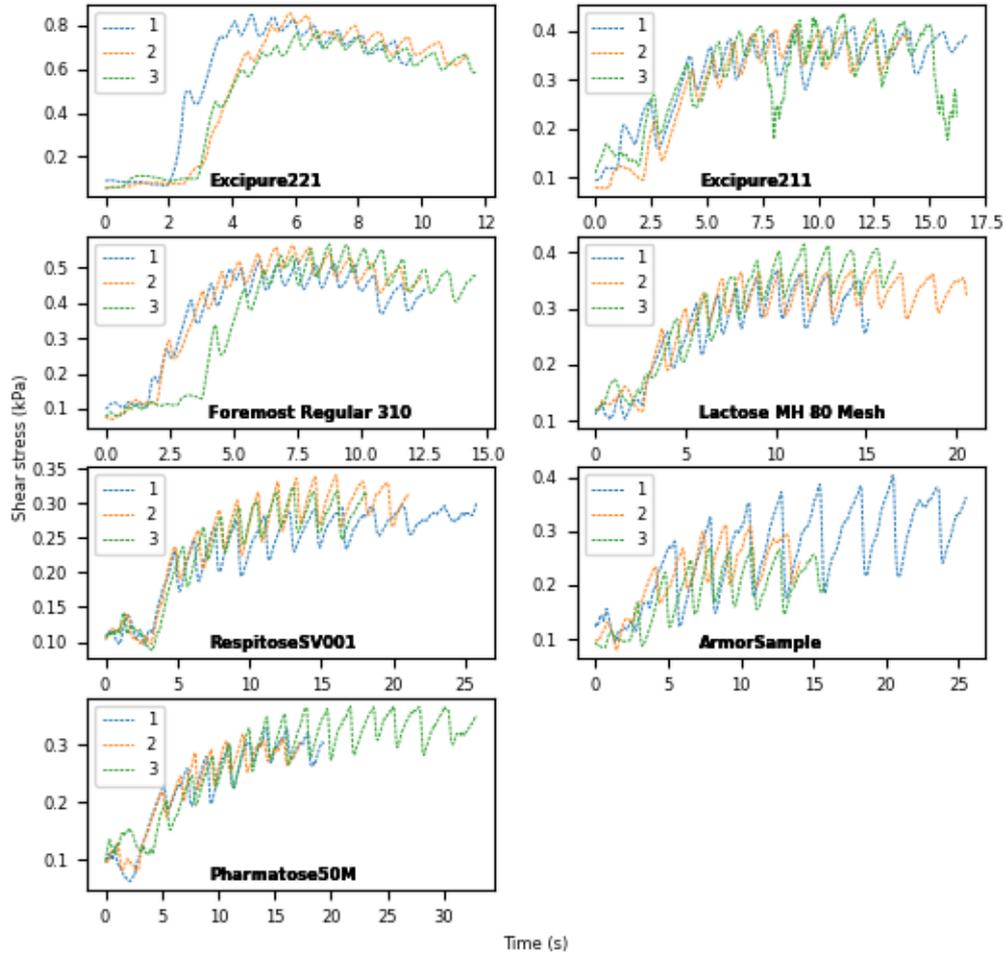

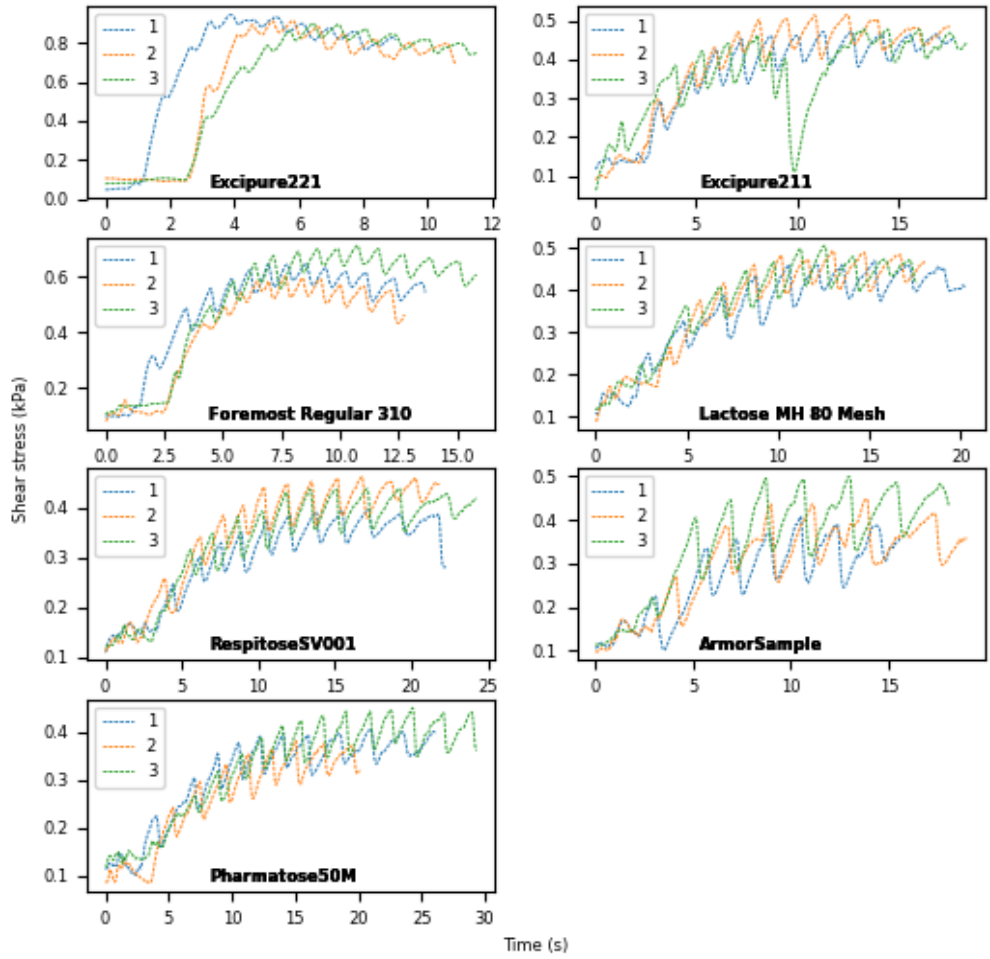

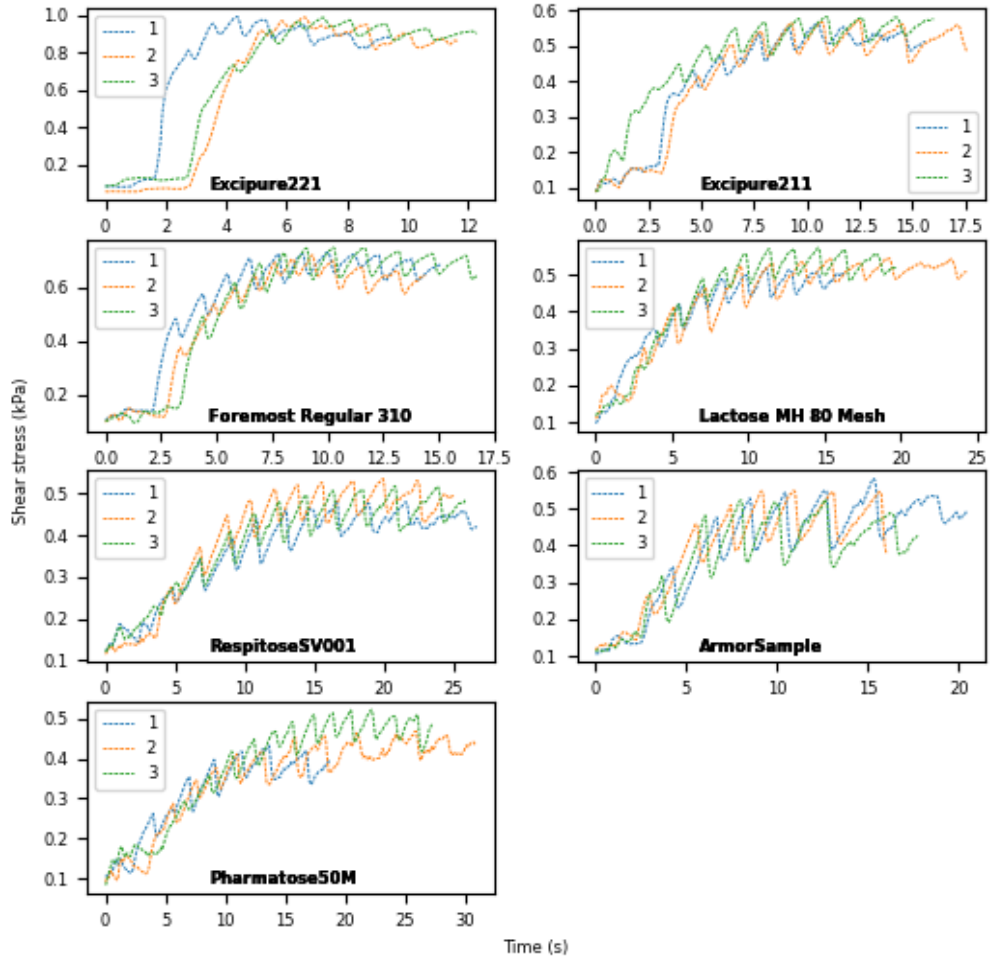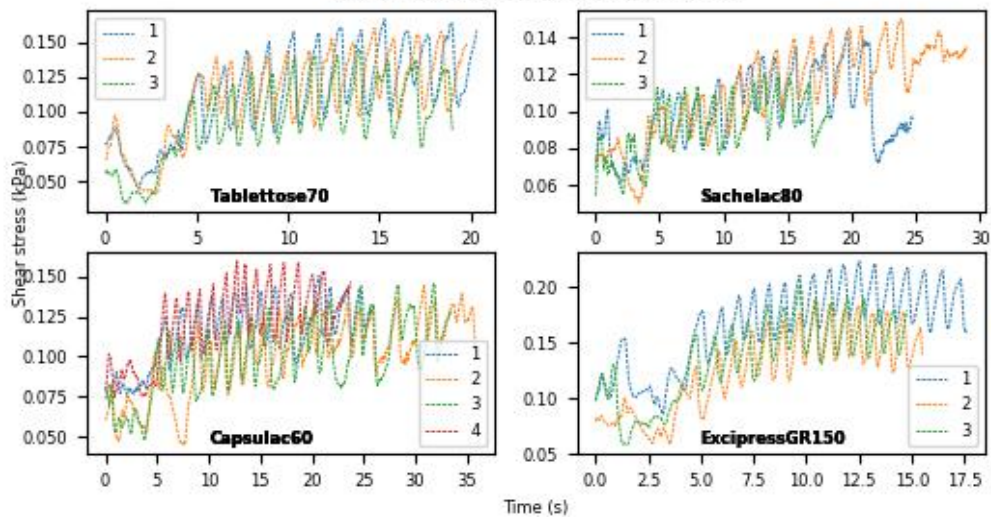

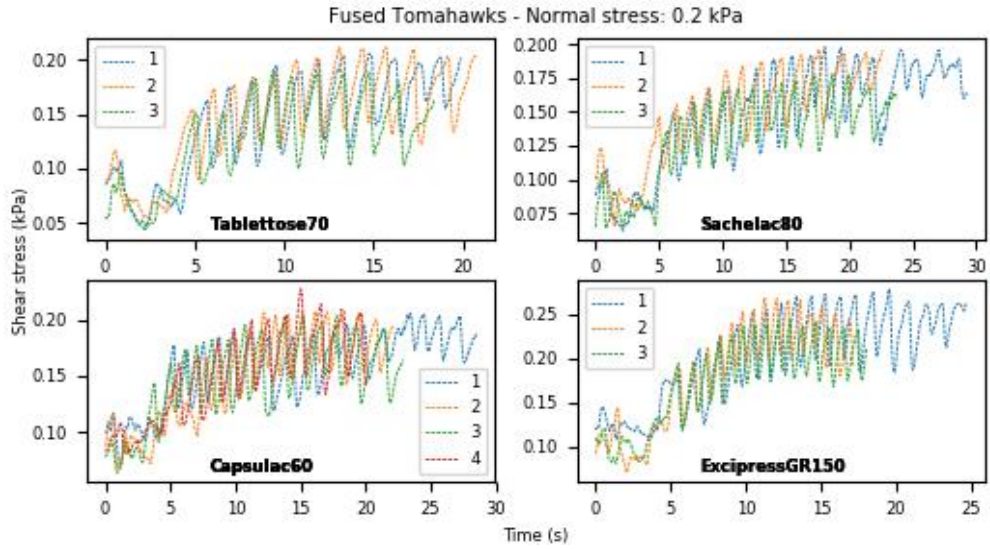
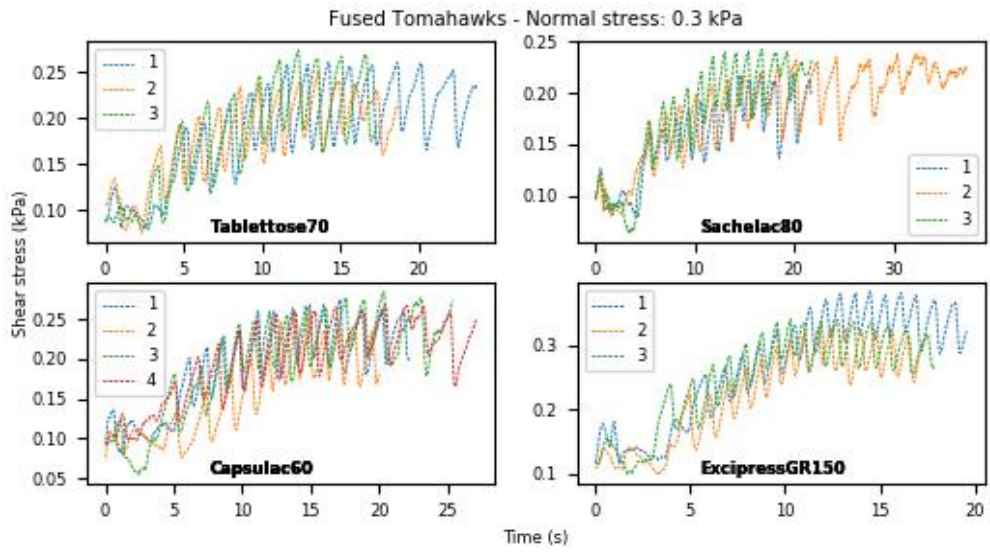
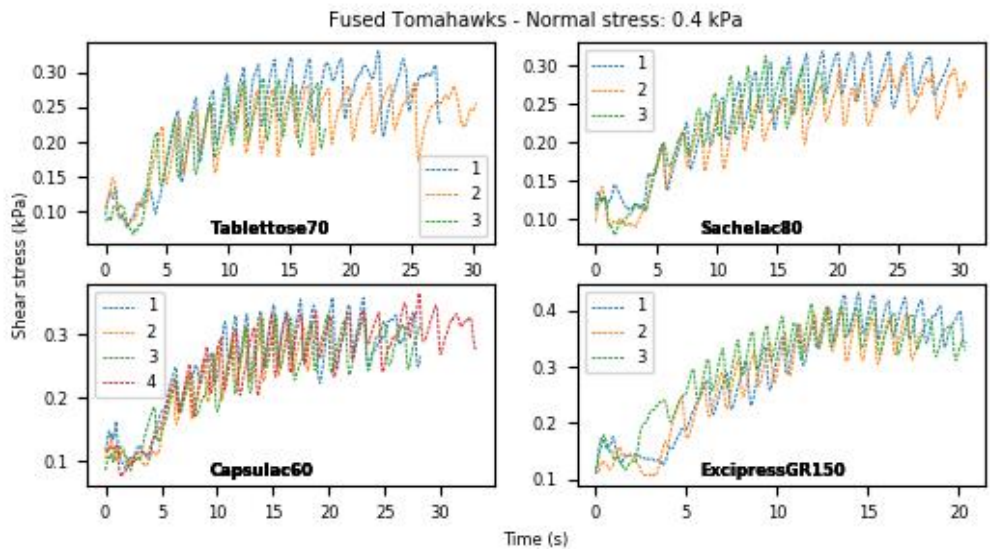

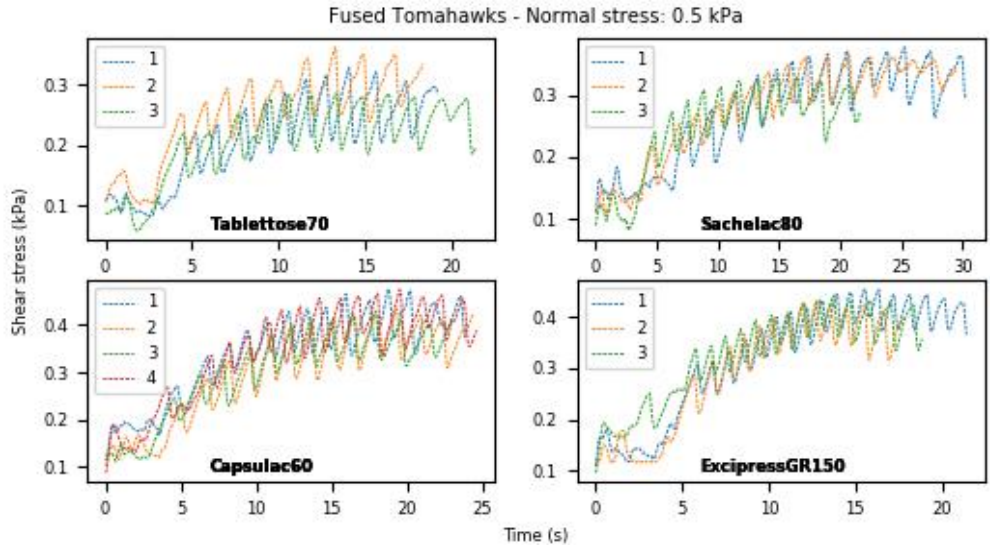
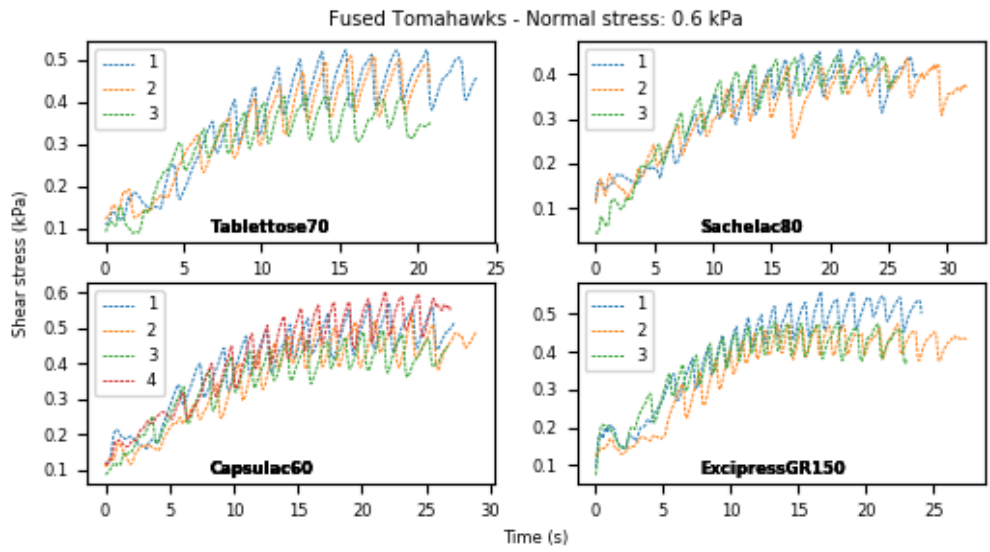
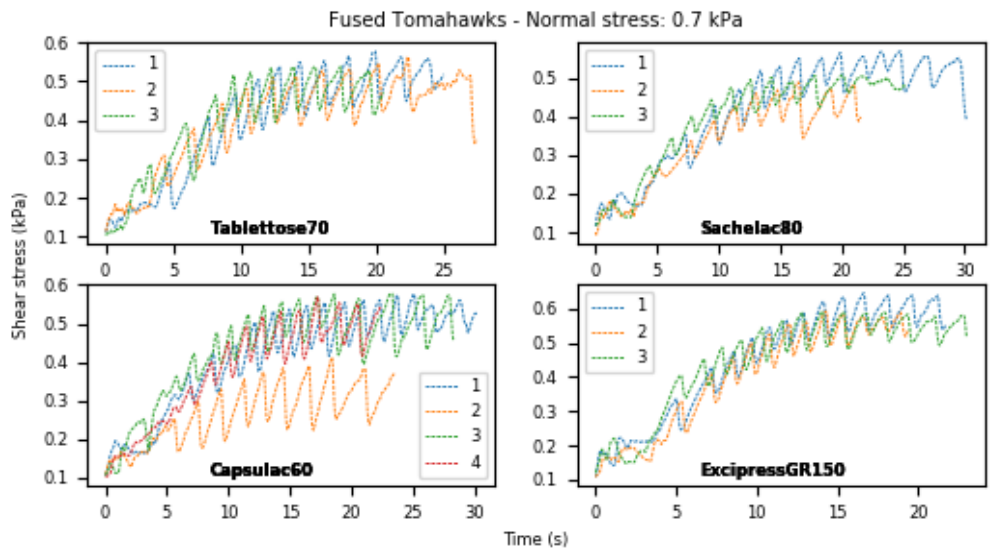

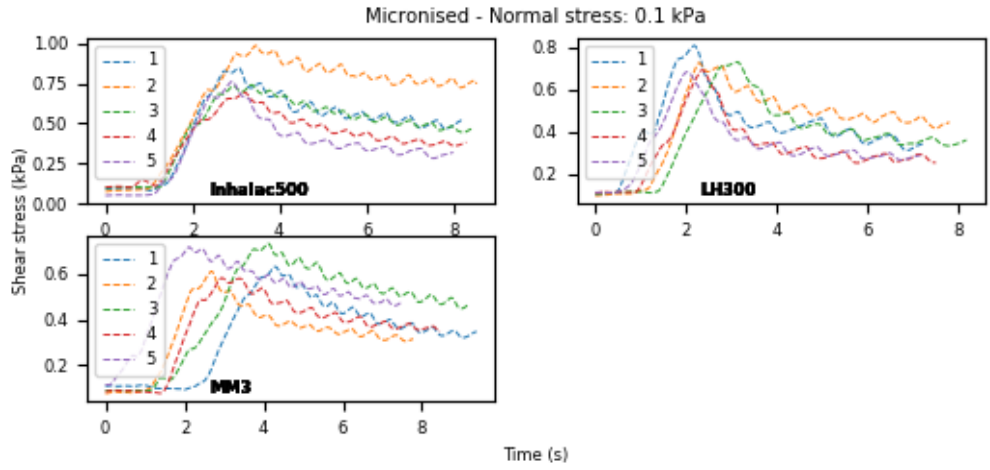

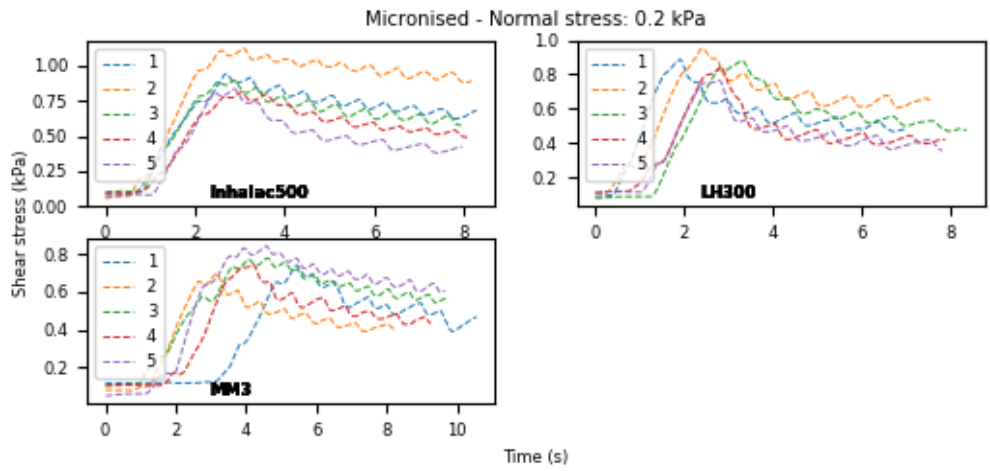

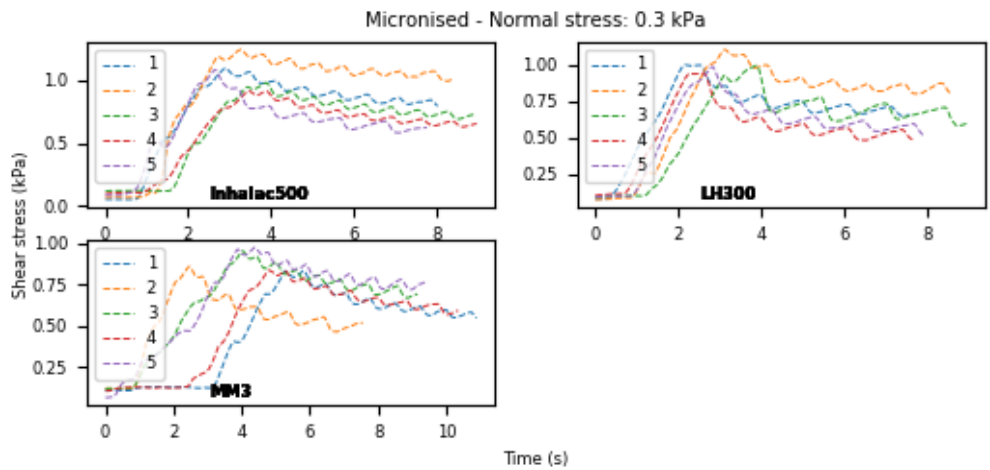

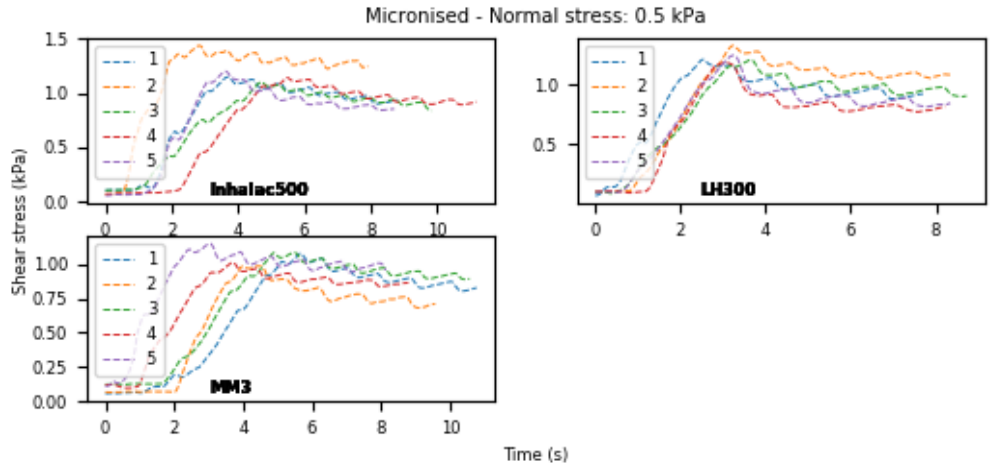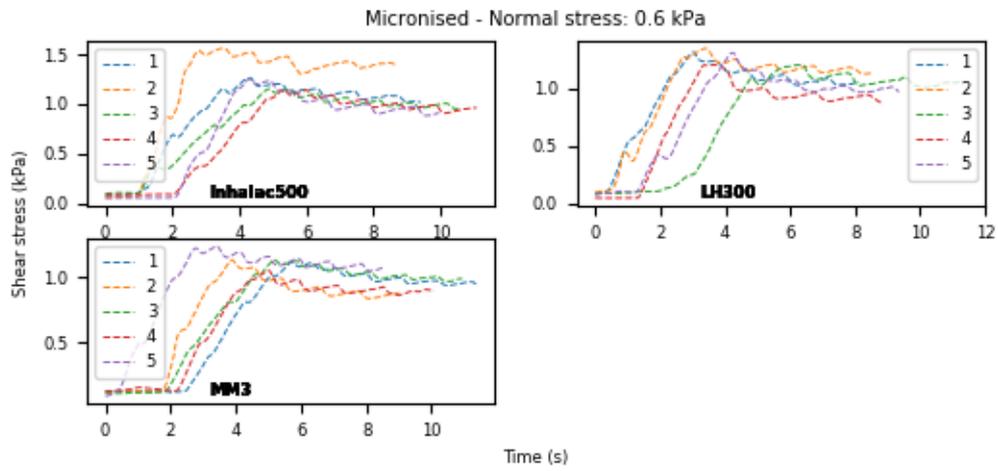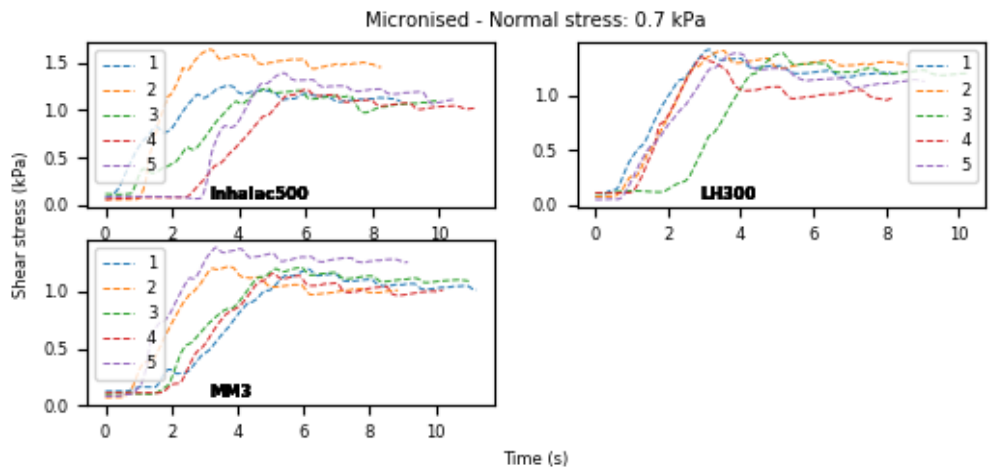

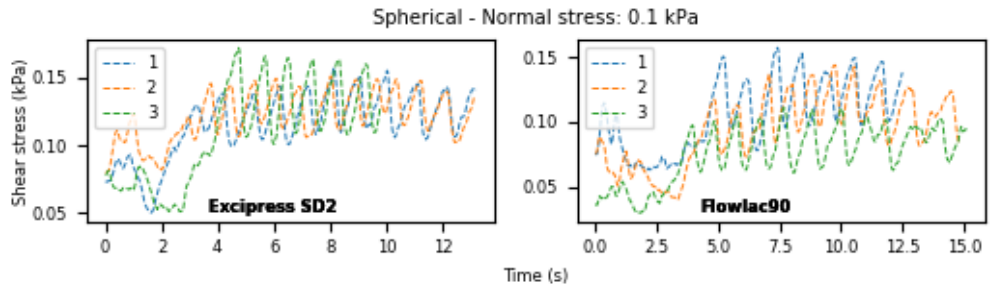
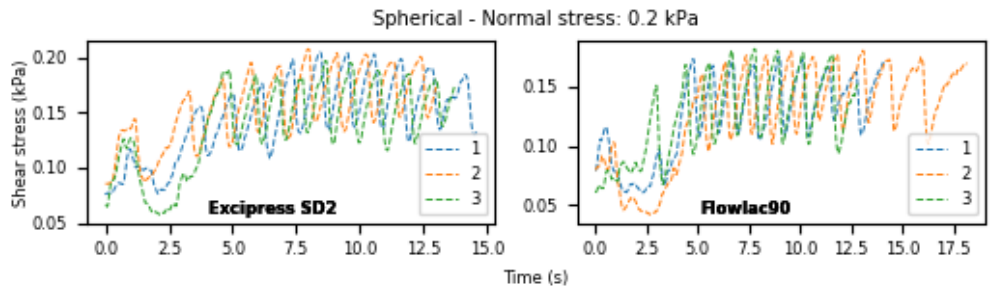
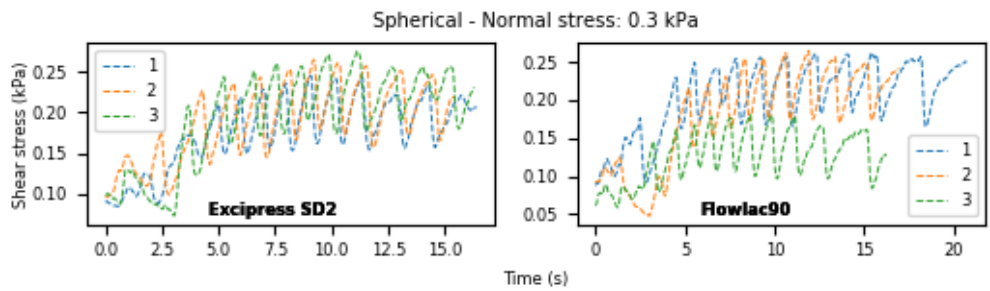
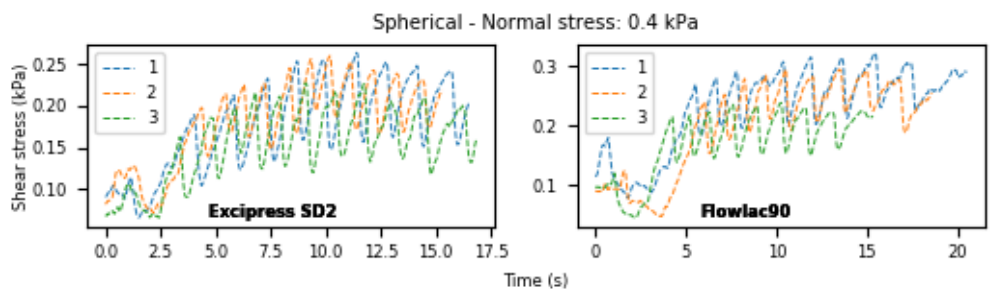
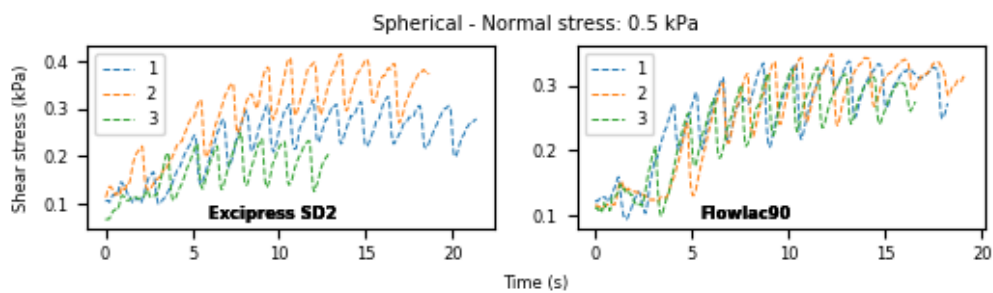

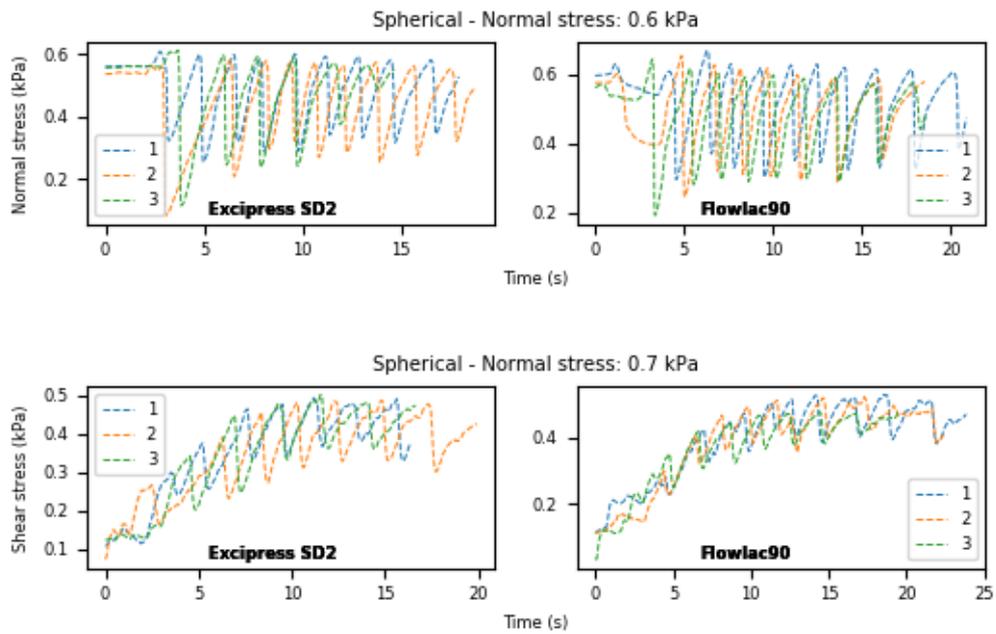

## Yield loci

Yield loci are plotted for all powders investigated and are shown grouped by shape categories (single tomahawks, fused tomahawks, spherical, sieved). Yield loci obtained from different samples of the same batch are numbered progressively.

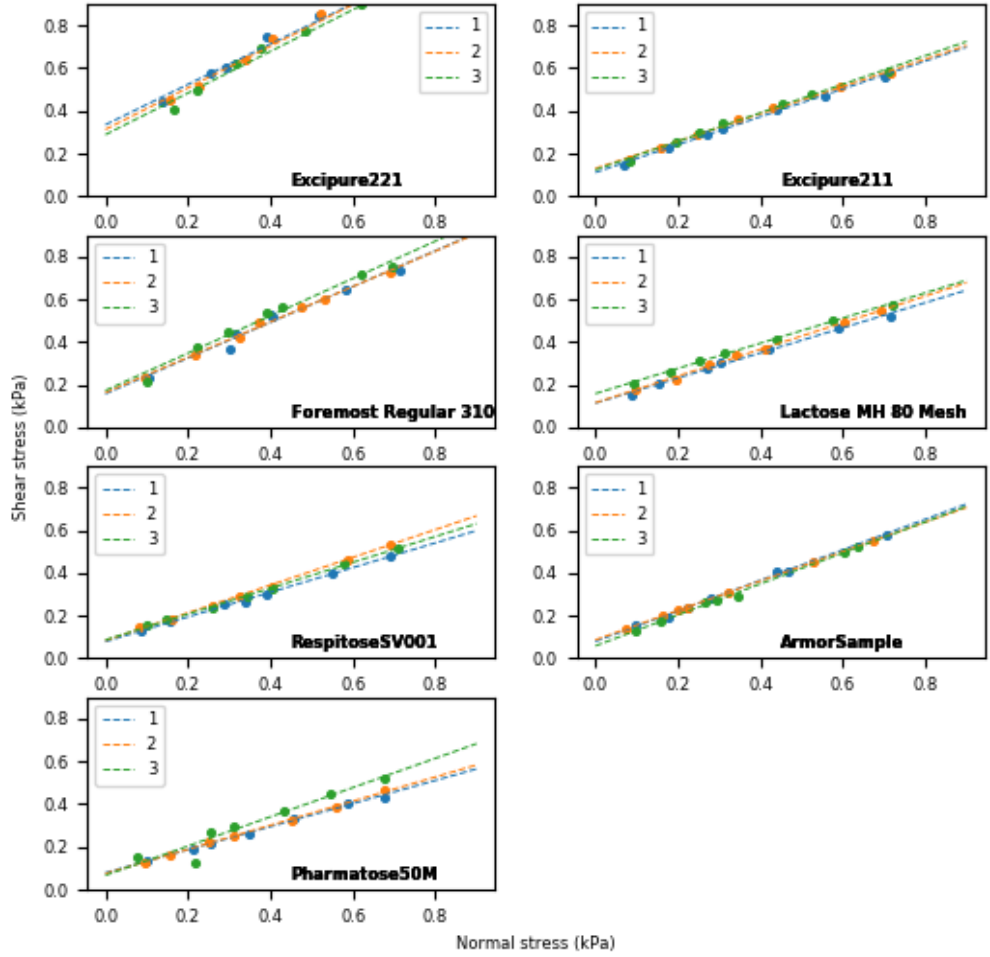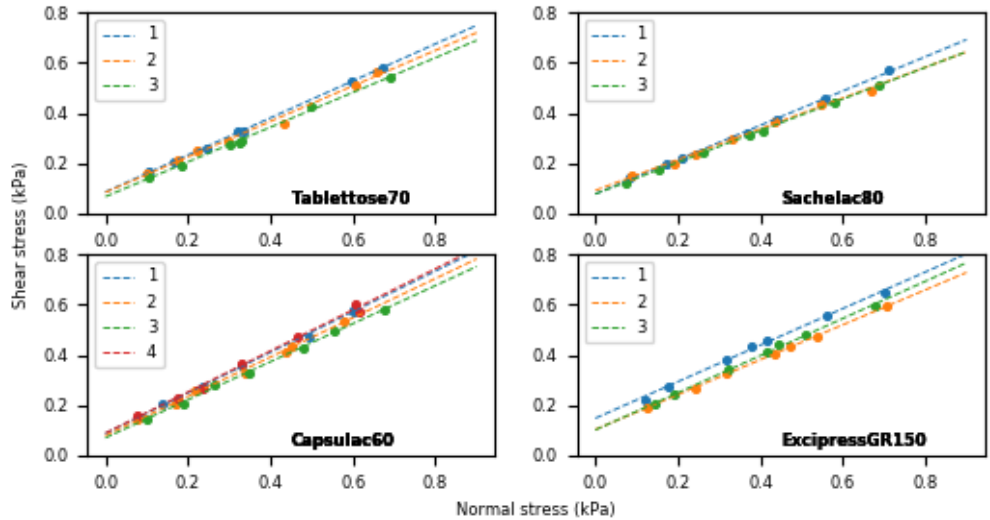

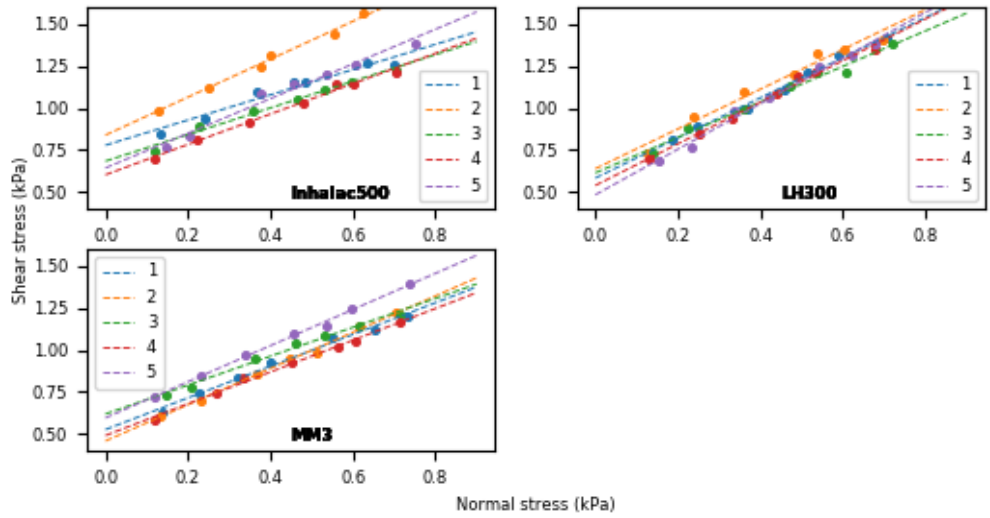
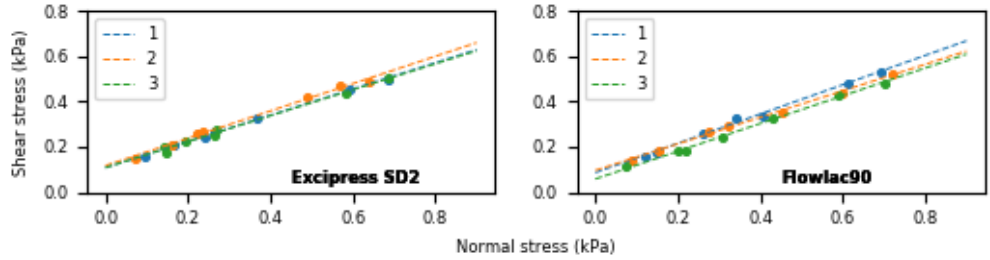
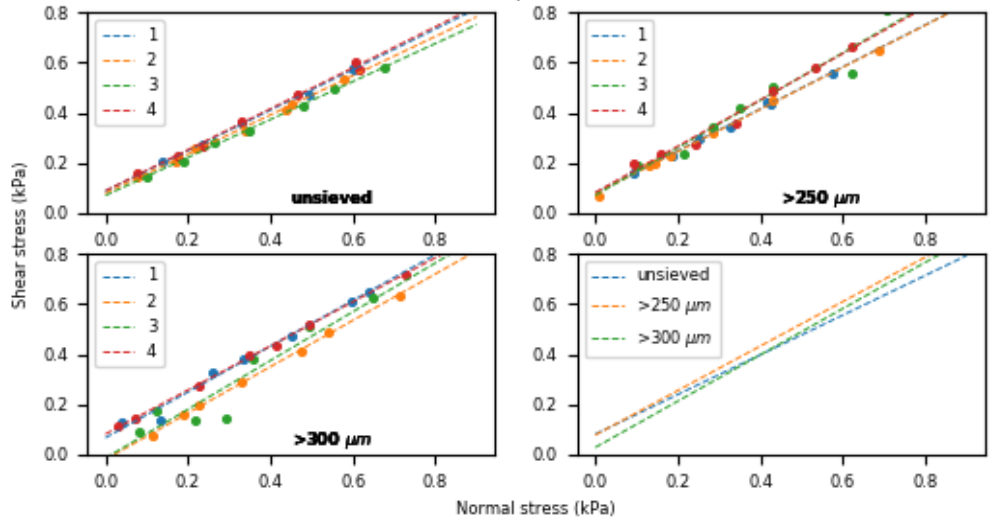

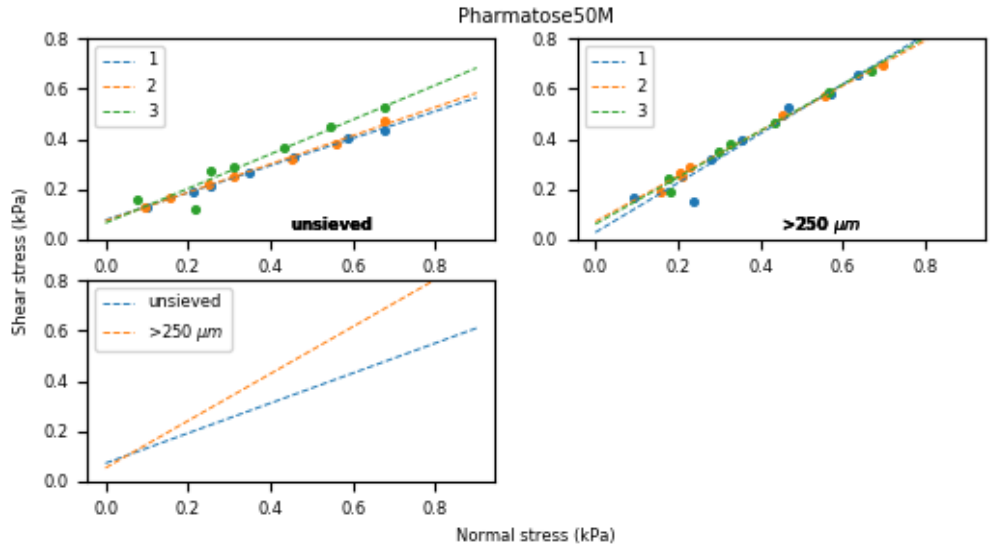

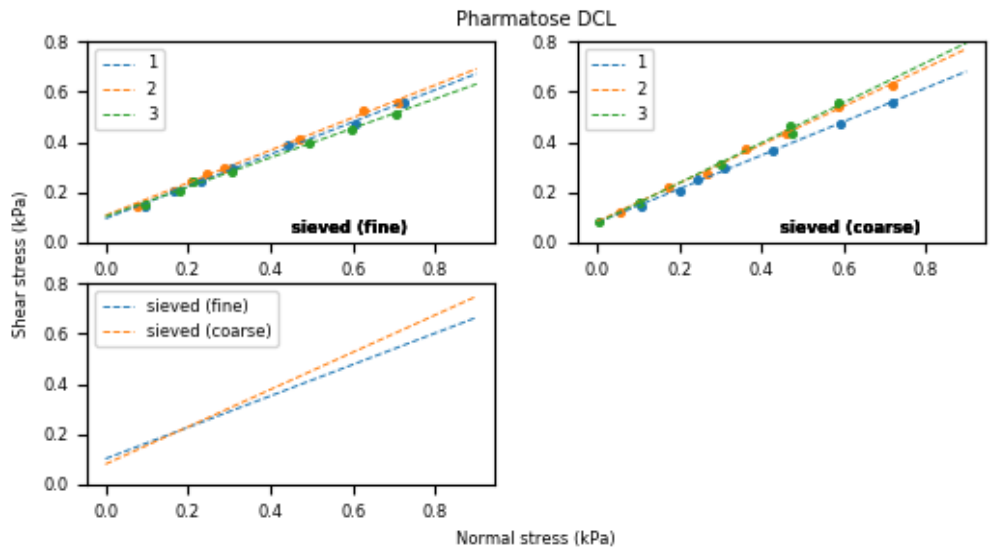

## Fitting functions

This section collects the fitting parameters obtained when using equations (3) and (4), described in the paper, to fit the yield loci of the single tomahawk lactose powders. Writing explicitly the T and C functions in equation (3) one gets:

$$\tau(\sigma) = \left(\frac{\beta}{dv_{50}^{\alpha}} + \gamma\right)\left[1 + \sigma / \left(\frac{\beta'}{dv_{50}^{\alpha'}} + \gamma'\right)\right] + \vartheta\, \sigma_{pre} \qquad (1)$$

| $\alpha$ | $\beta$ | $\gamma$ | $\vartheta$ | $\alpha'$ | $\beta'$ | $\gamma'$ |
|---|---|---|---|---|---|---|
| 0.368136 | 0.938869 | -0.087286 | 0.120542 | 0.046275 | 2.899765 | -2.174810 |

The adherence of the fitting model to the experimental data can be estimated by the sum of squared residues, the smaller the better. For the model in equation (3) such sum is 37.71, most of the residues coming from the high pre-shear curves. With equation (4) one has:

$$\tau(\sigma) = \left(\frac{\beta}{dv_{50}^{\alpha}} + \gamma\right)\left[1 + \sigma / \left(\frac{\beta'}{dv_{50}^{\alpha'}} + \gamma'\right)\right] + \left(\frac{\vartheta}{dv_{50}}\right)^{\chi} \sigma_{pre} \qquad (2)$$

| $\alpha$ | $\beta$ | $\gamma$ | $\vartheta$ | $\alpha'$ | $\beta'$ | $\gamma'$ | $\chi$ |
|---|---|---|---|---|---|---|---|
| 0.24118 | 0.75337 | -0.09510 | 0.07696 | 0.01925 | 5.08256 | -4.4.2017 | 0.48904 |

Here the sum of squared residues is 24.37 due to a better fit of the high pre-shear data, the residues of the low pre-shear data rise a little. Notice that the fitting parameters for the two intercepts *C* and *T* are in close agreement with the ones obtained in section 3.3 using only the data for $\sigma_{pre} = 1$ kPa.

# Stick-slip parameters profiles

## Stick time vs time

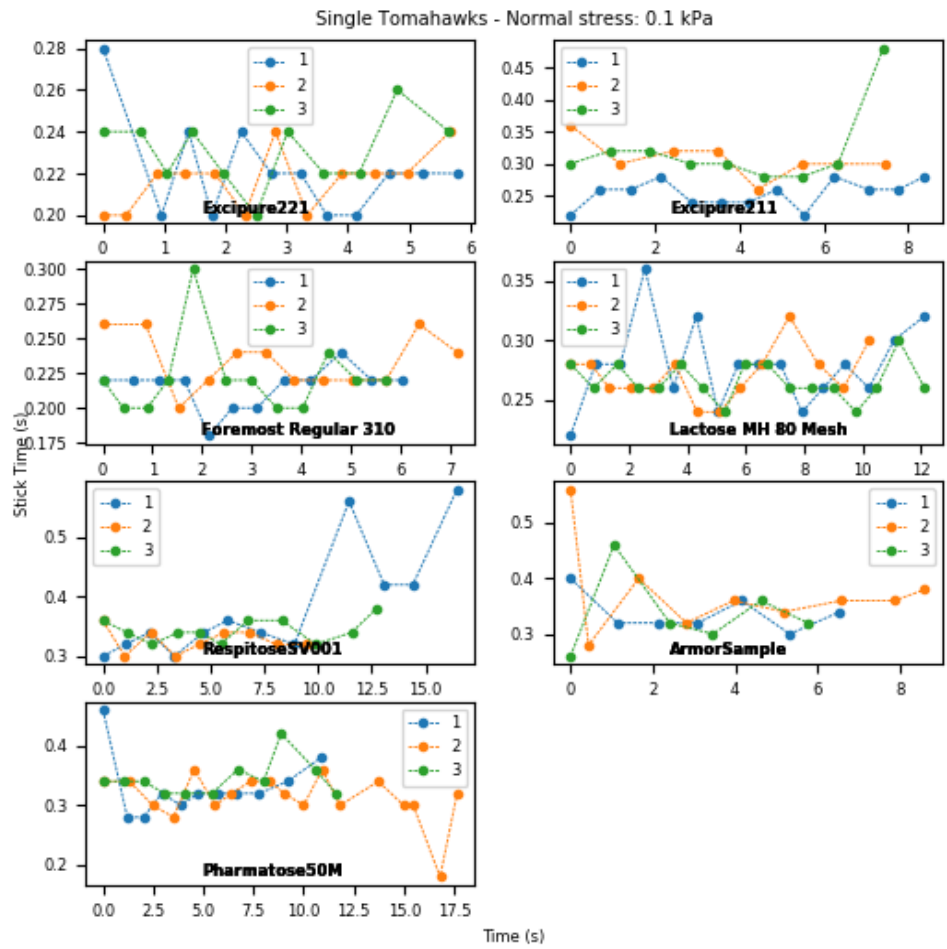

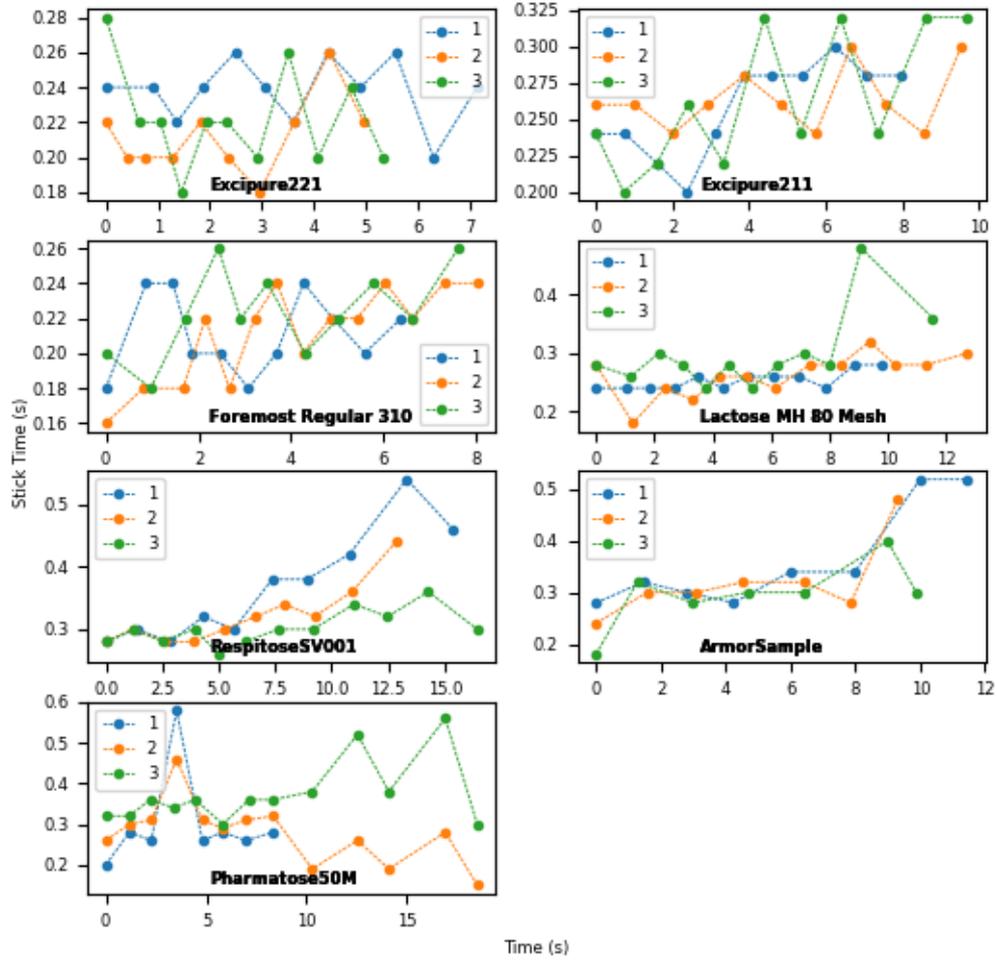

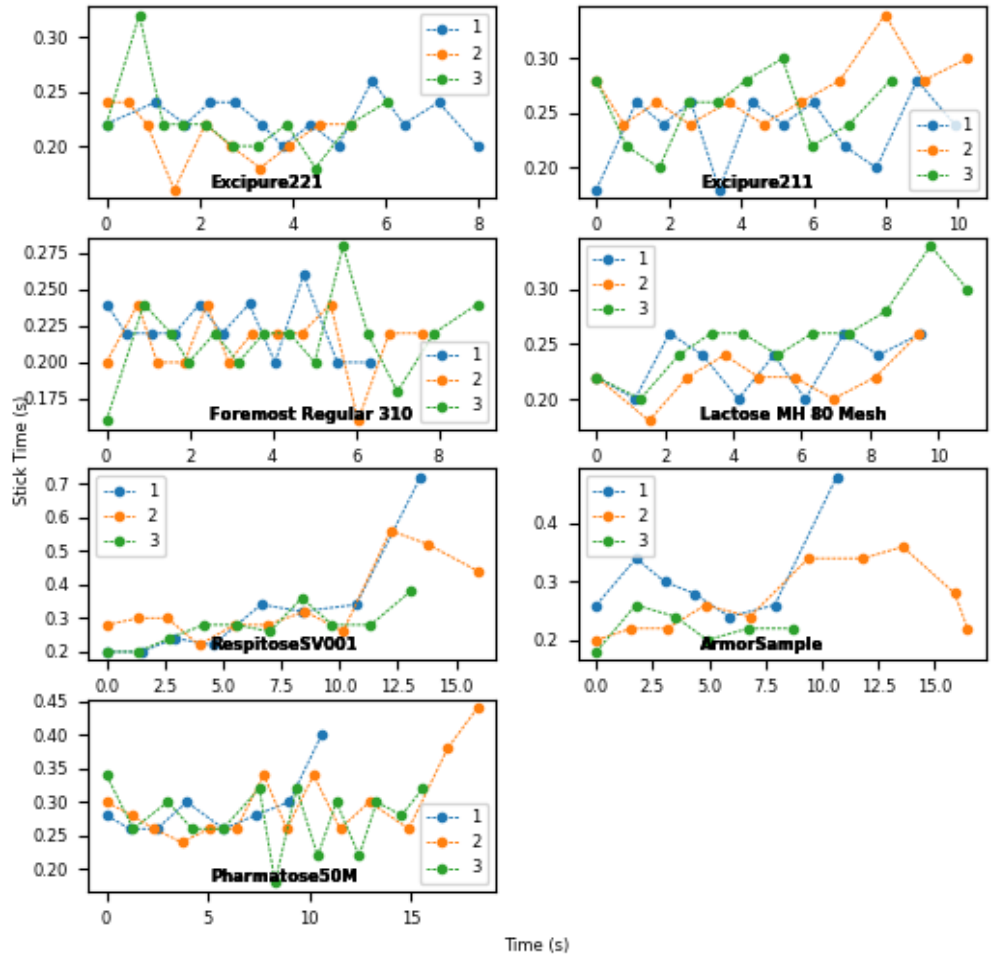

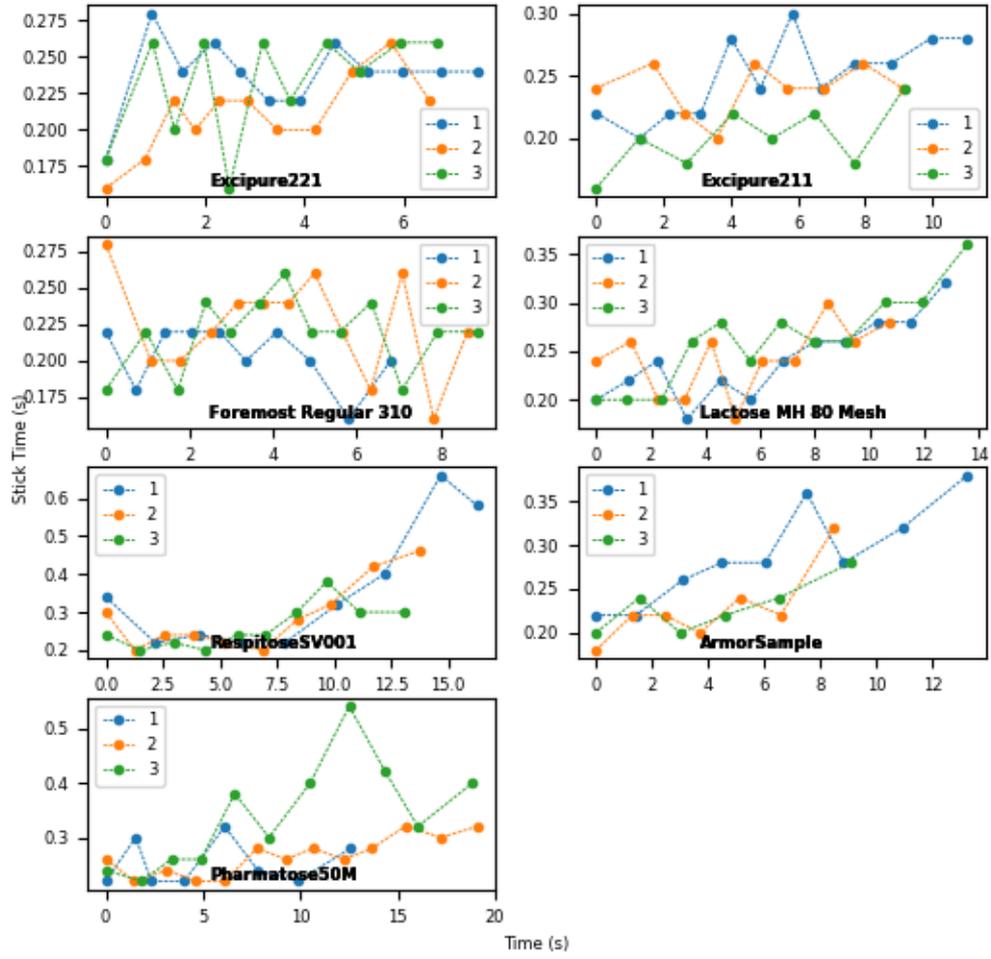

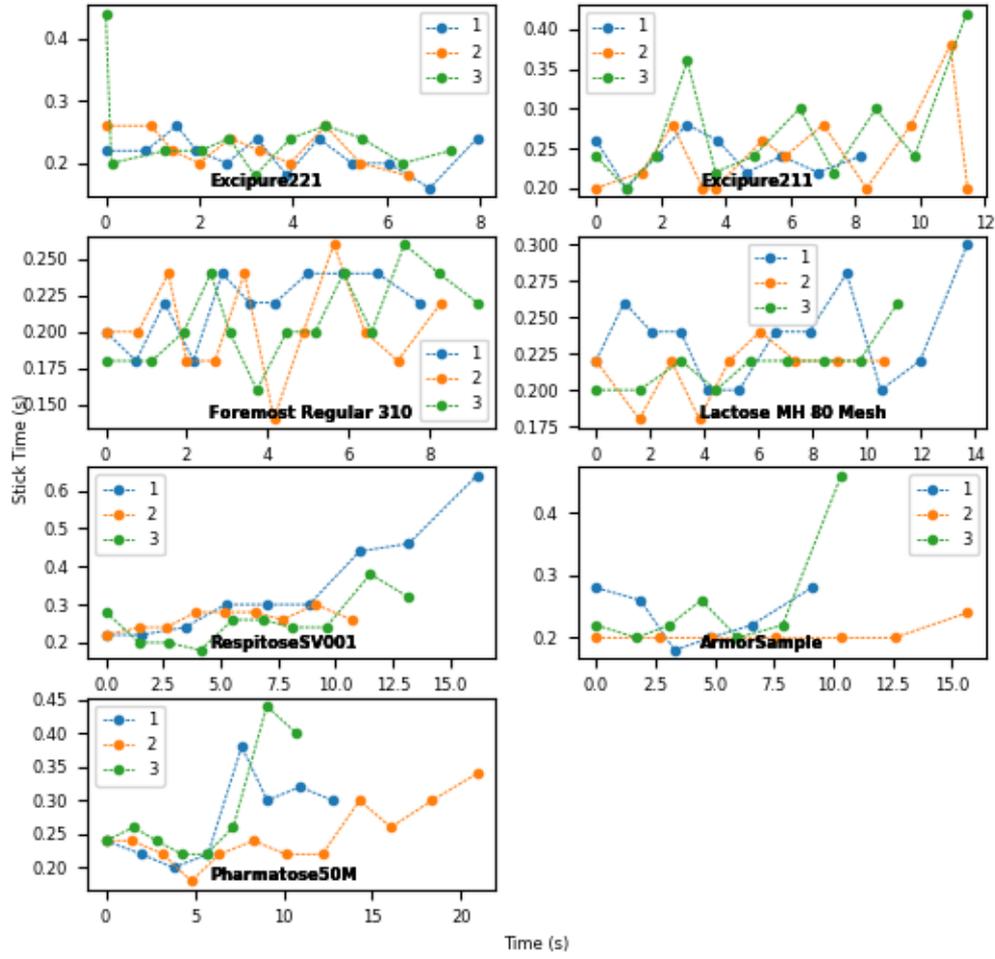

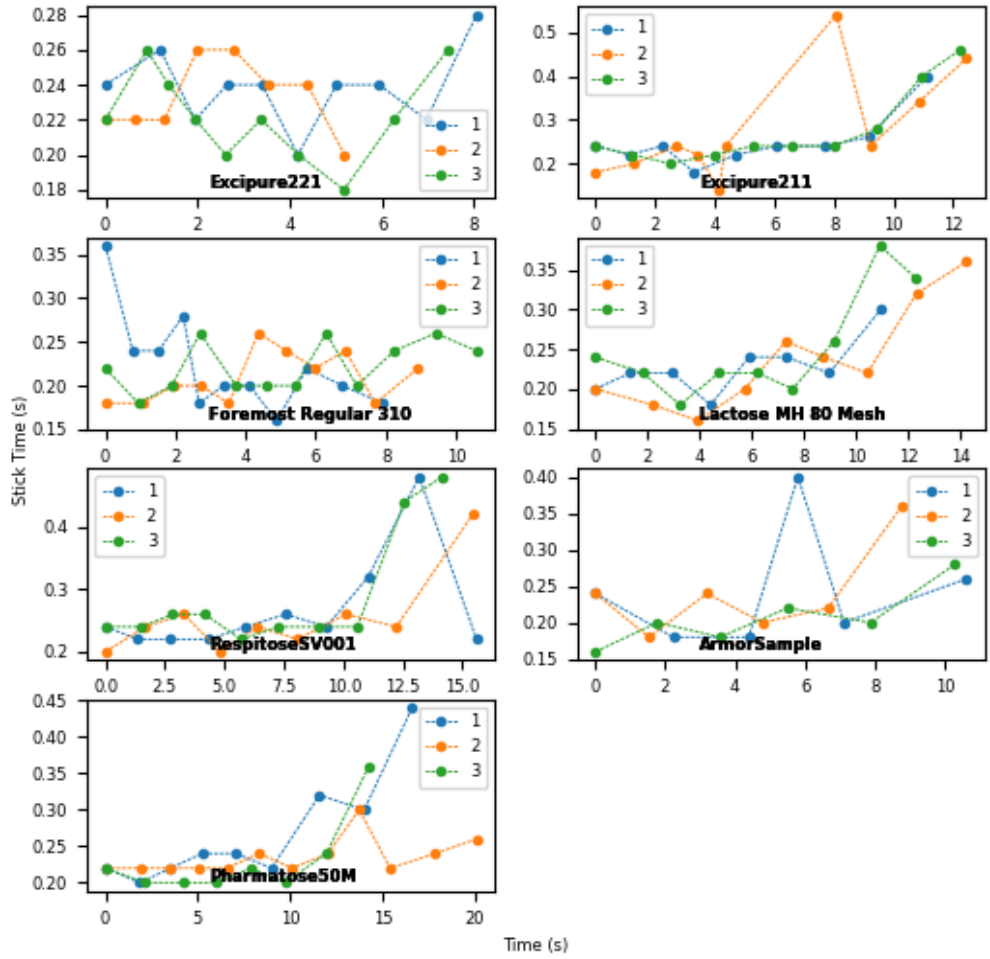

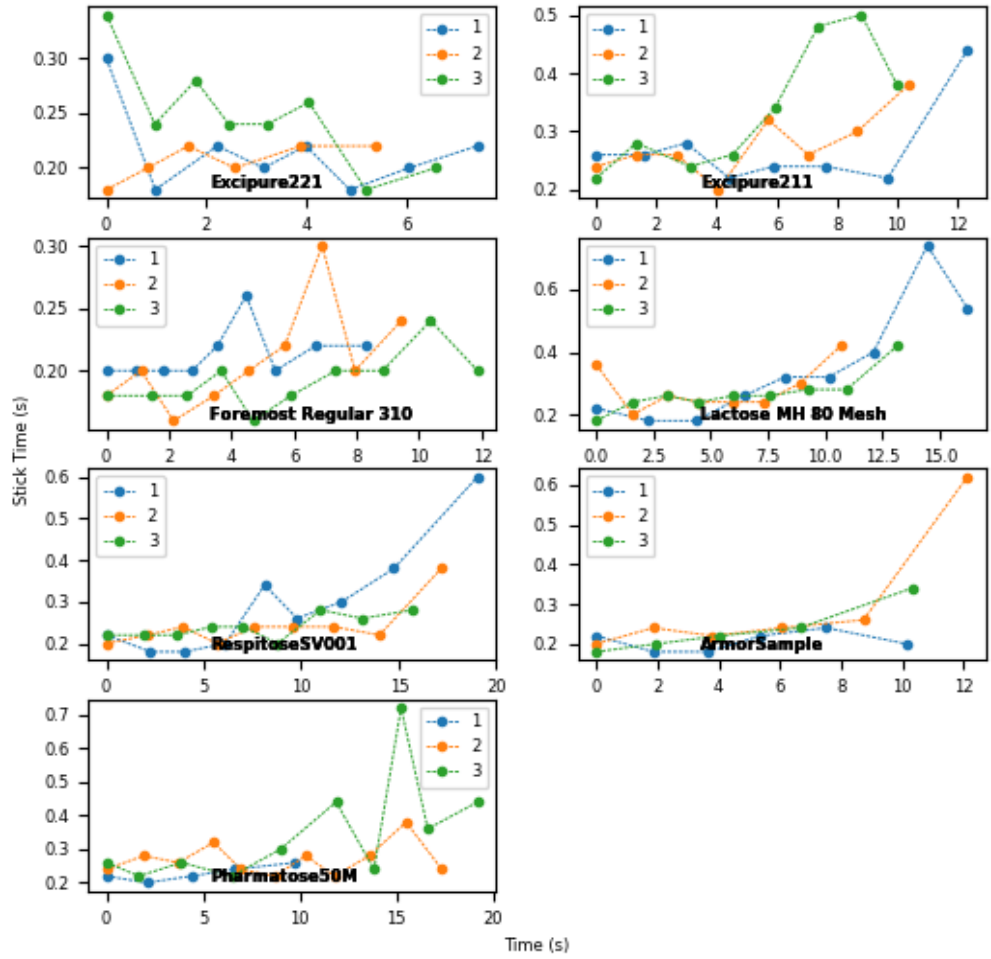

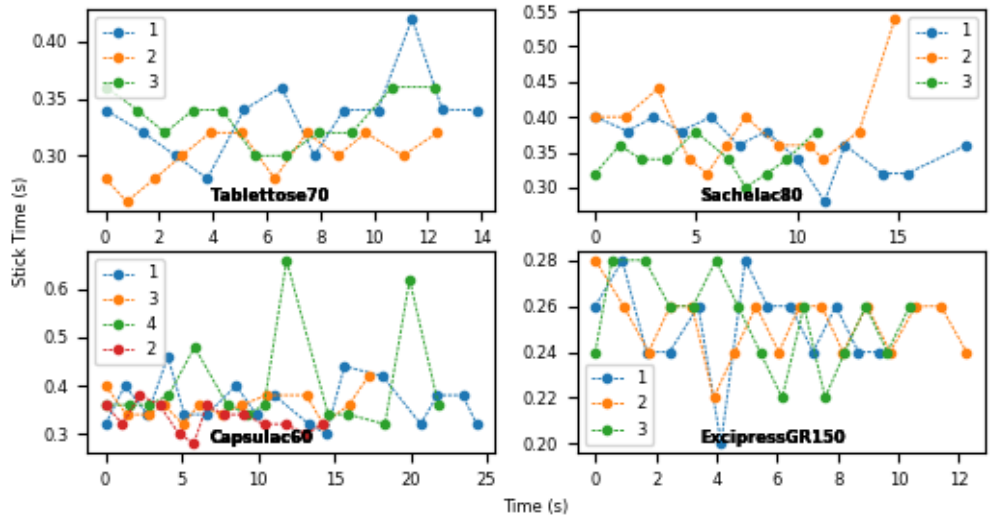
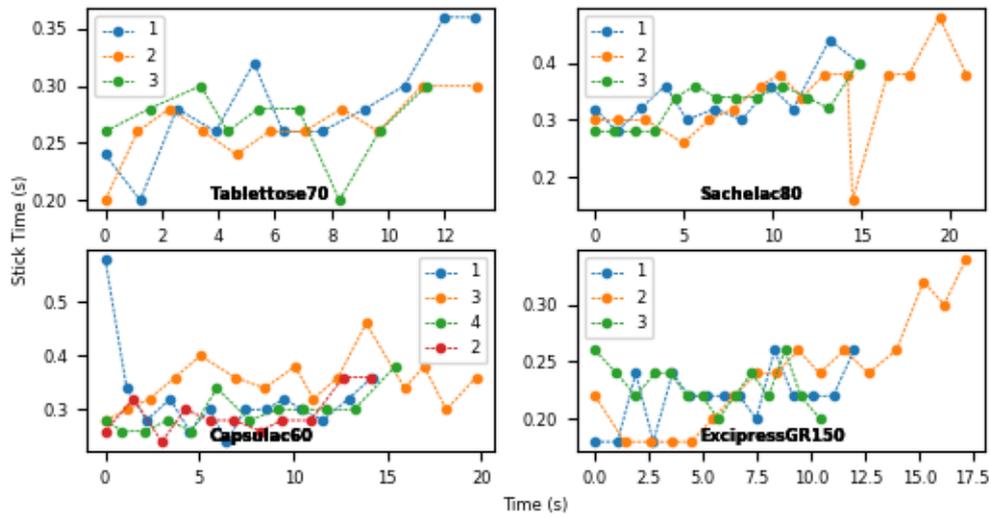
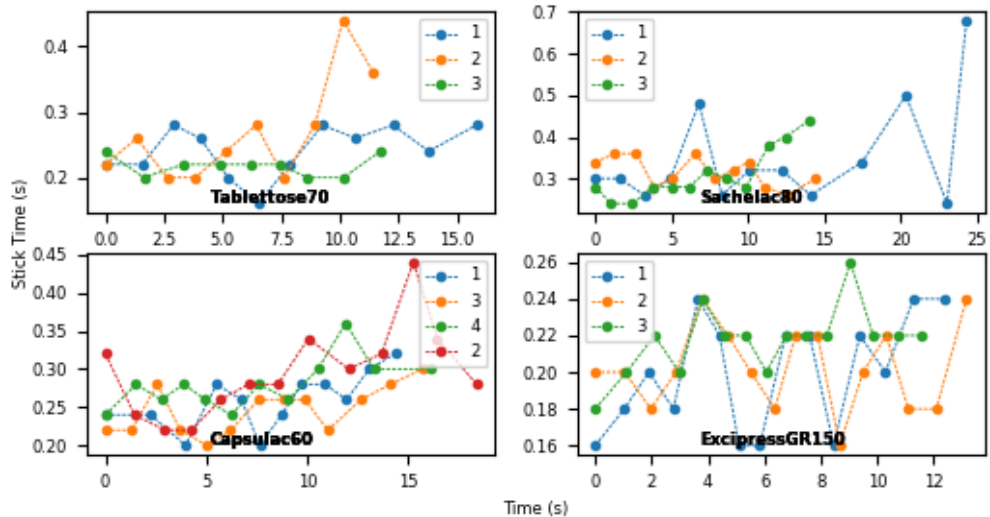

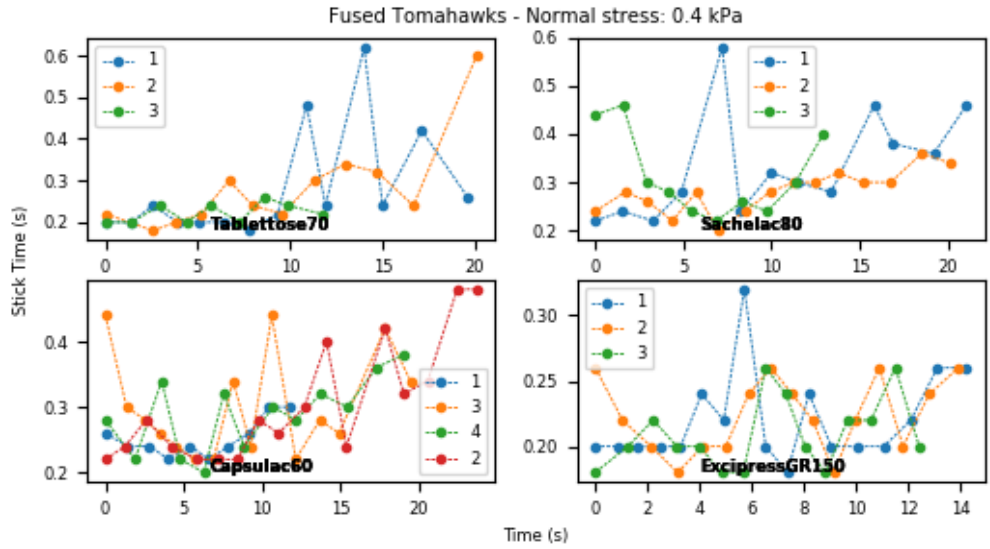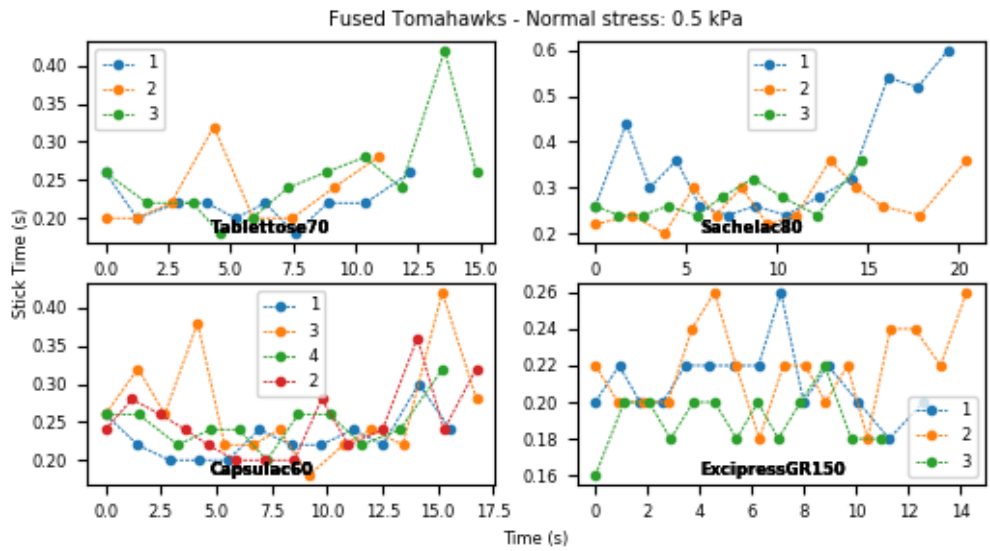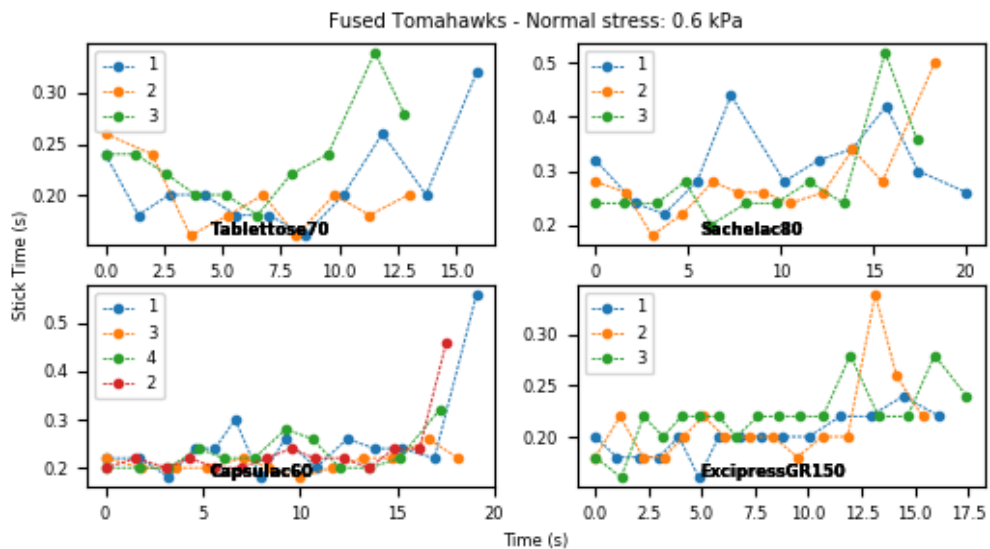

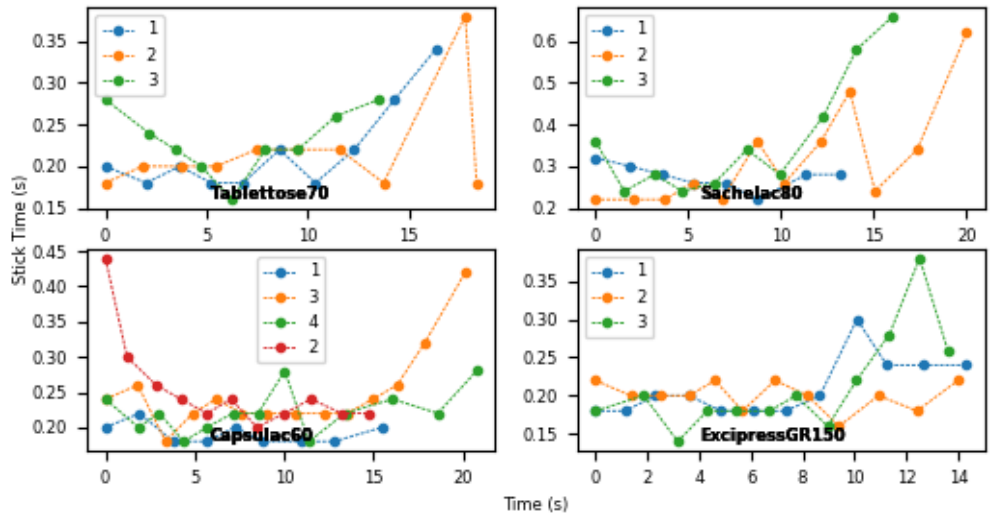
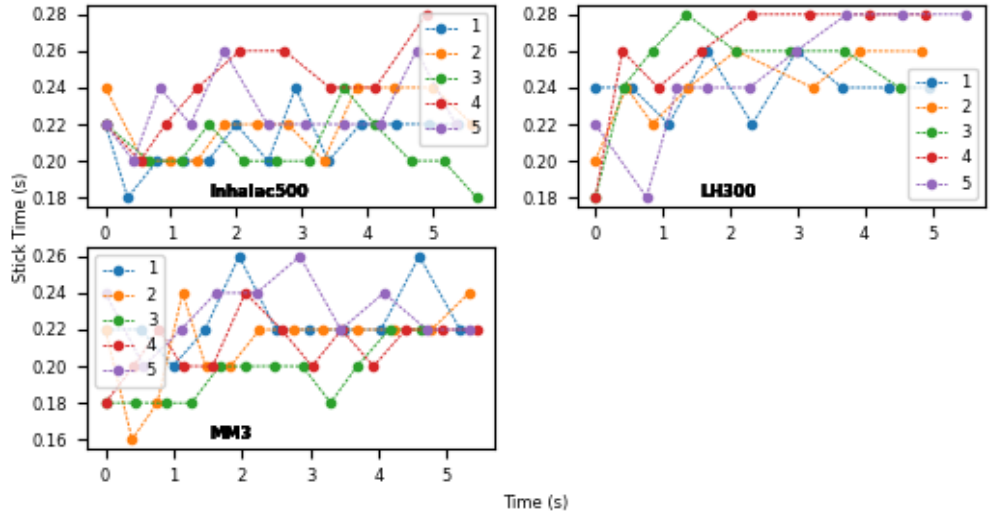
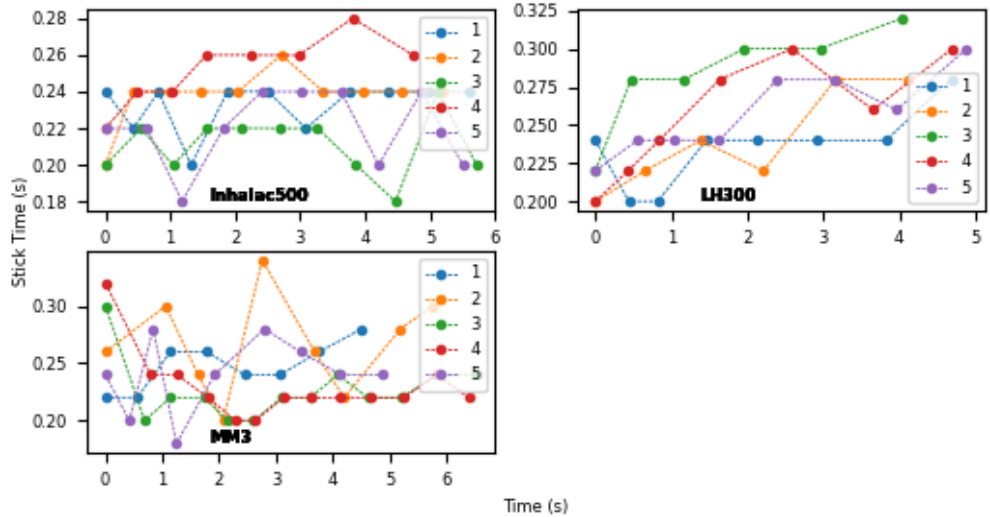

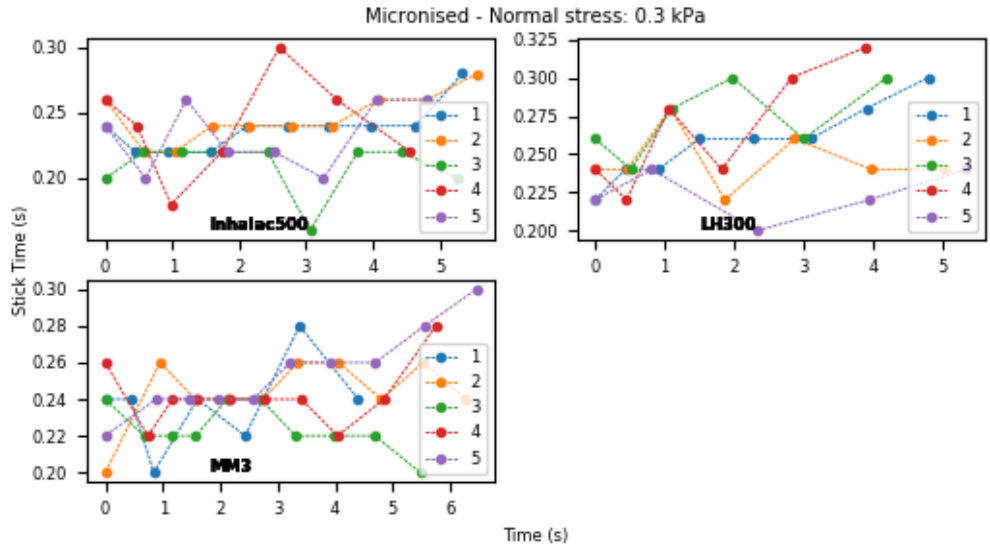
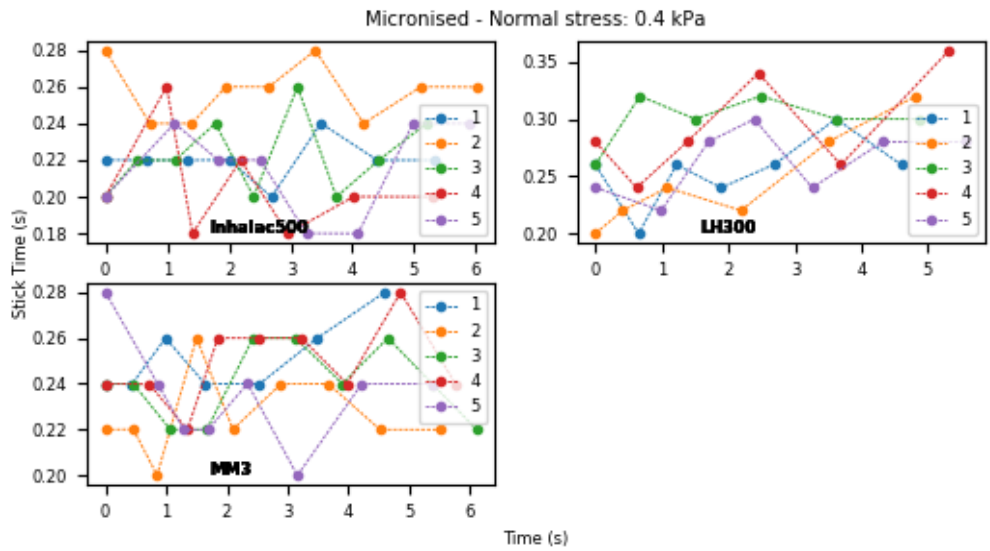
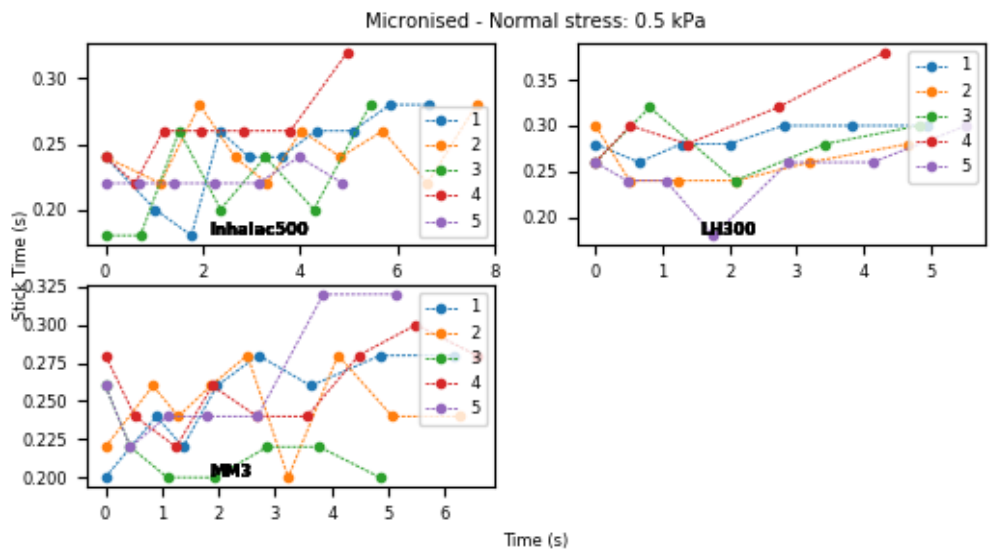

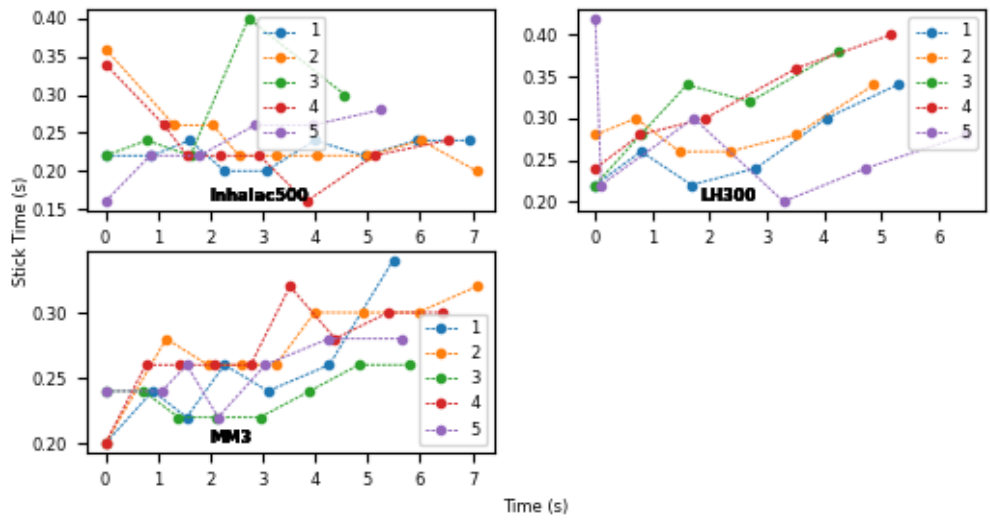
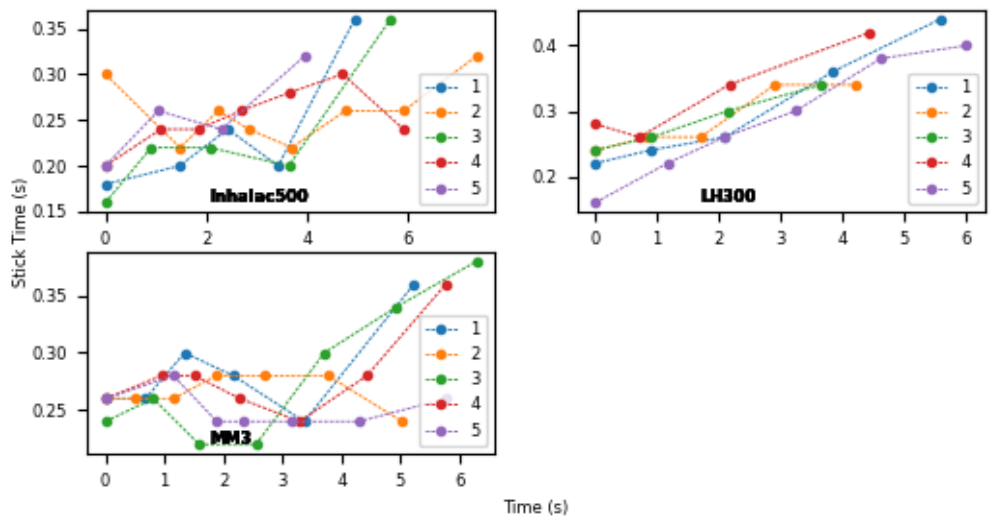
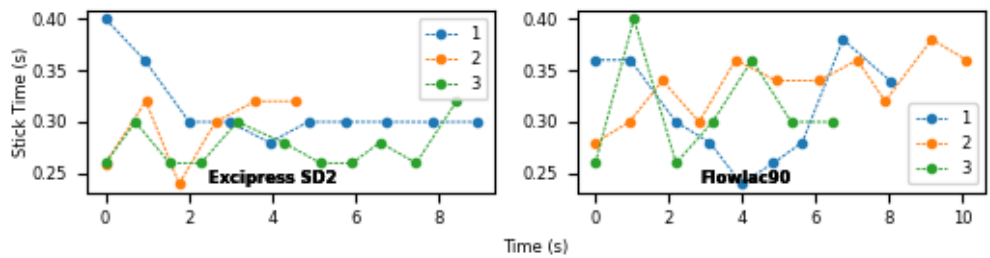
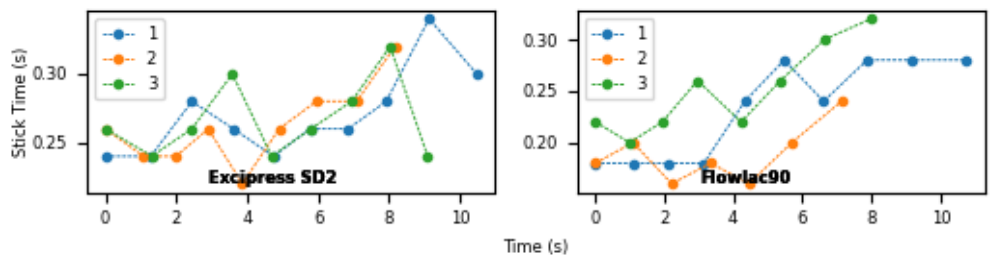

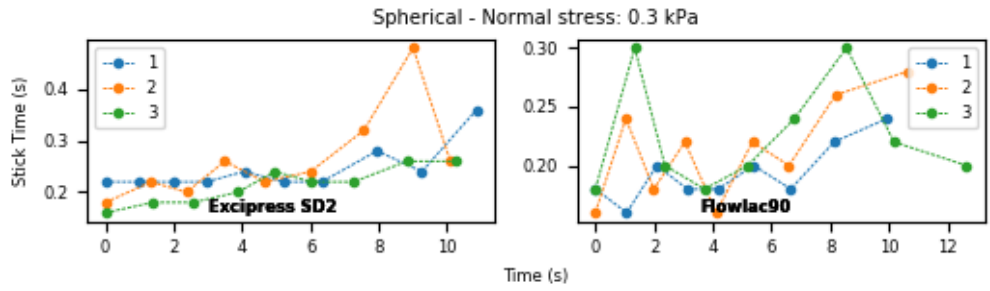
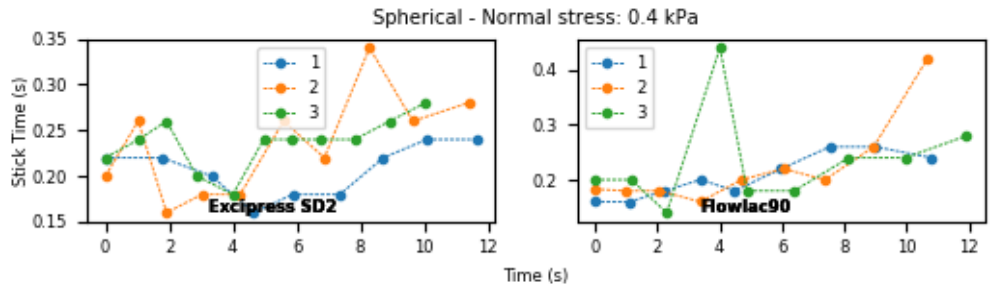
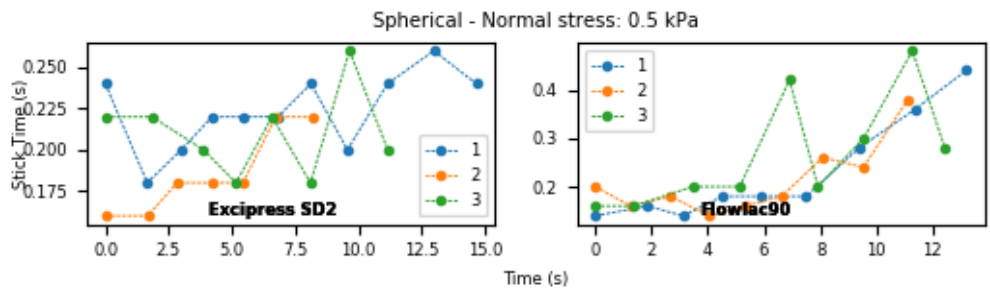
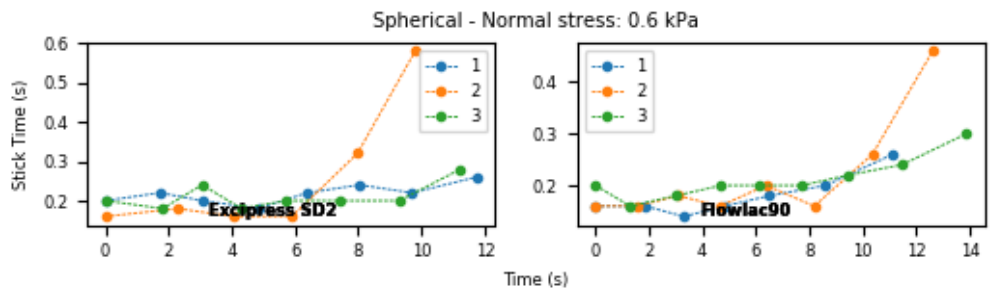
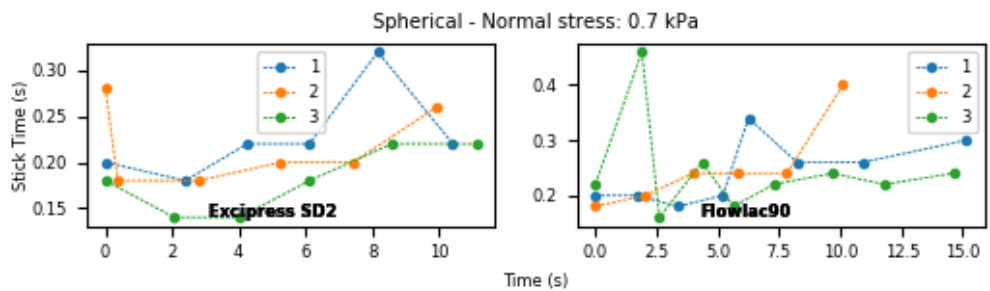

Stick-slip amplitude vs normal stress

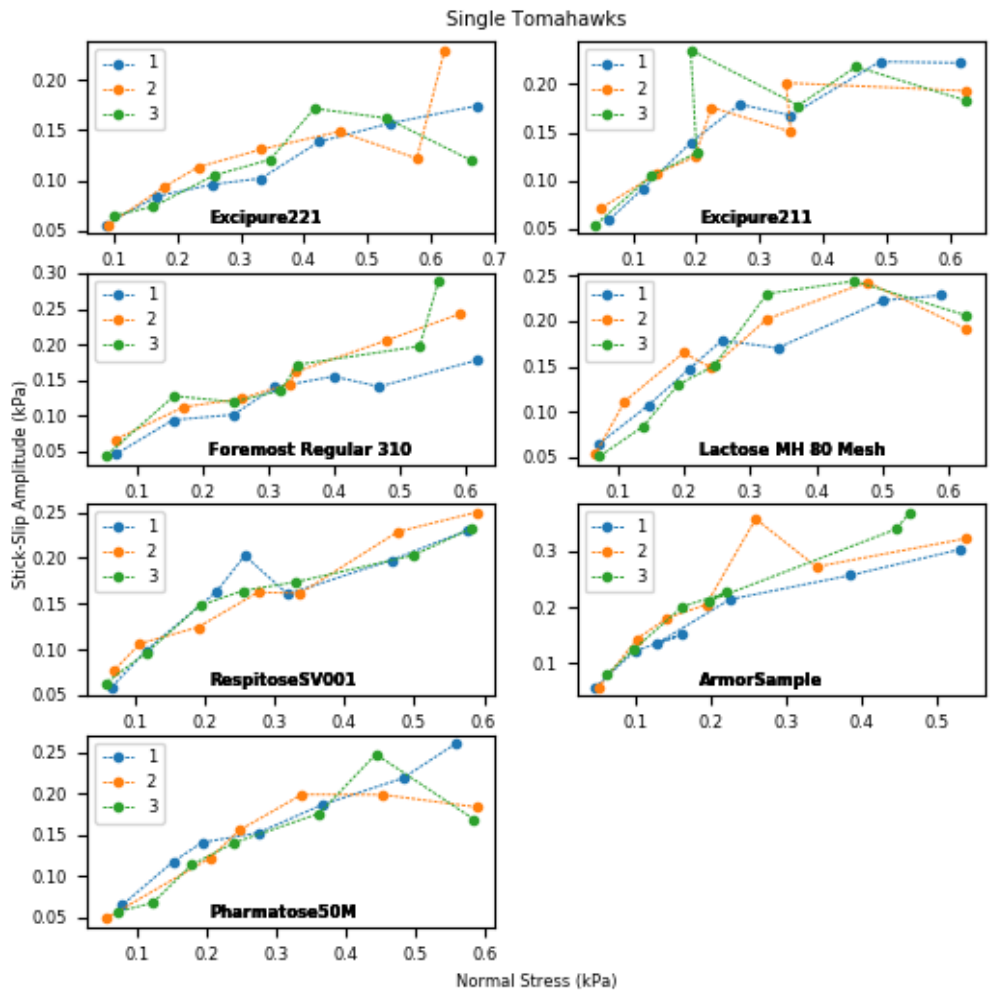

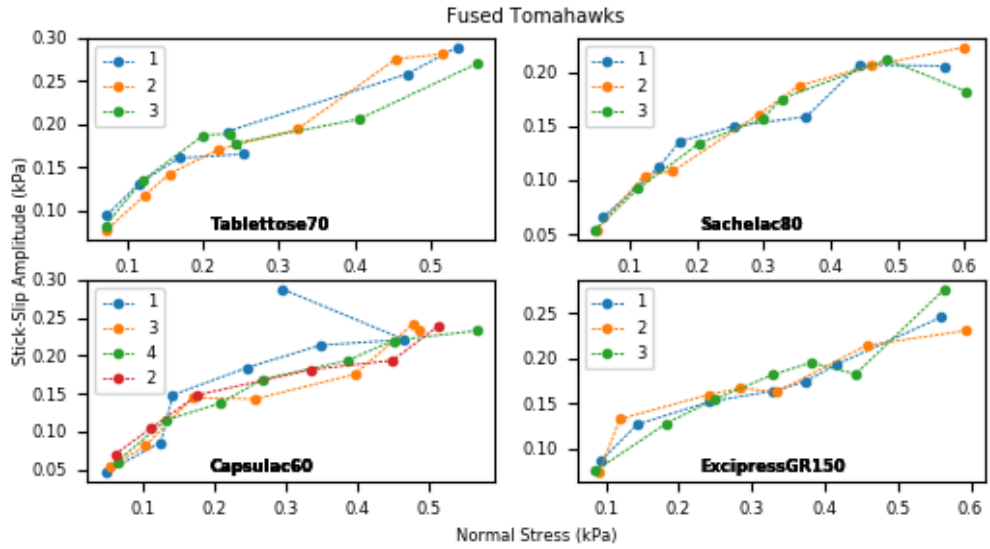
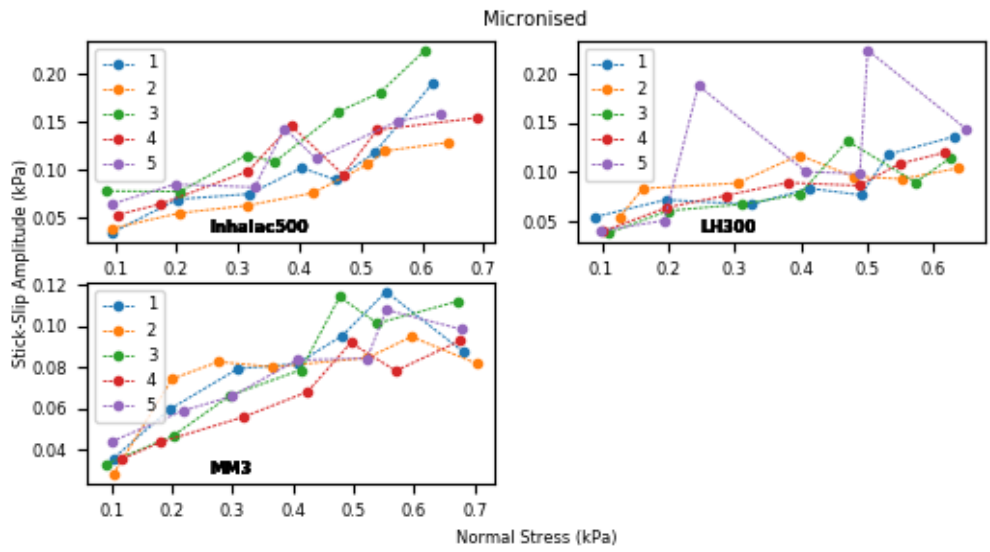
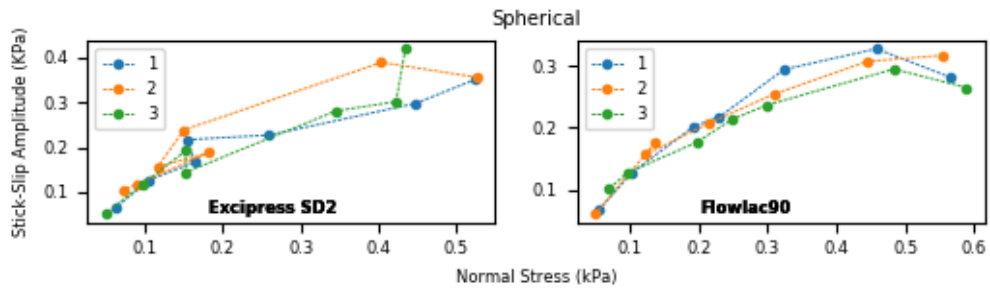

# Stick time vs normal stress

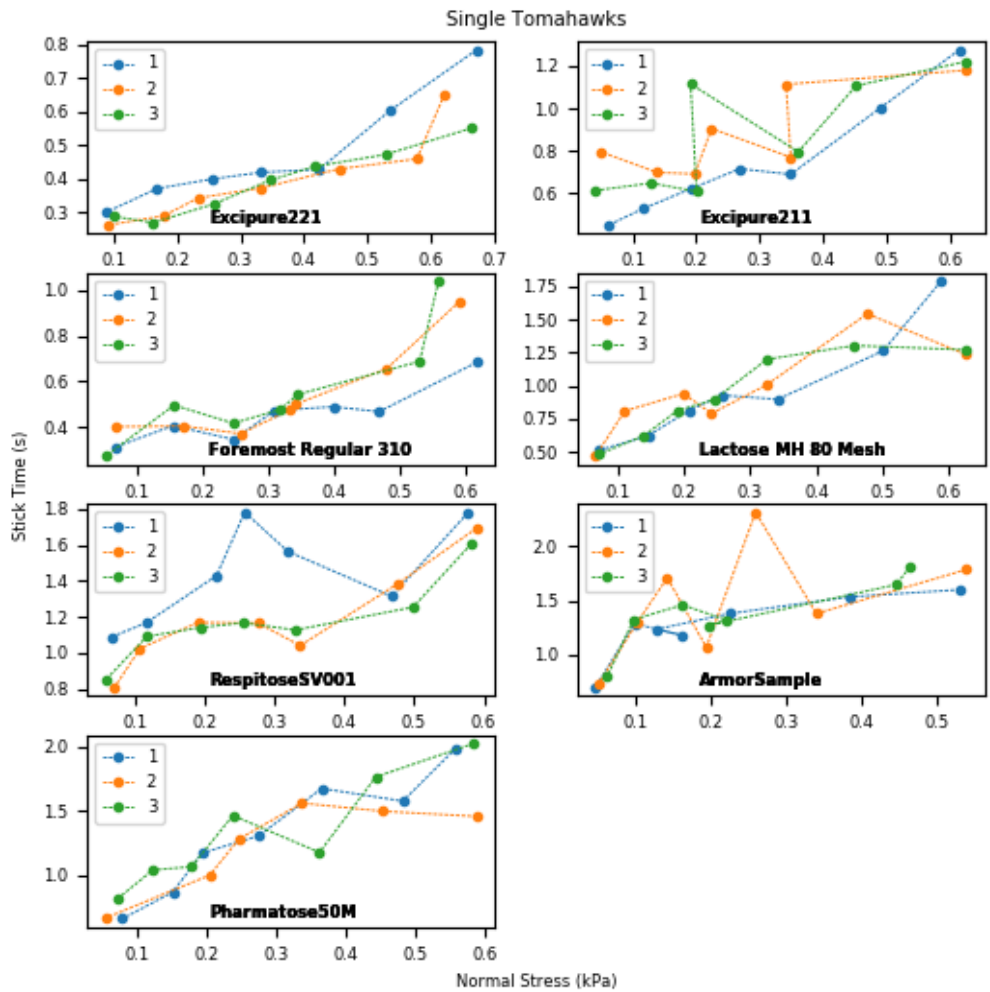

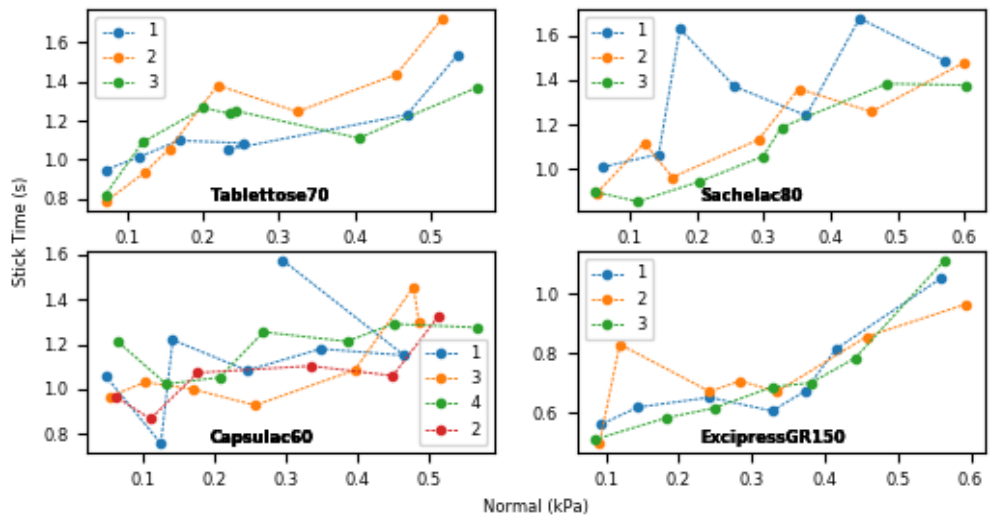
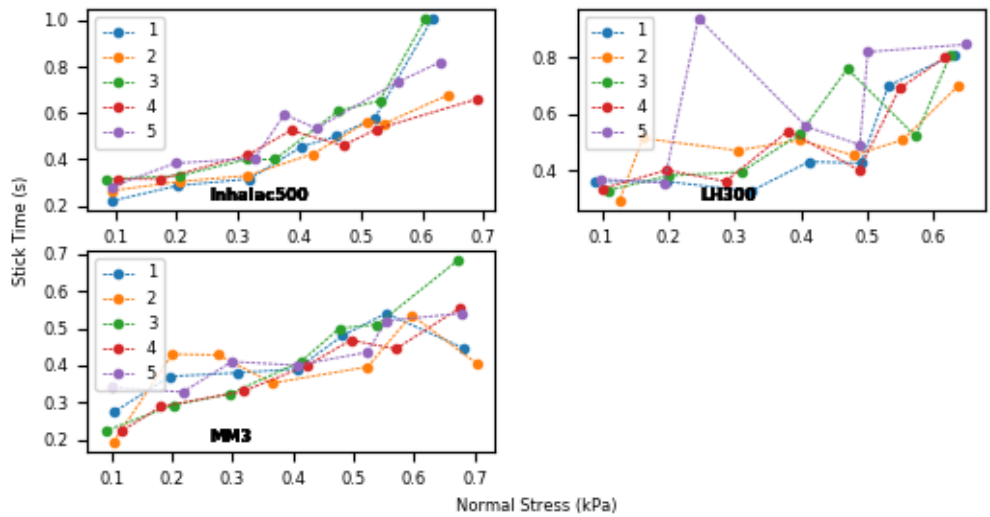
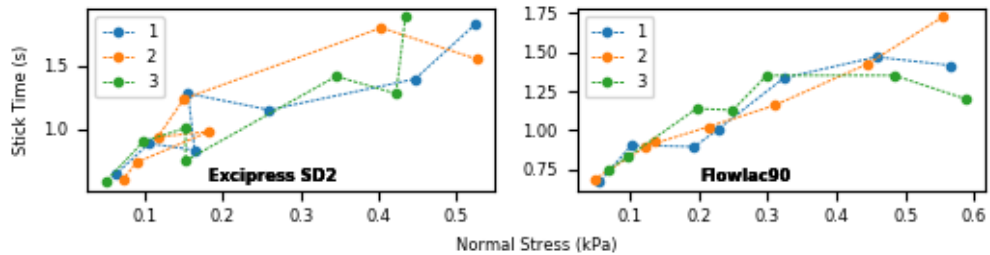

# Slip time vs normal stress

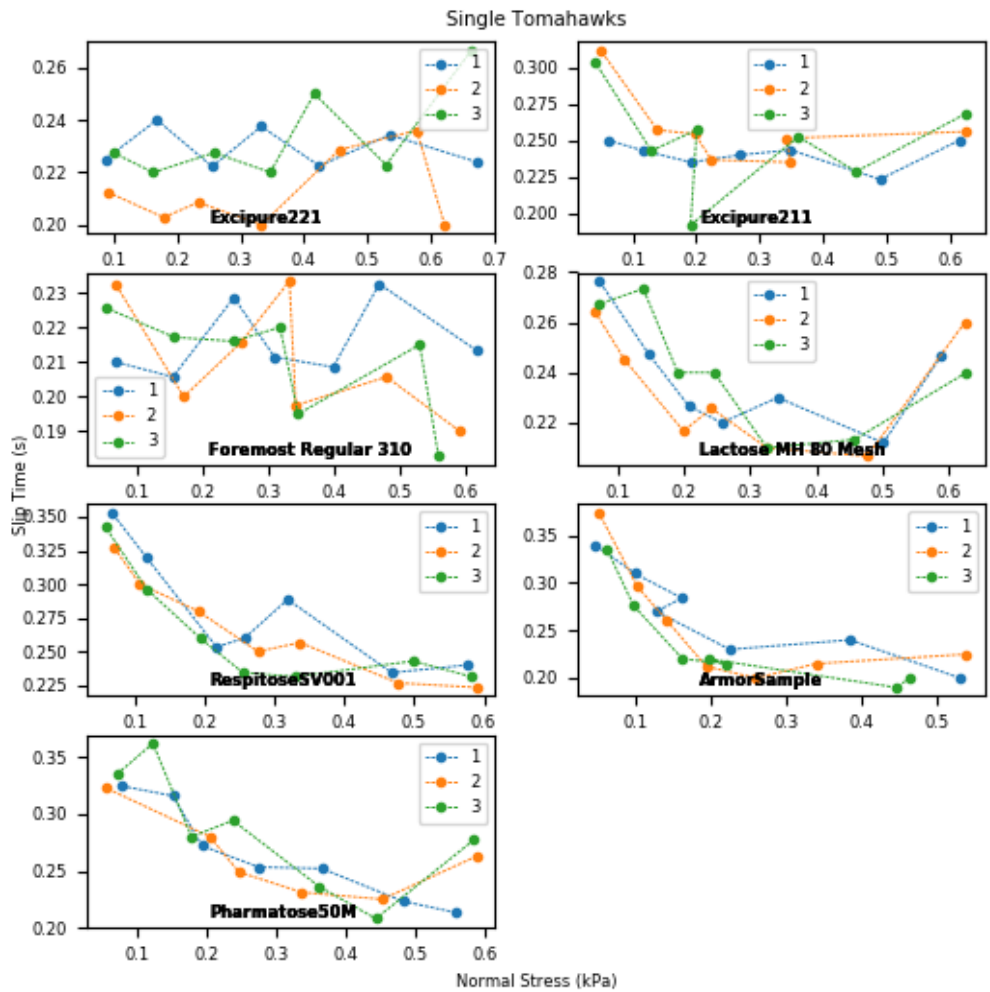

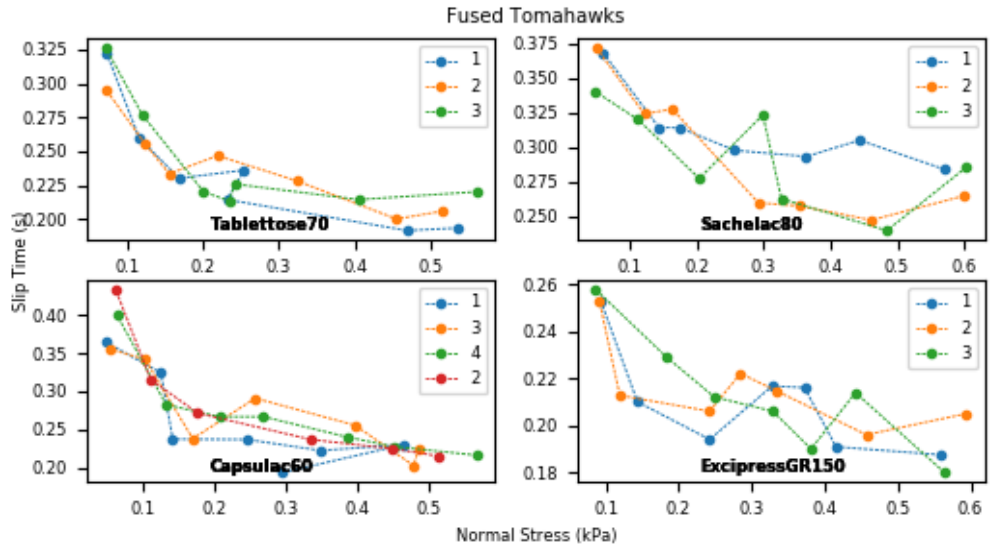
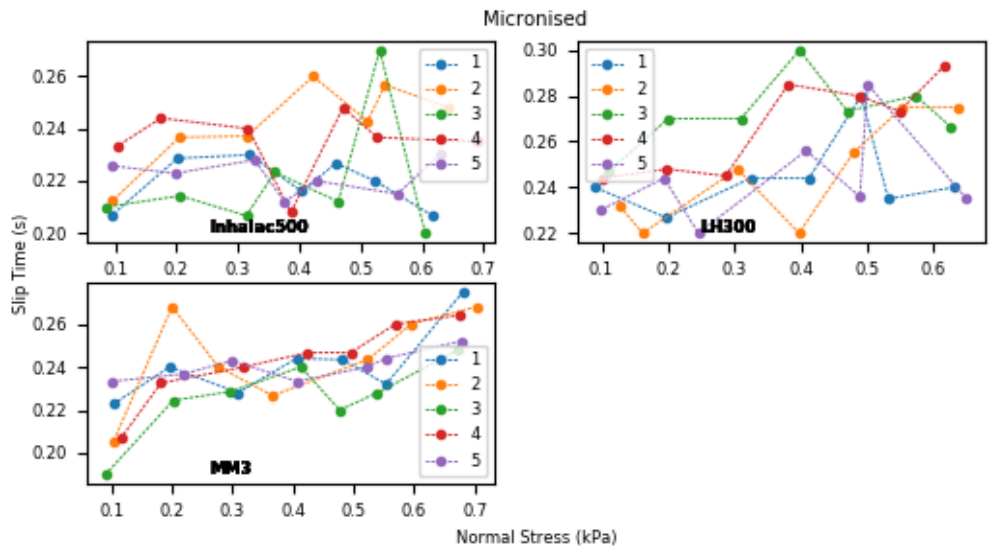
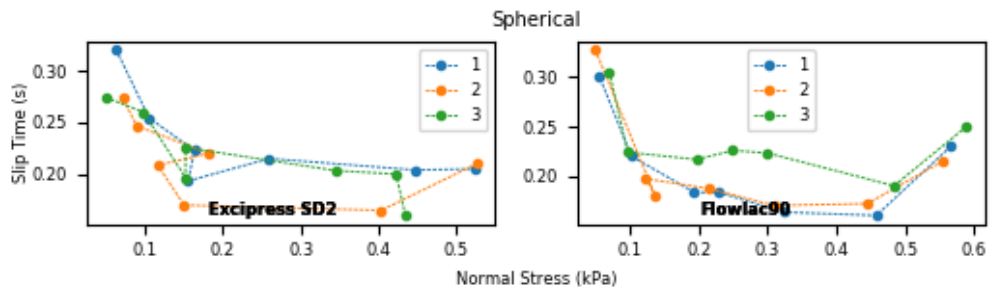

## Mohr circles and related flowability parameters

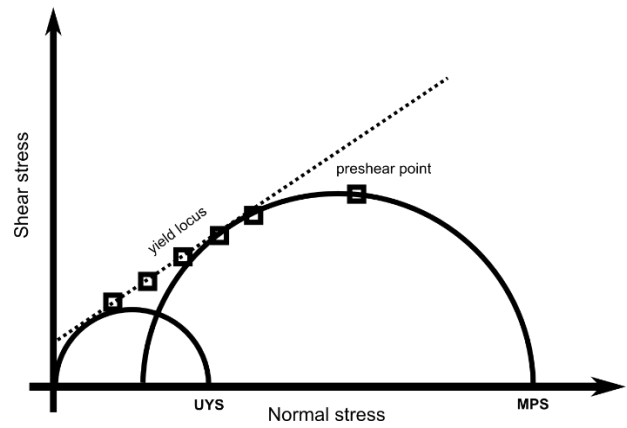

| Product Name | UYS (kPa) | MPS (kPa) | FF |
|---|---|---|---|
| Inhalac500 | 2.75 ± 0.03 | 2.59 ± 0.02 | 0.94 ± 0.02 |
| Lacto-Hale 300 | 2.75 ± 0.03 | 2.98 ± 0.06 | 1.08 ± 0.03 |
| Lacto-Sphere MM3 | 2.40 ± 0.03 | 2.71 ± 0.01 | 1.13 ± 0.02 |
| Excipure DPI 221 | 1.56 ± 0.02 | 2.41 ± 0.01 | 1.54 ± 0.02 |
| Foremost 310 regular NFLM | 0.68 ± 0.02 | 2.07 ± 0.01 | 3.04 ± 0.07 |
| Excipure DPI 211 | 0.47 ± 0.02 | 1.83 ± 0.02 | 3.89 ± 0.21 |
| Excipress SD2 150 | 0.40 ± 0.01 | 1.74 ± 0.03 | 4.35 ± 0.36 |
| FlowLac 90 | 0.34 ± 0.03 | 1.75 ± 0.02 | 5.15 ± 0.39 |
| Excipress GR 150 | 0.42 ± 0.04 | 2.07 ± 0.02 | 4.93 ± 0.33 |
| Tablettose 70 | 0.35 ± 0.01 | 2.20 ± 0.30 | 6.28 ± 1.28 |
| Lactose Monohydrate 80 Mesh | 0.42 ± 0.02 | 1.79 ± 0.02 | 4.27 ± 0.25 |
| Respitose SV 001 | 0.33 ± 0.01 | 2.00 ± 0.03 | 6.06 ± 0.58 |
| Excipure DPI 111 | 0.41 ± 0.02 | 2.04 ± 0.03 | 4.97 ± 0.41 |
| Capsulac 60 | 0.17 ± 0.04 | 2.54 ± 0.03 | 14.94 ± 2.87 |
| SacheLac 80 | 0.32 ± 0.02 | 2.01 ± 0.04 | 6.28 ± 0.58 |
| Pharmatose 50M | 0.27 ± 0.03 | 1.80 ± 0.05 | 6.67 ± 1.35 |

Unconfined yield strength (UYS), major principal stress (MPS) and flow function (FF) set of data at 1 kPa powder pre-shear.

## Additional evidence of stick-slip oscillations

The existence of a genuine stick-slip effect was challenged in several ways. First, to ensure that the latter is indeed triggered by the powder shearing and not by a mechanical regulation mechanism of the vertical motion of the piston/shear-head itself, a purely compressive test was performed and compared with a standard shear test. Capsulac 60 has been selected for this purpose. The compression was performed at two different normal stress levels typically employed for the shear stress tests.

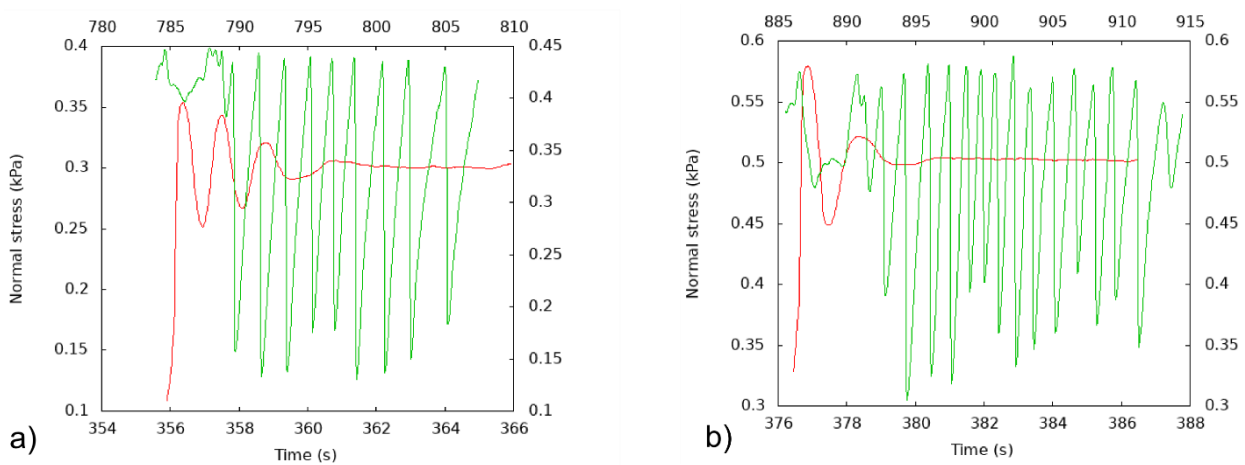

*Figure 1. Normal stress recorded during a compressibility test (red curve) and during a shear test (green curve) for Capsulac 60 at two different normal loads. Right and top tics refer to the shear profile, while left and bottom ticks refer to the compressibility test.*

From Figure 1 it can be evinced that, during compression at a given load, the piston of the FT4 is able to reach and maintain such load value after an initial transient due to the adjustment phase. By contrast, the shear stress imposed during a shear test generates a well recognizable sawtooth profile.

Moreover, a shear stress test was performed on limestone CRM116, a reference powder with well-known behavior. This powder was studied for example in ref. "The study of pharmaceutical powder mixing through improved flow property characterization and tomographic imaging of blend content uniformity" (by Brian Armstrong, university of Birmingham). In this study no significant slip-stick motion was detected. We performed shear tests at pre-shearing 9 kPa and 3 different yield locus points (4,5,6 kPa normal load), as in the reference paper. Typical results are reported in Figure 2: in this case, as expected, sawtooth features are absent in the shear response pattern. This is a further evidence that the experimental setup used in this work correctly describes the shear behavior of those powders not expected to undergo stick-slip motion.

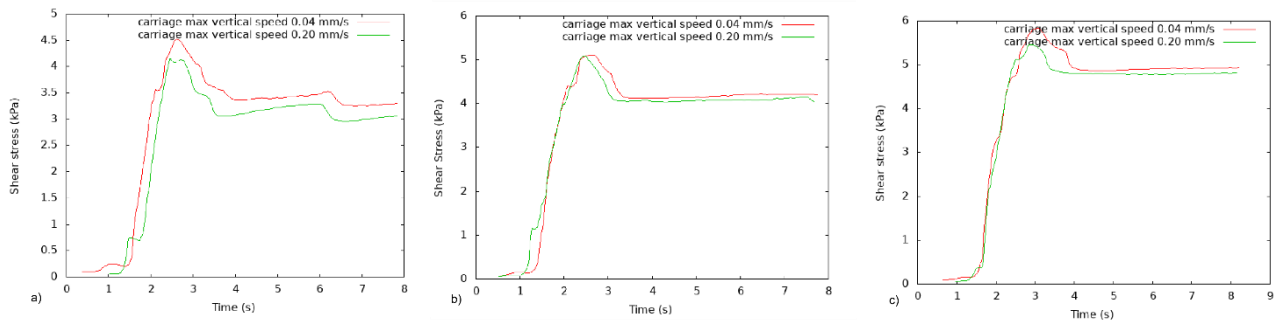

*Figure 2. Shear stress profile on CRM116 limestone at 4 (a), 5 (b), 6 (c) kPa target normal stress with pre-shearing at 9kPa. The influence of the shear head speed on the measurements can be appreciated comparing red and green lines.*

Only occasionally (see fig. 2a) a step in the profile appears after failure, but this is ascribed to local failures of the powder rather than to a true slip-stick oscillating mechanism.

Finally, we tested the influence of the maximum normal motion speed of the shear head on the stress profile. This test is meant to assess a possible influence of the adjustment time of the shear head during the experiment. It appears that the shear head adjustment speed has very little influence on the curves, pointing to the fact that the stick-slip behavior reported in this paper is rather linked to an actual powder property than to spurious effect of the powder rheometer.